\documentclass[fleqn,usenatbib]{mnras}

\usepackage{newtxtext,newtxmath}

\usepackage[T1]{fontenc}

\DeclareRobustCommand{\VAN}[3]{#2}
\let\VANthebibliography\thebibliography
\def\thebibliography{\DeclareRobustCommand{\VAN}[3]{##3}\VANthebibliography}

\usepackage{graphicx}	
\usepackage{amsmath}	
\usepackage{subcaption}
\usepackage{xcolor}
\title[Evolutionary sequence in IRAS\,00183-71]{The link between galaxy merger, radio jet expansion and molecular outflow in the ULIRG IRAS\,00183-7111}

\author[I. Ruffa et al.]{
Ilaria Ruffa,$^{1,2,3}$\thanks{E-mail: ruffai@cardiff.ac.uk}
Marilena Spavone,$^{4}$
Enrichetta Iodice,$^{4}$
Santiago Garcia-Burillo,$^{5}$
Timothy A. Davis,$^{1}$
\newauthor
Kazushi Iwasawa,$^{6,7}$
Henrik W. W. Spoon,$^{8}$
Rosita Paladino,$^{3}$
Michele Perna,$^{9}$
Cristian Vignali,$^{10,11}$
\newauthor
and Stanislav S. Shabala$^{12}$
\\
$^{1}$Cardiff Hub for Astrophysics Research \&\ Technology, School of Physics \&\ Astronomy, Cardiff University, Queens Buildings, The Parade, Cardiff, CF24 3AA, UK\\
$^{2}$INAF, Arcetri Astrophysical Observatory, Largo Enrico Fermi 5, I-50125 Florence, Italy\\
$^{3}$INAF - Istituto di Radioastronomia, via P.\ Gobetti 101, 40129 Bologna, Italy\\
$^{4}$INAF–Astronomical Observatory of Capodimonte, Salita Moiariello 16, 80131 Naples, Italy\\
$^{5}$Observatorio de Madrid, OAN-IGN, Alfonso XII, 3, 28014 Madrid, Spain\\
$^{6}$Institut de Ciències del Cosmos (ICCUB), Universitat de Barcelona (IEEC-UB), Martí i Franquès, 1, 08028 Barcelona, Spain\\
$^{7}$ICREA, Pg. Lluís Companys 23, 08010 Barcelona, Spain\\
$^{8}$Cornell Center for Astrophysics and Planetary Science, Ithaca, NY 14853, USA\\
$^{9}$Centro de Astrobiología (CAB), CSIC-INTA, Departamento de Astrofísica, Cra. de Ajalvir Km. 4, 28850 Torrejón de Ardoz, Madrid, Spain\\
$^{10}$Dipartimento di Fisica e Astronomia ``Augusto Righi", Università degli Studi di Bologna, via Gobetti 93/2, 40129 Bologna, Italy\\
$^{11}$INAF - Osservatorio di Astrofisica e Scienza dello Spazio di Bologna, via Gobetti 93/3, 40129 Bologna, Italy\\
$^{12}$School of Natural Sciences, University of Tasmania, Private Bag 37, Hobart, Tasmania 7001, Australia
}

\date{Accepted XXX. Received YYY; in original form ZZZ}

\pubyear{2015}

\begin{document}
\label{firstpage}
\pagerange{\pageref{firstpage}--\pageref{lastpage}}
\maketitle

\begin{abstract}
The ultraluminous infrared galaxy (ULIRG) IRAS 00183-7111 ($z=0.328$) is one of the three ULIRGs that are currently known to host an active galactic nucleus (AGN) with small-scale radio jets. We present a detailed study of the link between galaxy merger, AGN ignition, radio jet expansion and kpc-scale molecular outflow in IRAS 00183-7111, using high-resolution Atacama Large Millimeter/sub-millimeter Array (ALMA) observations of the $^{12}$CO(1-0) and $^{12}$CO(3-2) lines and very deep $i$-band VLT Survey Telescope (VST) imaging. The latter allows us to put constraints on the assembly history of the system, suggesting that it formed through a major merger between two gas-rich spirals, likely characterised by a prograde encounter and no older than $\approx2$~Gyr. The recent merger channelled about $(1.5\pm0.3)\times10^{10}$~M\textsubscript{$\odot$} of molecular gas in the central regions of the remnant, as traced by the CO detections. The spatial correlation between the CO distribution and the radio core suggests that this gas likely contributed to the ignition of the AGN and thus to the launch of the radio jets. Furthermore, by comparing the relative strength of the two CO transitions, we find extreme gas excitation (i.e.\,$T_{\rm ex}\gg50$~K) around the radio lobes, supporting the case for a jet-ISM interaction. A qualitative study of the CO kinematics also demonstrates that, despite the overall disturbed dynamical state with no clear signs of regular rotation, at least one non-rotational kinematic component can be identified and likely associated to an outflow with $v_{\rm out}\approx439$~km~s$^{-1}$ and $\dot{M_{\rm out}}\approx 609$~M$_{\odot}$~yr$^{-1}$.
\end{abstract}




\section{Introduction}\label{sec:intro}
Galaxy formation theories still struggle to shed light on the mechanisms that determine the observed properties and scaling relations of early-type galaxies (ETGs) in the local Universe (see, e.g., \citealp{Ruffa24}, for a recent review). Fundamental insights into this matter can come from the study of a particular class of objects, known as ultra-luminous infrared galaxies (ULIRGs). ULIRGs are so-called due to their huge infrared luminosity \citep[$L_{\rm 8-1000\mu m}>10^{12}$~L$_{\odot}$;][]{Sanders96}. The most widely accepted scenario suggests that these objects represent a specific stage in an evolutionary sequence which starts from the merger of two gas-rich spirals \citep[see e.g.][]{Alexander12}. The merger channels large masses of gas and dust towards the nuclei, triggering both a powerful starburst and a deeply obscured active galactic nucleus (AGN). The radiation from starburst and super-massive black hole (SMBH) accretion is reprocessed by the surrounding dust, making the system extremely bright at infrared wavelengths and thus giving rise to the ULIRG phase. The AGN successively starts blowing away the surrounding material, emerging as a bright quasar and eventually quenching both star formation and SMBH accretion. The galaxy then evolves as a passive spheroid \citep[e.g.][]{Lipari03}. The tight fundamental scaling relations of today's ETGs \citep[such as the Faber-Jackson relation;][]{Faber76} are believed to originate from this evolutionary process \citep[e.g.][]{Murray05}. ULIRGs (whose high-redshift counterparts are known as submillimeter galaxies, SMGs; \citealp[e.g.][]{Swinbank14}) are thus likely the progenitors of local ETGs, and also provide an ideal laboratory to study the mutual interplay between central SMBHs and their host galaxies. However, the many details of how these sources form and evolve have yet to be fully understood. 

IRAS F00183-7111 (I00183, hereafter) is a unique example of an ULIRG located at redshift $z=0.328$ \citep[][]{Calderon16}. It is one of the most luminous known (with $L_{\rm bol}=9\times10^{12}$~L$_{\odot}$, mostly radiated in the infrared; \citealp{Spoon09}), and is optically classified as a type 2 (i.e. obscured) Seyfert \citep{Armus89}. Detailed studies show that its mid-infrared (MIR; i.e. 5$-$37~$\mu$m) spectrum is heavily extincted, thus indicating the presence of dense material surrounding the nucleus \citep{Spoon04}. Nevertheless, the $11.3$~$\mu$m PAH feature is clearly detected and suggests that $\sim$10\% of the infrared luminosity arises from star formation \citep[][]{Spoon09}. This indicates that the primary nuclear powering source of I00183 is an AGN \citep{Spoon07}. The presence of an AGN in this source is confirmed by the overall X-ray (2-10~keV) spectral features, including the detection of a strong 6.7~keV iron emission line \citep[see][]{Nandra07,Ruffa18}. The MIR spectral properties, when compared with the so-called {\it Spoon} (or {\it Fork}) diagnostic diagram \citep{Spoon07}, suggest that the source has been caught in the brief transition period between a starburst-dominated ULIRG and the unobscured quasar phase. Further evidence in this regard comes from the detection of prominent gas outflows, which are believed to trace the disruption of the thick nuclear obscuring layers in ULIRGs. These have been already confirmed for I00183 in a number of ionized gas components (e.g.\,[Ne{\sc ii}], [Ne{\sc iii}] and [O{\sc iii}]$\lambda5007$), showing strongly asymmetric and blueshifted line profiles (with velocities up to $\approx3000$~km~s$^{-1}$) and extending up to 10$''$ ($\approx$50~kpc) to the east of the nucleus \citep[see e.g.][]{Drake04,Spoon09,Iwasawa17}. \citet{Calderon16} reported also the presence of a molecular gas outflow, as traced by a 119~$\mu$m OH doublet absorption with typical high-velocity (i.e.\,$v\approx1500$~km~s$^{-1}$) blueshifted wings. Furthermore, multi-frequency radio continuum observations of I00183 show the presence of a core-double lobe radio source with a total linear extent of $\approx 2$~kpc, a high radio power ($L_{\rm 2.3~GHz}\sim10^{25.8}$~W~Hz$^{-1}$), and a very steep 1-80~GHz spectrum ($\alpha=-1.4$, for $S\propto \nu^{\alpha}$; \citealp{Norris12}). Based on well-known correlations between the radio source size, spectral properties and age \citep[e.g.][]{ODea20}, the radio source in I00183 may thus likely be in an early phase of its evolution. This further confirms the transitional state of this system. The rarity of objects like I00183 attests the brevity of this period: to date, there are only other two ULIRGs known to host small-scale (likely young) radio sources, i.e.\,PKS1345+12 \citep{Lister03,Fotopoulou19}, and PKS1549-79 \citep{Holt06}. However, I00183 is the most extreme among them in terms of both infrared and radio luminosity. It thus provides a unique chance to test theoretical models of galaxy formation and evolution. 
A summary of the general properties of I00183 is provided in Table~\ref{tab:I00183_properties}.

In this paper, we present newly-acquired Atacama large millimeter/submillimeter array (ALMA) observations of the $^{12}$CO(1-0) line and 94~GHz continuum, along with very deep $i$-band VLT Survey Telescope (VST) imaging of I00183. These data have resolutions of $0.22\arcsec$ ($\approx1$~kpc) and $0.85\arcsec$ ($\approx4$~kpc), respectively, and are analysed in combination with archival ALMA data of the $^{12}$CO(3-2) line and 250~GHz continuum. Such multi-wavelength analysis allows us to explore the link between galaxy merger, AGN activity, radio jet expansion and galactic-scale (i.e. $\sim$kpc-scale) molecular outflows, for the first time with a high level of detail in an ULIRG at $z>0.3$. The paper is structured as follows. In Section~\ref{sec:obs} we describe ALMA and VST observations and the data reduction. We present the CO analysis in Section~\ref{sec:cube_analysis}, and the photometric analysis of the VST data in Section~\ref{sec:phot}. We discuss the results in Section~\ref{sec:discussion}, before summarising and concluding in Section~\ref{sec:conclusions}.

To allow a coherent comparison with the work of \citet{Ruffa18}, where lower-resolution ALMA $^{12}$CO(1-0) data of I00183 were presented, throughout our analysis we assume a $\Lambda$CDM cosmology with H$_{\rm 0}=70$\,km\,s$^{-1}$\,Mpc$^{\rm -1}$, $\Omega_{\rm \Lambda}=0.7$ and $\Omega_{\rm M}=0.3$. This gives a scale of $4.733$~kpc/$''$ at the redshift assumed for I00183 (see Table~\ref{tab:I00183_properties}).

\begin{table}
\centering
\caption{General properties of I00183.}
\label{tab:I00183_properties}
\begin{tabular}{l l c c}
\hline
\multicolumn{1}{c}{ ID } &
\multicolumn{1}{c}{ Parameter} &
\multicolumn{1}{c}{ Value } \\
\hline
(1) & Type & ULIRG\\
(2) & RA (ICRS) & 00$^{\rm h}$20$^{\rm m}$34$^{\rm s}$.69\\
(3) & DEC (ICRS) & -70$^{\circ}$55$^{'}$26$^{''}$.69\\
(4) & Redshift &  $0.3282\pm0.0005$ \\
(5) & D$_{\rm L}$ (Mpc) &  1722 \\
 (6) & log~$P_{\rm 2.3GHz}$ (W\,Hz$^{-1}$) & 25.8   \\
 (7) & $v_{\rm opt}$ (km~s$^{-1}$) & $98022\pm20$ \\
\hline
\end{tabular}
\parbox[t]{8.5cm}{ \textit{Notes.} $-$ Rows: (1) Galaxy type. (2) and (3) Preferred sky coordinates of the optical galaxy centre as reported in the NASA/IPAC Extragalactic Database (NED; see also Section~\ref{sec:VST_obs}). (4) Galaxy redshift from \citet[][]{Calderon16}. (5) Luminosity distance derived from the tabulated redshift and assuming the cosmology stated in Section~\ref{sec:intro}. (6) Radio power at 2.3~GHz taken from \citet{Norris12}, and including all the radio emission associated with the source. (7) Optical systemic velocity of the galaxy.}
\end{table}

\begin{table*}
\begin{small}
\setlength{\tabcolsep}{5pt} 
\centering
\caption{Main properties of the ALMA observations presented in this paper.}
\label{tab:ALMA observations summary}
\begin{tabular}{l l c c c c c c c c c c}
\hline
\multicolumn{1}{c}{ ALMA } &
\multicolumn{1}{c}{ Observation } &
\multicolumn{1}{c}{ Targeted } & 
\multicolumn{1}{c}{ $\nu$\textsubscript{rest}} &
\multicolumn{1}{c}{ $\nu$\textsubscript{sky}} &
\multicolumn{1}{c}{ Time } & 
\multicolumn{1}{c}{   MRS } & 
\multicolumn{1}{c}{   FOV } & 
\multicolumn{1}{c}{   $\theta$\textsubscript{maj} } & 
\multicolumn{1}{c}{   $\theta$\textsubscript{min}  } & 
\multicolumn{1}{c}{   PA$_{\rm beam}$} & 
\multicolumn{1}{c}{   Scale }\\        
\multicolumn{1}{c}{ Band } &  
\multicolumn{1}{c}{ date } &  
\multicolumn{1}{c}{ line} &   
\multicolumn{1}{c}{    } &      
\multicolumn{1}{c}{   } &          
\multicolumn{1}{c}{  } &          
\multicolumn{1}{c}{  } &
\multicolumn{1}{c}{  } &
\multicolumn{2}{c}{  } &
\multicolumn{1}{c}{    } &
\multicolumn{1}{c}{   } \\
\multicolumn{1}{c}{ } &  
\multicolumn{1}{c}{ } &  
\multicolumn{1}{c}{ } &   
\multicolumn{1}{c}{   (GHz)  } &      
\multicolumn{1}{c}{   (GHz) } &          
\multicolumn{1}{c}{   (min)} &          
\multicolumn{1}{c}{   (kpc) (arcsec)} & 
\multicolumn{1}{c}{   (kpc) (arcsec)} & 
\multicolumn{2}{c}{ (arcsec) } &   
\multicolumn{1}{c}{   (deg) } &
\multicolumn{1}{c}{   (kpc)} \\
\multicolumn{1}{c}{   (1) } &   
\multicolumn{1}{c}{   (2) } &
\multicolumn{1}{c}{   (3) } &
\multicolumn{1}{c}{   (4) } & 
\multicolumn{1}{c}{   (5) } &   
\multicolumn{1}{c}{   (6) } &
\multicolumn{1}{c}{   (7) } &               
\multicolumn{1}{c}{   (8) } &
\multicolumn{1}{c}{   (9) } &
\multicolumn{1}{c}{   (10) } &
\multicolumn{1}{c}{   (11) } &
\multicolumn{1}{c}{   (12) } \\

\hline
 Band 3  &  2021-07-09/16 & $^{12}$CO(1-0)  & 115.271 & 86.735 & 166 &  33 (6.8)  &  294 (62) &  0.4  & 0.3  & -18 & 1.9 \\ 
 \hline
 Band 6 &    2015-09-01    & $^{12}$CO(3-2)  & 345.796  & 260.192  &    34     &    13 (2.7)    &  109 (23) &  0.3    &    0.2    &    21    &    1.4   \\
\hline
\end{tabular}
\parbox[t]{1\textwidth}{ \textit{Notes.} $-$ Columns: (1) ALMA frequency band. (2) Date of the observation. (3) Targeted molecular transition. (4) Rest-frame central line frequency. (5) Redshifted (sky) central frequency of the line estimated from the redshift listed in row (4) of Table~\ref{tab:I00183_properties}. (6) Total time-on-source of the observation. (7) Maximum recoverable scale (MRS) in kiloparsec, and corresponding scale in arcseconds in parentheses. (8) Field-of-view (FOV; i.e.\,primary beam FWHM) in kiloparsec, and corresponding scale in arcseconds in parentheses. (9) and (10) Major and minor axis FWHM of the synthesized beam. (11) Position angle of the synthesized beam. (12) Physical scale corresponding to the major axis FWHM of the synthesized beam.}
\end{small}
\end{table*}

\begin{figure}
\begin{subfigure}[t]{0.3\textheight}
\caption{Band 3 continuum}\label{fig:cont_band3}
\includegraphics[clip=true, trim={30 30 40 30}, scale=0.24]{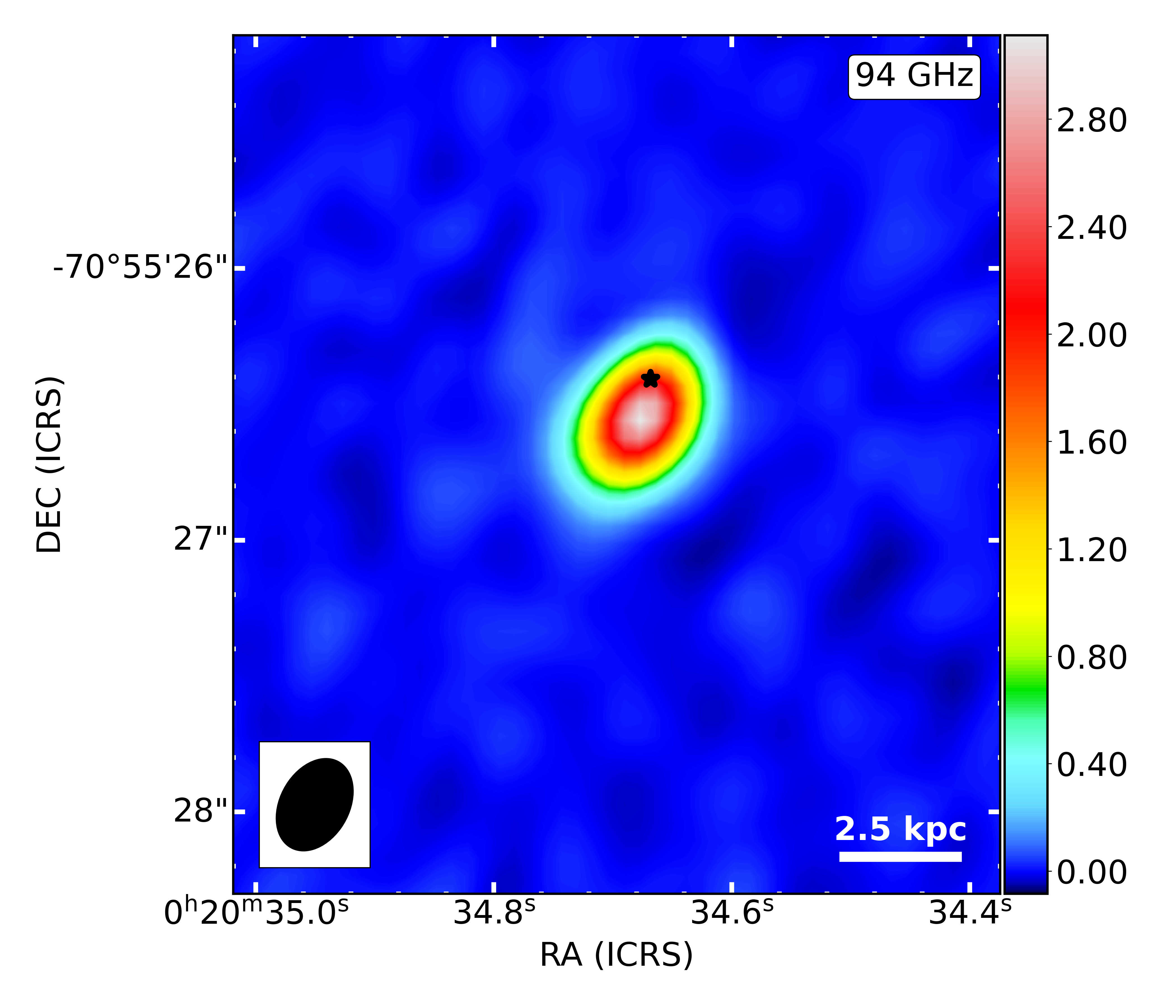}
\end{subfigure}
\medskip
\begin{subfigure}[t]{0.3\textheight}
\caption{Band 6 continuum}\label{fig:cont_band6}
\includegraphics[clip=true, trim={30 30 40 30}, scale=0.24]{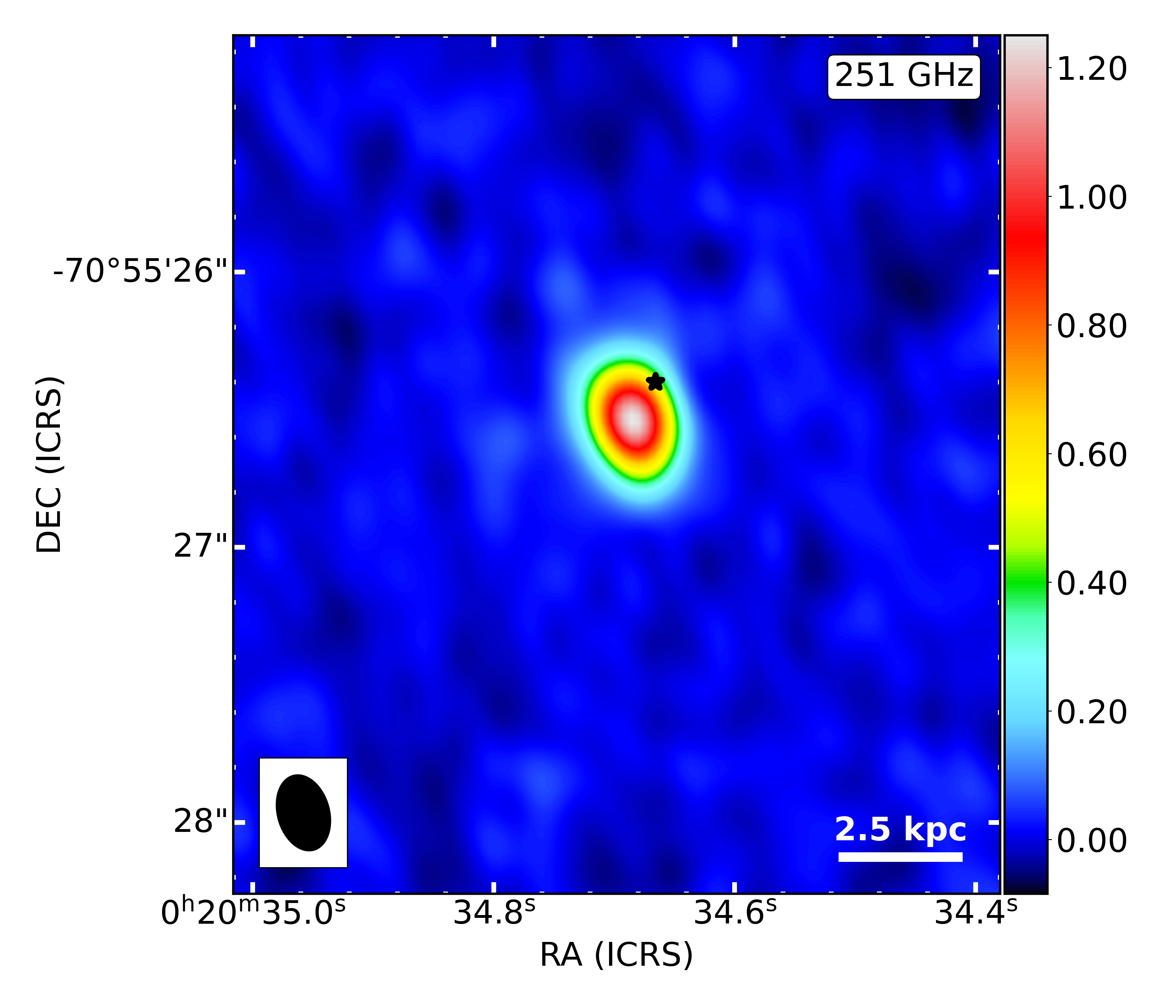}
\end{subfigure}
\caption[]{Continuum maps of I00183 in ALMA Band 3 (panel a) and 6 (panel b). The reference frequency of each map is indicated in its top-right corner. The bar to the right of each panel shows the colour scale in mJy~beam$^{-1}$. The synthesised beam and a scale bar are shown in the bottom-left and bottom-right corners, respectively. In both panels, the image centre is set to the preferred sky coordinates of the optical galaxy centre, which are reported in Table~\ref{tab:I00183_properties}; East is to the left and North to the top. The black star indicates the position of the radio core, as inferred from the very-long baseline interferometry (VLBI) observations of I00183 presented in \citet[][see also \citealp{Ruffa18} and Section~\ref{sec:ratios_results}]{Norris12}. The properties of the continuum maps are summarised in Section~\ref{sec:ALMA_imaging} and discussed in Section~\ref{sec:cont_discuss}}\label{fig:continuum}
\end{figure}

\section{Observations and data reduction}\label{sec:obs}
\subsection{ALMA observations}
The Band 3 ALMA observations presented in this paper were carried out during Cycle 7, between 9 and 16 July 2021 (project code: 2019.1.00851.S; PI: I.\,Ruffa). I00183 was observed for a total time-on-source of 166~min, targeting the $^{12}$CO(1-0) transition (rest frequency, $\nu_{\rm rest}=115.271$~GHz) and the $\sim100$~GHz continuum. The spectral configuration consisted of a total of four spectral windows (SPWs): one centred at the line redshifted frequency ($\nu_{\rm sky}=86.735$~GHz) and divided into 1920 (0.976~MHz-wide) channels, the other three used to map the continuum emission and divided into 480 (3.906-MHz-wide) channels. A total of 44 12-m antennas were used, with a maximum baseline length of 3.6~km.
 
To get a more detailed view of the sub-mm line and continuum properties, we include in this work also unpublished Band 6 ALMA observations of the $^{12}$CO(3-2) transition ($\nu_{\rm rest}=345.796$~GHz) and $\sim250$~GHz continuum in I00183 (project code: 2013.1.00826.S; PI: R.\,Norris). These data were obtained during ALMA Cycle 2, and acquired on 1 September 2015. The total time-on-source is 34~min. The spectral configuration consisted of a total of four SPWs: one centred at the line redshifted frequency ($\nu_{\rm sky}=260.192$~GHz), the other three used to map the continuum emission. All the SPWs were divided into 128 (16~MHz-wide) channels. A total of 35 12-m antennas were used, with a maximum baseline length of 1.6~km. A summary of the basic properties of the ALMA observations presented in this paper is reported in Table~\ref{tab:ALMA observations summary}.
 
 For both the Band 3 and 6 data, a standard calibration strategy was adopted, using a single bright quasar as both flux and bandpass calibrator, and a second one as phase calibrator. The calibrated Cycle 7 (Band 3) data were provided upon request by the ALMA Regional Centre (ARC) at the European Southern Observatory (ESO) Headquarters. We instead manually calibrated the Cycle 2 (Band 7) data using the Common Astronomy Software Application \citep[{\sc CASA};][]{McMullin07} package (version 4.7.2), with standard data reduction scripts. We carried out the imaging and analysis of all the calibrated data as described below, using the more recent {\sc CASA} version 6.1.2.

 \subsubsection{Continuum and line imaging}\label{sec:ALMA_imaging}
 The continuum SPWs and the line-free channels of the line SPWs were used to produce the continuum maps, using the \texttt{tclean} task in multi-frequency synthesis (MFS) mode \citep{Rau11}. In all the cases, Briggs weighting with a robust parameter of 0.5 was adopted, thus obtaining the best trade-off between sensitivity and resolution. This allowed us to achieve root-mean square (rms) noise levels of 18~$\mu$Jy~beam$^{-1}$ and 20~$\mu$Jy~beam$^{-1}$ on the 100~GHz and 250~GHz maps, respectively, for synthesized beam sizes of $0.38\arcsec \times 0.28\arcsec$ ($\approx 1.80 \times 1.30$~kpc; Band 3) and $0.29\arcsec \times 0.21\arcsec$ ($\approx 1.40 \times 0.99 $~kpc; Band 6) full-width at half-maximum (FWHM). The cleaned, primary-beam corrected continuum maps are shown in Figure~\ref{fig:continuum} and discussed in Section~\ref{sec:cont_discuss}.

\begin{figure*}
    \centering
    \includegraphics[clip=true, trim={15 20 15 15}, scale=0.65]{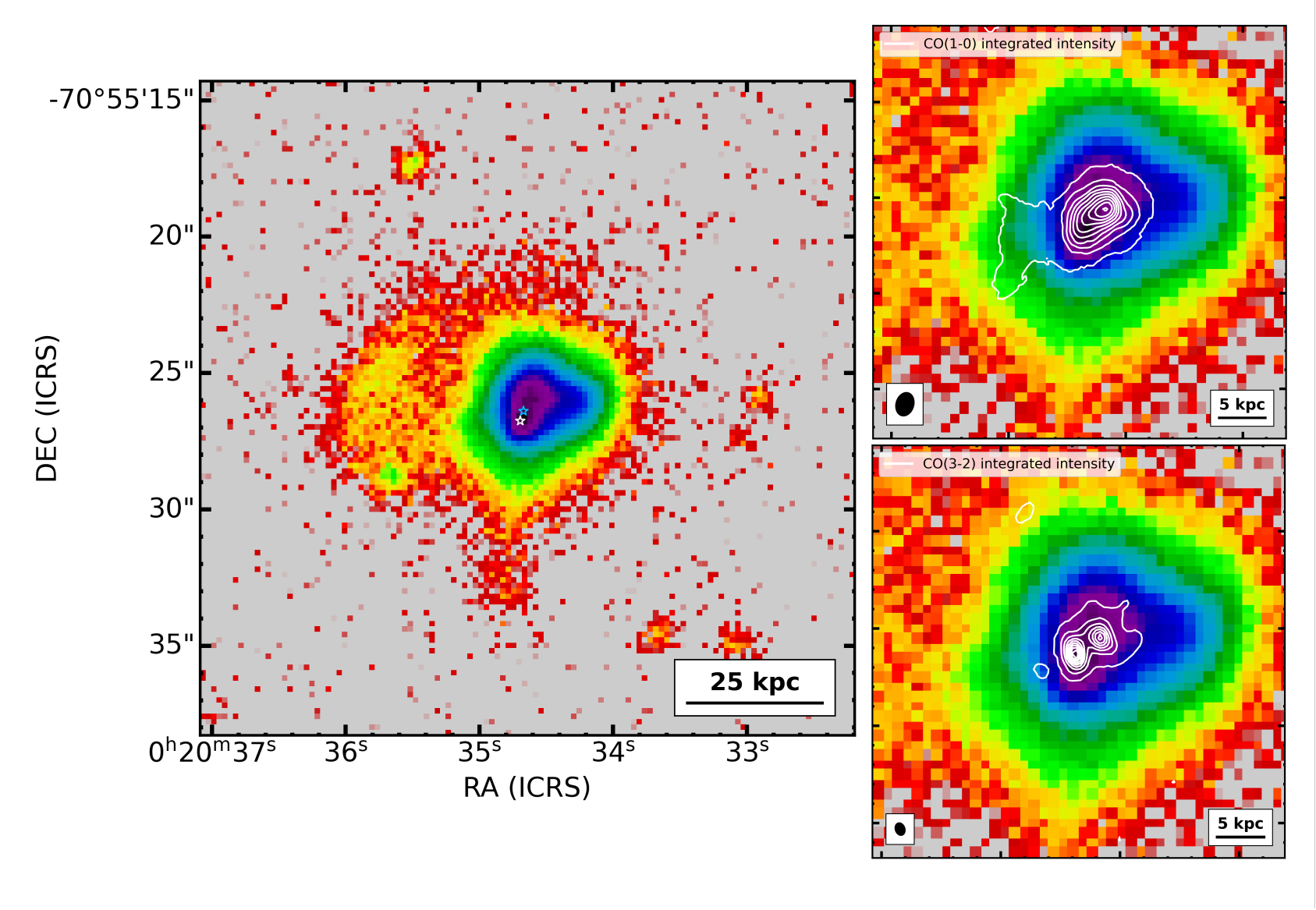}
    \caption{{\bf Left panel:} VST {\it i}-band ($770$~nm) sky-subtracted image of the $25\arcsec \times 25\arcsec$ ($\approx120\times120$~kpc$^{2}$) around I00183. The image resolution (i.e.\,seeing FWHM) is $\approx0.85\arcsec$, and the reached surface brightness depth is $\mu_{i} \sim$ 29.5 mag~arcsec$^{-2}$. The white star indicates the preferred position assumed so far for the optical galaxy centre (corresponding to the coordinates reported in Table~\ref{tab:I00183_properties}), while the light-blue star indicates the position of the radio core as inferred from the very-long baseline interferometry (VLBI) observations presented in \citet{Norris12} (see also Section~\ref{sec:ratios_results}). {\bf Right panel, top:} Zoom-in of the VST $i$-band map in the central $8\arcsec \times 8\arcsec$ ($\approx38\times38$~kpc$^{2}$) with overlaid in white CO(1-0) integrated intensity contours from the Cycle 7 ALMA observations presented in this paper. {\bf Right panel, bottom:} As above, but with overlaid CO(3-2) integrated intensity contours from the Cycle 2 ALMA observations presented in this paper. A scale bar is shown in the bottom-right corner of each panel, the CO synthesised beam sizes are shown in the bottom-left corners of the right panels. For each CO transition, we show 10 contours, which are equally spaced between the minimum and the maximum significant values of the integrated intensity map (as illustrated by the colour scale in the left-hand panels of Figure~\ref{fig:CO_moments}).}
    \label{fig:VST_map}
\end{figure*}

The CO line emission was isolated in the $uv$-plane using the {\sc casa} task {\tt uvcontsub}. This estimates a continuum model through a linear fit in line-free channels, and then subtracts this model from the line visibilities. The data cubes were created using the {\tt tclean} task with Briggs weighting (robust=0.5) and a channel width of 20~km~s$^{-1}$ for both the CO(1-0) and CO(3-2) transitions. The channel velocities were computed in the source frame, so that the systemic velocity of each CO data cube corresponds to the respective redshifted central frequency ($\nu_{\rm sky}$; see Table~\ref{tab:ALMA observations summary}). The continuum-subtracted dirty cubes were cleaned in regions of line emission (identified interactively) to a threshold roughly equal to 1.5 times the rms noise level (determined from line-free channels), and then primary-beam corrected. CO(1-0) and CO(3-2) emission is clearly detected (see Figure~\ref{fig:VST_map} and Section~\ref{sec:cube_analysis} for details). Table~\ref{tab:line images} summarises the properties of the obtained data cubes.

\begin{table}
\centering
\caption{Properties of the imaged data cubes.}
\label{tab:line images}
\begin{tabular}{ l c c c c c c}
\hline
\multicolumn{1}{c}{ Transition } &
\multicolumn{1}{c}{ rms } &
\multicolumn{1}{c}{ Peak flux } &
\multicolumn{1}{c}{ S/N  } &
\multicolumn{1}{c}{ $\Delta v_{\rm chan}$  } \\
\multicolumn{1}{c}{  } &
\multicolumn{1}{c}{ (mJy~beam$^{-1}$) } &
\multicolumn{1}{c}{ (mJy~beam$^{-1}$) } &
\multicolumn{1}{c}{  } &
\multicolumn{1}{c}{ (km~s$^{-1}$) } \\
\multicolumn{1}{c}{ (1) } &
\multicolumn{1}{c}{ (2)} &
\multicolumn{1}{c}{ (3) } &
\multicolumn{1}{c}{ (4) } &
\multicolumn{1}{c}{ (5) } \\
\hline
 $^{12}$CO(1-0)  &  0.22  &  1.96  &   9 &   20  \\
 $^{12}$CO(3-2)  &  0.23   &  7.58  &   33  &   20  \\
\hline
\end{tabular}
\parbox[t]{1\columnwidth}{ \textit{Notes.} $-$ Columns: (1) Line transition. (2) Rms noise measured from line-free channels with the widths listed in column (5). (3) Peak flux of the line emission. (4) Peak signal-to-noise ratio of the detection. (5) Channel width of the data cube (km\,s$^{-1}$ in the source frame).}
\end{table}

\subsection{VST observations}\label{sec:VST_obs}
The photometric data presented in this work are part of the VST Early-type GAlaxies Survey (VEGAS\footnote{\url{https://sites.google.com/inaf.it/vegas/home}}, P.I. E. Iodice, \citealt{Capaccioli2015, Iodice2021}). 
VEGAS is a multi-band {\it u}, {\it g}, {\it r} and {\it i} imaging survey carried out with the European Southern Observatory (ESO) VLT Survey Telescope (VST). 
The VST is a 2.6~m wide field optical telescope \citep{Schipani2012} equipped with OmegaCAM, a $1^{\circ} \times 1^{\circ}$ camera with a pixel size of $0.21$~arcsec~pixel$^{-1}$. 
The data we present in this work (run ID: 0110.A-6004(A)) were acquired in the {\it i}-band (covering the rest-frame wavelength range $\sim650-850$~nm), during dark time in photometric conditions, with an average seeing FWHM (i.e. averaged over the seeing FWHM of each observing night) of $\approx0.85\arcsec$. The total exposure time is 6 hours, and the observations were performed using the standard diagonal dithering strategy. The data were processed by using the dedicated {\it AstroWISE} pipeline developed to reduce OmegaCam observations \citep{McFarland2013}. The main steps of the data reduction with AstroWISE are extensively described in \citet{Venhola2017}, and we thus refer the reader to that work for details. We nevertheless briefly summarise in the following the procedure adopted for the astrometric registration, as this is essential information for the analysis carried out in this work. The first-order astrometric calibration was done by first matching the pixel coordinates to RA and Dec using the World Coordinate System (WCS) information from the fits header. Point source coordinates were then extracted using SExtractor \citep[][]{Bertin96} and associated with the 2 Micron All Sky Survey Point Source Catalog (2MASS PSC, \citealt{Skrutskie2006}). To make sure individual random positional errors are well smoothed out, the number of reference stars used ranges between 15 and 1200. The transformation was then extended by a second-order two-dimensional polynomial across the focal plane. SCAMP \citep{Bertin2006} was used for this purpose. The polynomial was fit iteratively five times, each time clipping the $2\sigma$ outliers. The astrometric solution gives typically rms errors of $0.3\arcsec$ (compared to 2MASS PSC) for a single exposure, and $0.1\arcsec$ for the stacked final mosaic.

Another critical step in deep photometry is the sky background subtraction, since it can affect the ability to detect faint structures. In the case presented in this work, since the angular extent of the target is much smaller than the VST field-of-view (FOV), the background subtraction has been performed by fitting a 2D polynomial to the pixel values of the mosaic unaffected by celestial sources or defects (see \citealt{Spavone2017}).

Thanks to the long integration time, the final stacked $i$-band image reaches a surface brightness depth of $\mu_{i} \sim$ 29.5 mag~arcsec$^{-2}$. 
This very deep VST image of I00183 is shown in Figure~\ref{fig:VST_map}, and illustrates with unprecedented detail the optical features characterising this complex system. Particularly interesting are those pointing at a disturbed dynamical state, such as the asymmetry in the South East-North West (SE-NW) direction, the prominent arc-like structure on the East side and the tail to the South. As extensively discussed in Section~\ref{sec:discussion}, all of these are clear signs of a recent merger. The high quality of the VST data also allows us to reveal, for the first time, additional interesting details about the nuclear region of the system. For instance, it is clear from the left panel of Figure~\ref{fig:VST_map} that those typically adopted as preferred sky coordinates for the optical centre of this galaxy (and reported in Table~\ref{tab:I00183_properties}) correspond indeed to the optical peak (as illustrated by the white star). This, however, is shifted slightly southward with respect to the position that - when looking at the purple area in Figure~\ref{fig:VST_map} - one would identify by-eye as the geometrical centre of the system. Nevertheless, the direction of such a shift is consistent with the southern tidal elongation of the system, as well as with the typical asymmetries observed in disturbed objects like ULIRGs \citep[see e.g.][]{Pereira18,Lamperti22}.

Furthermore, the left panel of Figure~\ref{fig:VST_map} illustrates for the first time that the position of the radio core (marked by the light-blue star and inferred from the very-long baseline interferometry (VLBI) observations of I00183 presented by \citealt{Norris12}; see also Section~\ref{sec:ratios_results}) is shifted northward with respect to the optical peak (white star). We estimate that the projected linear separation between the two positions is $\approx0.51\arcsec$ ($\approx2.4$~kpc). This may indicate the presence of two nuclei from the progenitor galaxies which have not yet completely merged. However, as discussed in detail in Section~\ref{sec:phot}, the discrepancy between these two positions and the results from the photometric analysis, as well as the lack of any other evidence for the presence of a double nucleus, make this hypothesis currently not easy to support. Another possible explanation is that such a displacement is caused by the extreme dust obscuration that affects the areas around the central SMBH even at near-infrared wavelengths (see e.g. \citealp{Spoon09,Ruffa18}), thus making the optical peak moving towards the outer shells of the nucleus. We further analyse and discuss these scenarios in Sections~\ref{sec:phot} and \ref{sec:discussion}. We note that, to allow a coherent comparison also with previous works, all the maps presented in this paper have been centred on the optical peak sky coordinates reported in Table~\ref{tab:I00183_properties} (white star in the left panel of Figure~\ref{fig:VST_map}).

\begin{figure*}
\centering
\begin{subfigure}[t]{1.0\textwidth}
\centering
\caption{\textbf{CO(1-0) moments}}\label{fig:CO10_moms}
\vspace{-0.1cm}
\includegraphics[clip=true, trim={30 5 10 30}, scale=0.32]{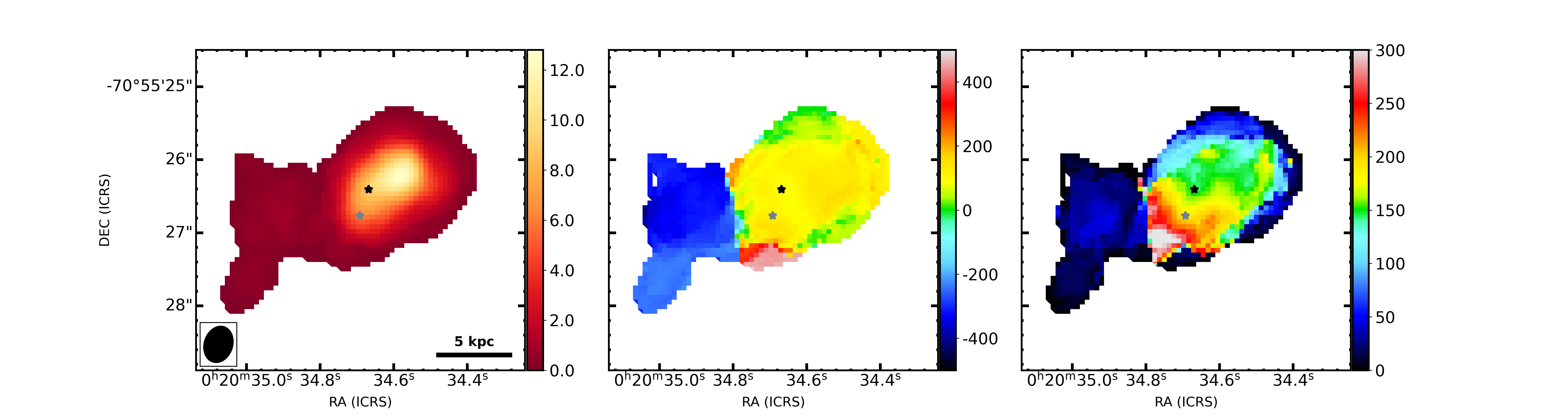}
\end{subfigure}

\medskip

\begin{subfigure}[t]{1.0\textwidth}
\centering
\caption{\textbf{CO(3-2) moments}}\label{fig:CO32_moms}
\vspace{-0.1cm}
\includegraphics[clip=true, trim={30 5 10 30}, scale=0.32]{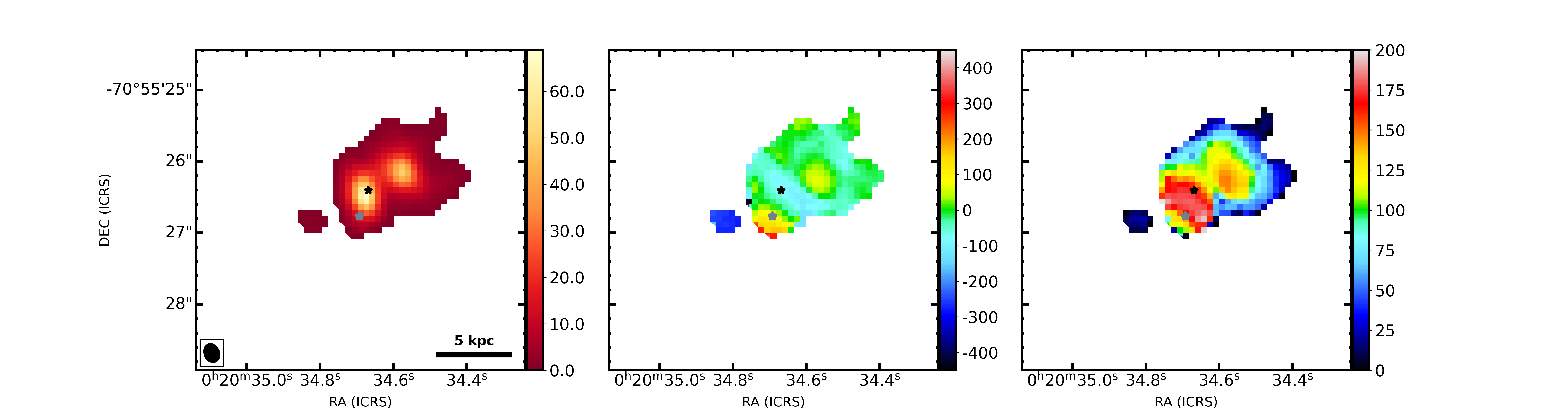}
\end{subfigure}
\caption[]{I00183 moment 0 (integrated intensity; left panels), moment 1 (intensity-weighted mean line-of-sight velocity; middle panels) and moment 2 (intensity weighted line-of-sight velocity dispersion; right panels) maps of the CO(1-0) and CO(3-2) transitions (top and bottom row, respectively). The maps were created using data cubes with a channel width of 20~km~s$^{-1}$ (see Table~\ref{tab:line images}). The synthesised beam and a scale bar are shown in the bottom-left and bottom-right corner, respectively, of each moment 0 map. The bar to the right of each map shows the colour scales (in mJy~beam$^{-1}$~km~s$^{-1}$ and km~s$^{-1}$ for moment 0 and moment 1/2 maps, respectively). The phase centre of each map is set to the sky coordinates reported in Table~\ref{tab:I00183_properties}, corresponding to the optical peak of the system (see also Section~\ref{sec:VST_obs} for details); East is to the left and North to the top. In each panel, this is also marked with a grey star, while the black star indicates the position of the radio core, as inferred with high precision from the VLBI observations of I00183 presented in \citet[][see also \citealp{Ruffa18} and Section~\ref{sec:ratios_results}]{Norris12}. Velocities are measured in the source frame, so that the systemic velocity of each CO line corresponds to the redshifted central frequency of the line ($\nu_{\rm sky}$, reported in Table~\ref{tab:ALMA observations summary}; see also Section~\ref{sec:ALMA_imaging}). The black dashed lines in the middle panel of the top row illustrate the direction of the ionised gas outflow, as inferred from the [OIII] data presented in \citet[][see also Spoon et al.\,in preparation]{Iwasawa17}. \label{fig:CO_moments}}
\end{figure*}

\section{Image cube analysis and results}\label{sec:cube_analysis}
\subsection{Moment maps}\label{sec:mom_maps}
Integrated intensity (moment 0), intensity-weighted mean line-of-sight velocity (moment 1) and velocity dispersion (moment 2) maps of the CO lines were created from the cleaned data cubes using the masked moment technique as described by \citet[][see also \citealt{Bosma81a,Bosma81b,Kruit82,Rupen99}]{Dame11}. In short, in this technique, a copy of the cleaned data cube is first Gaussian-smoothed spatially (with a FWHM that is $1.5x$ that of the synthesised beam) and then Hanning-smoothed in velocity. A three-dimensional mask is then defined by selecting all the pixels above a fixed flux-density threshold. This threshold is chosen so as to recover as much flux as possible while minimising the noise (higher thresholds usually need to be set for noisier maps; see also \citealp[][]{Ruffa19a,Ruffa22}). Given the significance of our CO detections (see Table~\ref{tab:line images}), a threshold of 0.9$\sigma$ (where $\sigma$ is the rms noise level measured in the un-smoothed data cube) has been used in all the cases. The moment maps were then produced from the un-smoothed cubes using the masked regions only, and the results are presented in Figure~\ref{fig:CO_moments}.

\begin{figure*}
    \centering
    \includegraphics[clip=true, trim={5 0 5 0}, scale=0.5]{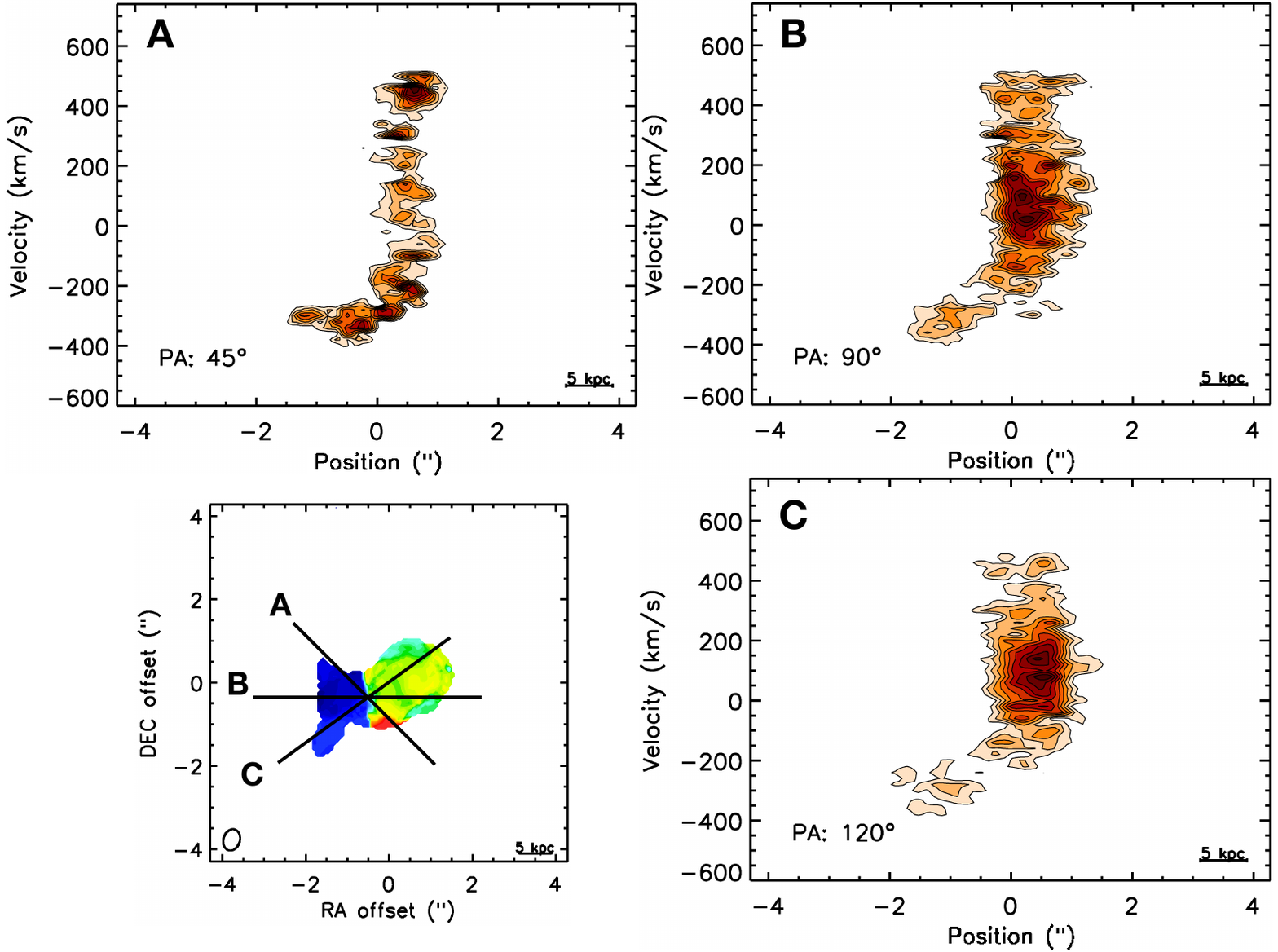}
    \caption{Position-velocity diagrams (PVDs) of the CO(1-0) transition extracted within rectangular areas whose long axes are orientated according to the position angles indicated in the bottom left corners of each panel, and illustrated by the mean velocity map in the bottom-left panel. The PA is measured counterclockwise from North through East. A scale bar is shown in the bottom-right corner of each PVD. Velocities are measured in the source frame, so that the systemic velocity (i.e. the point along the y-axis where $v=0$) corresponds to the redshifted central frequency of the CO line ($\nu_{\rm sky}$, reported in Table~\ref{tab:ALMA observations summary}; see also Section~\ref{sec:ALMA_imaging}). As detailed in Section~\ref{sec:line_profiles}, however, the observed centroids of both lines are clearly shifted with respect to those expected from $\nu_{\rm sky}$, thus explaining the slight shift observed in the PVDs between the estimated $v_{\rm sys}$ and the observed centroid of the emission. Coordinates are with respect to the image phase centre, which is set to the preferred sky coordinates of the optical galaxy centre reported in Table~\ref{tab:I00183_properties} (see also Section~\ref{sec:VST_obs}); East is to the left and North to the top.}
    \label{fig:CO10_PVDs}
\end{figure*}

\begin{figure*}
    \centering
    \includegraphics[clip=true, trim={5 0 5 0}, scale=0.5]{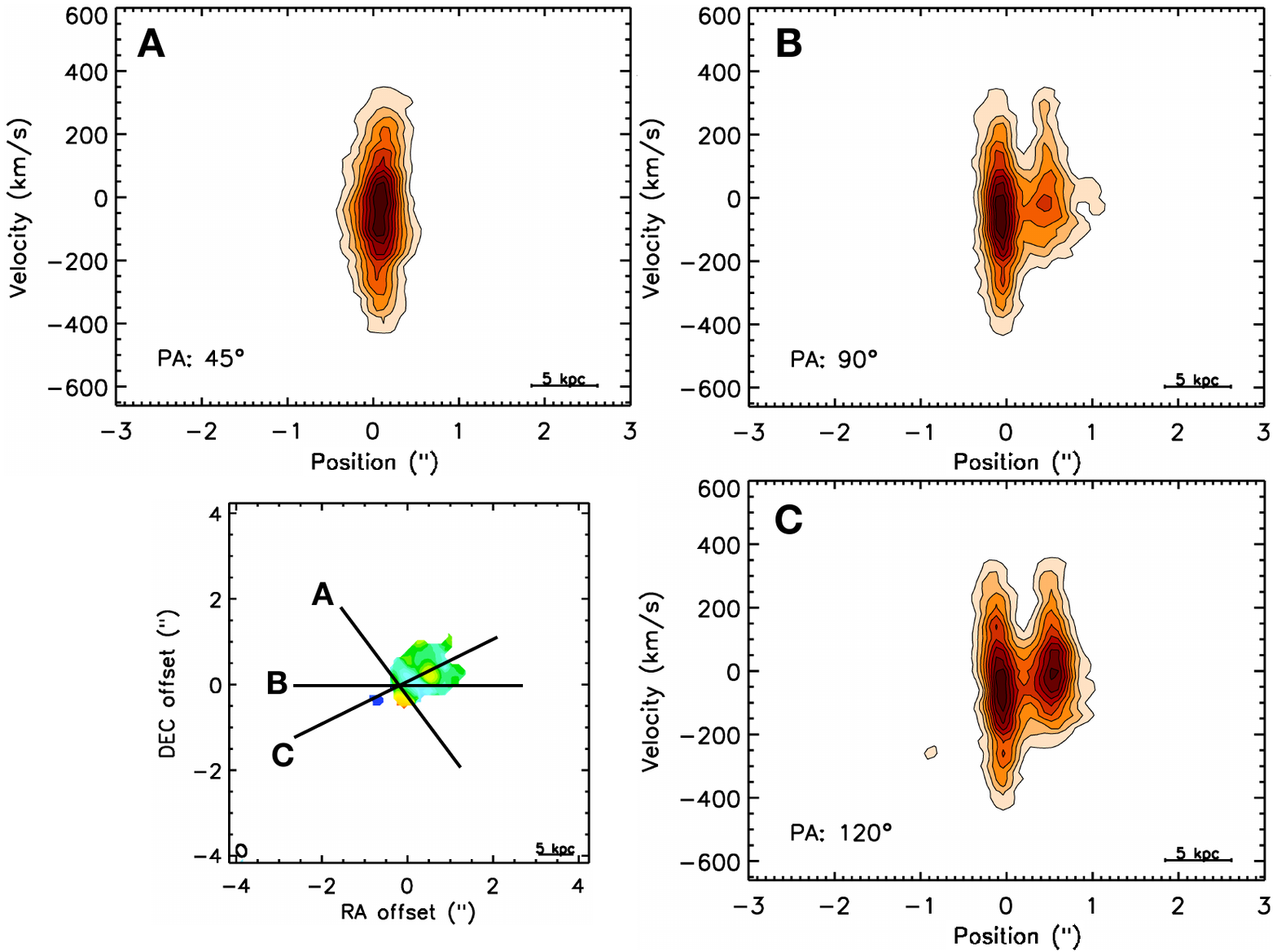}
    \caption{As in Figure~\ref{fig:CO10_PVDs}, but for the CO(3-2) transition.}
    \label{fig:CO32_PVDs}
\end{figure*}

The integrated intensity maps (left panels of Figure~\ref{fig:CO_moments}) illustrate some differences in the distribution of the two CO transitions. The CO(1-0) shows a bright area 
surrounded by emission which becomes progressively fainter (minimum ${\rm S/N}=3.5$), while extending up to $\approx2.0^{\prime \prime}$ ($\approx9.5$~kpc) to the East of the image centre in a sort of ``plume". This has a position angle (PA; measured counterclockwise from north to east) of $\approx90^{\circ}$, which is consistent with that of a similar (although more extended) feature observed in the ten times lower-resolution ALMA CO(1-0) detection of I00183 presented in \citet[][having a beam FWHM of $3.10\arcsec \times 2.20\arcsec$, equivalent to $\approx 14.7 \times 10.4$~kpc]{Ruffa18}. The entire CO(1-0) distribution is also clearly elongated in the same SE-NW direction of the galaxy's optical asymmetries visible in Figure~\ref{fig:VST_map}. Two regions of high surface brightness can be instead clearly distinguished in the moment 0 map of the CO(3-2) transition: the brightest one is roughly coincident with the position of the radio core (marked by the black star in the bottom-left panel of Figure~\ref{fig:CO_moments}), the faintest is located at a distance of about $1^{\prime \prime}$ ($\approx5$~kpc) from it, to the north-west (NW). It is plausible to believe that a similar, double-peaked surface brightness distribution could have been observed also in the CO(1-0), if detected at the same (or higher) resolution of the (3-2) transition (see Table~\ref{tab:ALMA observations summary}). This, however, does not affect the CO analysis presented in this work. The extended eastern feature, which is very prominent in the 1-0 transition, is instead barely detected in the 3-2. Although it is possible that (at least some of) the CO(3-2) emission in this area has been resolved out by the higher resolution of the Band 6 observations, similar extended features are typically somewhat diffuse (i.e. less dense) and thus only visible (or much brighter) in lower-J CO transitions \citep[see e.g.][]{Ruffa22}. In the regions where both transitions are detected, the CO(3-2) is mostly much brighter than the CO(1-0), clearly indicating high gas excitation around the radio source. This is further analysed in Section~\ref{sec:ratios_results} and discussed in Section~\ref{sec:discussion}.

The moment 1 maps of the two CO transitions (Figure~\ref{fig:CO_moments}, middle panels) reveal that the gas kinematics is only mildly resolved at the resolution of all our observations. This prevents us from carrying out a robust kinematic modelling. Overall, however, the detected features hint at a complex gas kinematics, with no clearly distinguishable patterns of (ordered) circular rotation. This is not surprising in disturbed objects like I00183, where the gas has likely been recently acquired through the merging process (see also Section~\ref{sec:phot}) and thus still in the process of settling onto the potential of the remnant system. 
It is nevertheless interesting to note the blueshifted velocities of the extended CO(1-0) plume, reaching values up to at least $\approx-400$~km~s$^{-1}$, much higher than those typically observed in regularly-rotating and compact (i.e. kpc-scale) molecular gas structures \citep[see e.g.][]{Oosterloo17,Ruffa19a,Ruffa24}. Similar features may trace residual gas perturbances due to the recent merger and/or the presence of non-circular motions such as a gas inflow/outflow. Based on the analysis carried out in this work, on the ancillary information available in the Literature and on the evolutionary stage of the system, we favour the hypothesis of an outflow (see Sections~\ref{sec:line_profiles}, \ref{sec:ratios_results} for further analysis, and Section~\ref{sec:discussion} for a detailed discussion). 

\begin{figure}
\centering
\vspace{-0.5cm}
\begin{subfigure}[t]{\columnwidth}
\centering
\caption{\textbf{CO(1-0) spectrum}}\label{fig:CO10_spectrum}
\includegraphics[clip=true, trim={20 10 10 20}, scale=0.38]{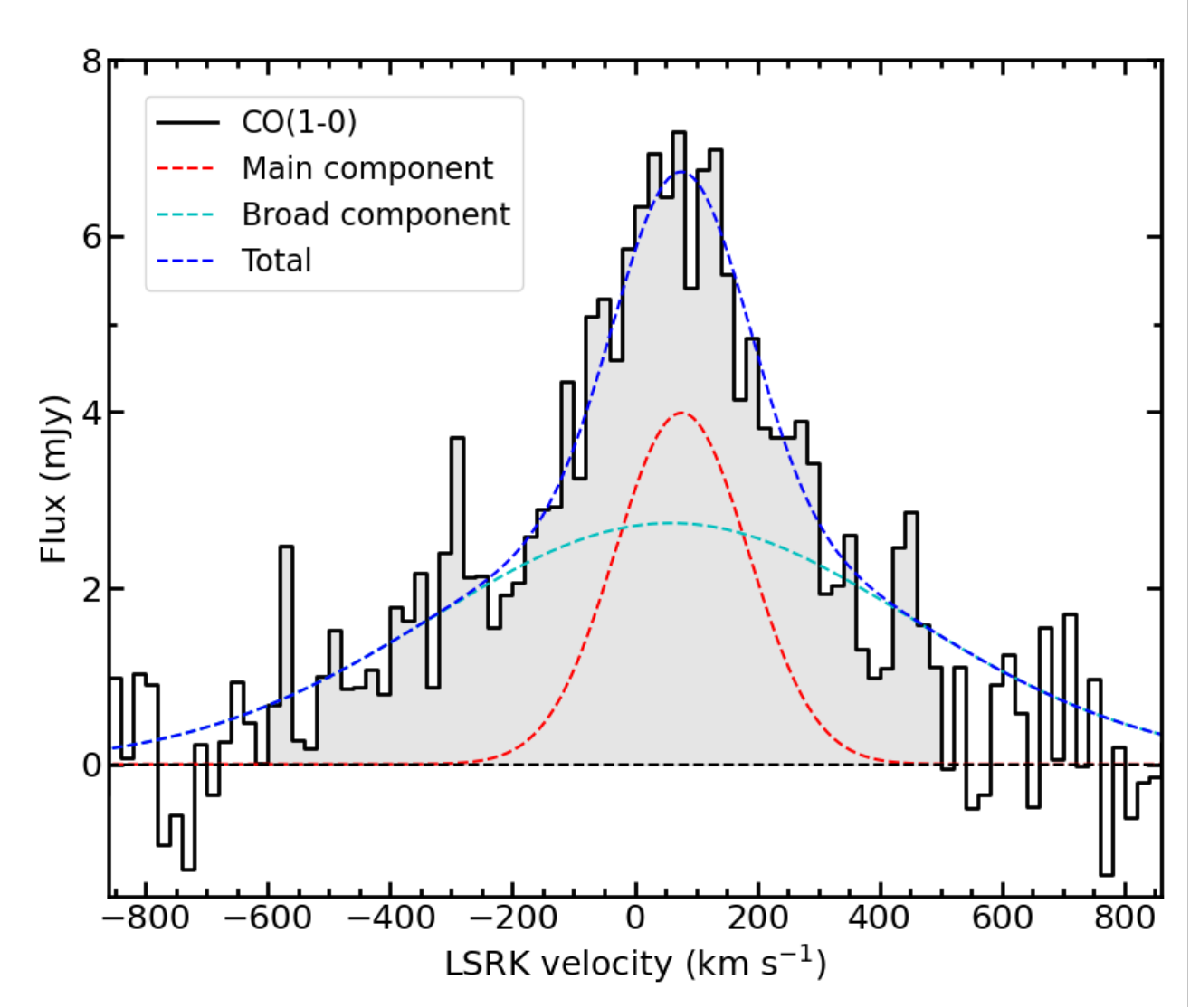}
\end{subfigure}
\begin{subfigure}[t]{\columnwidth}
\vspace{0.3cm}
\centering
\caption{\textbf{CO(3-2) spectrum}}\label{fig:CO32_spectum}
\includegraphics[clip=true, trim={25 20 30 50}, scale=0.38]{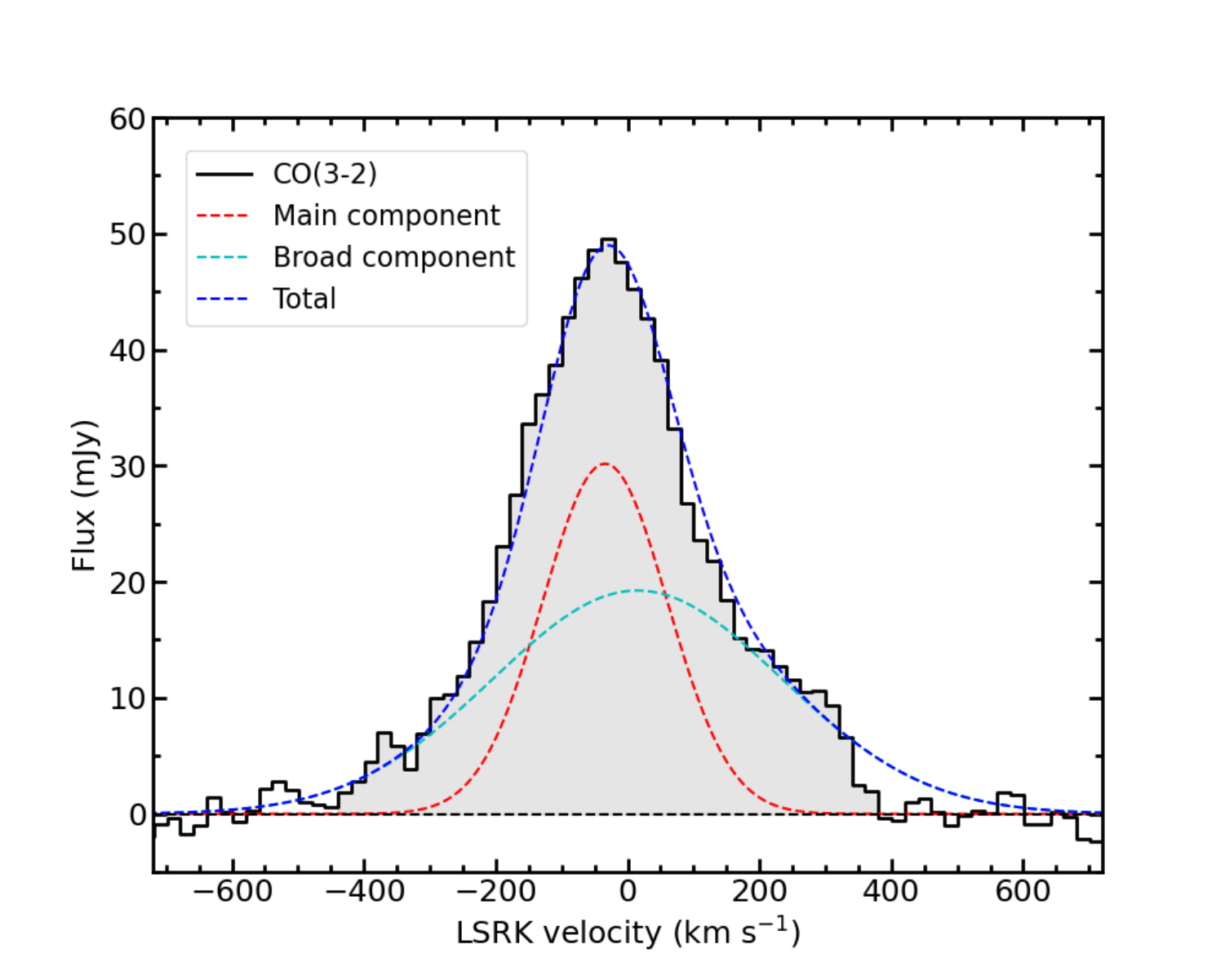}
\end{subfigure}
\caption[]{CO(1-0) (top panel) and CO(3-2) (bottom panel) spectral profiles extracted within polygonal areas drawn by-eye around all of the CO emission visible in the left panels of Figure~\ref{fig:CO_moments}. The best-fitting Gaussian profiles are overlaid as labelled in the top-left corner of each panel. In both panels, the black dashed horizontal line indicates the zero flux level. The gray shaded areas highlight the spectral channels included the line FWZI and thus used to estimate the integrated flux density of each line. Velocities are measured in the source frame, and the systemic velocity of each CO line has been assumed to correspond to the nominal central frequency of the line at the adopted source redshift ($\nu_{\rm sky}$; Table~\ref{tab:ALMA observations summary}). Both observed line centroids, however, are clearly shifted with respect to the latter. See Section~\ref{sec:line_profiles} for details.}\label{fig:CO_spectra}
\end{figure}

Further support to the presence of disturbed/non-circular kinematic components comes from the analysis of the CO position-velocity diagrams (PVDs) extracted along different PAs and illustrated in Figures~\ref{fig:CO10_PVDs} and \ref{fig:CO32_PVDs}. In first instance, the CO(1-0) PVDs extracted along all the directions resemble the ones of gas in circular rotation, with a steep velocity gradient at the centre, followed by signs of velocity flattening at larger scales. Only a central velocity gradient is instead visible in the CO(3-2) PVD extracted along the A direction (upper-left panel of Figure~\ref{fig:CO32_PVDs}). However, if we assume that the observed velocity curves arise from gas in circular motions, the dynamical mass ($M_{\rm dyn} = \dfrac{v_{\rm rot}^{2}r}{G}$) in the ``flat" part of the rotation curve would be $\approx10^{11}$~$M_{\odot}$, which is unrealistically large. We therefore conclude that the observed features do not trace regular circular rotation, but rather arise from a complex kinematics which is made of multiple sub-components. In particular, based on the hypothesis outlined above, the high blueshifted velocities in the "flat" part of the CO(1-0) PVDs should trace the outflow component (corresponding to the blueshifted plume in the moment 1 map). Moving along the directions labelled as B and C in Figures~\ref{fig:CO10_PVDs} and \ref{fig:CO32_PVDs}, a second high-velocity component is clearly visible (especially in the 3-2 transition), extending up to a RA offset of $1^{\prime \prime}$ ($\approx5$~kpc) to the north-west (NW) and with projected velocities up to at least $\approx+400$~km~s$^{-1}$. If we assume a bi-conical geometry for the putative outflow, with the same axis for both gas phases, this component may represent the redshifted part of the outflow \citep[e.g.][and references therein]{Garcia14,Dominguez20,Murthy22,Ruffa22,Audibert23,Ruffa23}. This is not clearly distinguishable (by-eye) from the moment 1 maps in Figure~\ref{fig:CO_moments} because it is likely embedded within the complex (unresolved) kinematics of the main CO structure (to the NW of the blueshifted plume).

The line-of-sight gas velocity dispersion ($\sigma_{\rm gas}$; Figure~\ref{fig:CO_moments}, right panels) varies significantly across the gas sky distributions, with values ranging from $\approx20$~km~s$^{-1}$ at the outskirts of the gas distributions, to $>200$~km~s$^{-1}$ around the centre of the system. The highest $\sigma_{\rm gas}$ values of the CO(1-0) transition are concentrated around the location of the galaxy's optical peak (marked by the grey star in Figure~\ref{fig:CO_moments}), whereas they are located around both the radio core (black star in Figure~\ref{fig:CO_moments}) and the optical peak in the 3-2 transition. Such large $\sigma_{\rm gas}$ values would imply that an ongoing perturbation (such as non-axisymmetric perturbations or shocks) is affecting the dynamical state of the gas, inducing turbulence within the gas clouds. While this is extremely plausible to occur in a recent merger like I00183, where both a vigorous central star formation activity and an AGN co-exist \citep[see e.g.][]{Ruffa18}, caution is needed in drawing conclusions from the observed line-of-sight velocity dispersions. Indeed, these can be significantly overestimated due to observational and instrumental effects. Among these, beam smearing (i.e.\,contamination from partially resolved velocity gradients within the host galaxy) is usually the dominant one, especially in regions of large velocity gradients and when the gas distribution is not fully resolved.

\subsection{Line widths and profiles}\label{sec:line_profiles}
The spectral profiles of the CO(1-0) and CO(3-2) lines are shown in Figure~\ref{fig:CO_spectra}. These were extracted from the cleaned (un-smoothed) data cubes within polygonal areas drawn by-eye around the line structures observed in Figure~\ref{fig:CO_moments}. Both spectra present an overall Gaussian-like shape with a single main peak. As described in Section~\ref{sec:ALMA_imaging}, velocities are measured in the source frame, and the systemic velocity of each CO line has been assumed to correspond to the nominal central frequency of the line at the adopted source redshift ($\nu_{\rm sky}$, reported in Table~\ref{tab:ALMA observations summary}). In other words, we built both data cubes by assuming that the systemic velocity of each CO line corresponds to the optical systemic velocity of the system. Both observed line centroids, however, are clearly shifted with respect to this latter, as reported in Table~\ref{tab:line_parameters}. Such shifts are consistent with the line peak positions observed in the moment 0 maps in Figure~\ref{fig:CO10_moms} and \ref{fig:CO32_moms} (and described in Section~\ref{sec:mom_maps}), and are also commonly observed in disturbed systems like ULIRGs, where indeed the systemic velocity may be always poorly defined \citep[e.g.][]{Kimball15}. We note, however, that such velocity shifts in the line centroids do not affect in any way the analysis and results presented in this paper.

In addition to the above, broad line components are clearly visible in both line profiles, with velocities up to $\approx +500$~km~s$^{-1}$ and $\approx -600$~km~s$^{-1}$ for the CO(1-0)\footnote{We note that the discrepancy between the maximum CO(1-0) velocities observed in the moment 1 map and PVD (Figures~\ref{fig:CO10_moms} and \ref{fig:CO10_PVDs}), and those inferred from the integrated spectral profile (Figure~\ref{fig:CO10_spectrum}) is likely due to the different methods used to derive these data products. In particular, the masked-moment technique used to produce the former relies on a smoothed data cube. This approach is extremely useful for suppressing noise and recovering most of the signal above a given threshold (see Section~\ref{sec:mom_maps}); however, as discussed in e.g. \citet[][]{Dame11,Leroy21}, it may also result in the loss of information from very faint and/or spatially extended structures confined to only a few velocity channels (like the highest-velocity structures observed in the CO(1-0) spectrum of I00183). The spectrum shown in Figure~\ref{fig:CO10_spectrum}, on the other hand, was extracted from the \textit{unsmoothed} cube (as mentioned in the text), thereby avoiding the potential issues described above. } and within $\pm400$~km~s$^{-1}$ for the CO(3-2). 
Similar spectral features (often called ``wings") are classic signatures attesting the presence of high-velocity, non-circular kinematic components, and are thus typically associated with outflows \citep[see e.g.][and references therein]{Feruglio10,Cicone14,Feruglio15,Pereira18,Dominguez20,Lamperti22,Holden24}. To rigorously demonstrate the presence of such secondary spectral components and determine the overall properties of the line profiles, we performed Gaussian fits on the integrated spectra using the least-squares minimization Python routine \texttt{lmfit} \citep{Newville14}. For both CO lines, we tested two models: one with a single Gaussian component and one with two Gaussian components: a main, narrow one including the bulk of the line emission and a broader one, extending up to higher velocities. We also performed a BIC (Bayesian Information Criterion) test \citep{Liddle07} to verify the statistical significance of one model with respect to the other. In both cases, the model with two Gaussians (overlaid as labelled in Figure~\ref{fig:CO_spectra}) resulted as preferred and the evidence for the need of the second component as very strong, with  $\Delta$BIC$ = 21.8$ for the CO(1-0) and $\Delta$BIC$ = 52.2$  for the CO(3-2). We thus confidently conclude that two components are needed in both cases to reproduce most of the observed spectral features. The mean, sigma and normalisation of each Gaussian function were left free to vary in the fitting process, and the resulting best-fitting values of the first two parameters are reported in Table~\ref{tab:line_parameters}, together with the corresponding statistical uncertainties. All of this provides support to the scenario disclosed in Section~\ref{sec:mom_maps}, which is also extensively discussed in Section~\ref{sec:discuss_outflow}.

We note that line widths were also measured as full-width at zero-intensity (FWZI), which was defined as the full velocity range covered by channels (identified interactively in the channel maps) with clear line detections with intensities $\geq3\sigma$. These channels are highlighted by the gray shaded areas in Figure~\ref{fig:CO_spectra}. The integrated flux density of each CO transition is calculated over this range, and reported together with the corresponding FWZI in Table~\ref{tab:line_parameters}.

\begin{table*}
\centering
\caption{Main integrated parameters of the observed molecular transitions.}
\label{tab:line_parameters}
\begin{tabular}{l c c  c  c c c}
\hline
\multicolumn{1}{c}{Transition} &
\multicolumn{1}{c}{ Component }&
\multicolumn{1}{c}{ Centroid }&
\multicolumn{1}{c}{ FWHM }&
\multicolumn{1}{c}{  FWZI  }&
\multicolumn{1}{c}{ $S_{\rm CO}$}&
\multicolumn{1}{c}{ R\textsubscript{J1}  } \\
\multicolumn{1}{c}{  } &
\multicolumn{1}{c}{  } &
\multicolumn{1}{c}{ (km~s$^{-1}$) } &
\multicolumn{1}{c}{ (km~s$^{-1}$) } &
\multicolumn{1}{c}{ (km~s$^{-1}$) } &
\multicolumn{1}{c}{ (Jy km~s$^{-1}$) } &
\multicolumn{1}{c}{  } \\
\multicolumn{1}{c}{ (1) } &
\multicolumn{1}{c}{ (2) } &
\multicolumn{1}{c}{ (3) } &
\multicolumn{1}{c}{ (4) } &
\multicolumn{1}{c}{ (5) } &
\multicolumn{1}{c}{ (6) } &
\multicolumn{1}{c}{ (7) } \\
\hline
 $^{12}$CO(1-0)  &  \begin{tabular}[]{@{}c@{}}  Main \\ Broad \end{tabular} & \begin{tabular}[]{@{}c@{}}  $+76\pm10$ \\ $+107\pm26$ \end{tabular} & \begin{tabular}[]{@{}c@{}} $109\pm15$ \\ $392\pm73$ \end{tabular} &  1100  &  2.3$\pm0.2$  &  $-$  \\ 

$^{12}$CO(3-2) &  \begin{tabular}[]{@{}c@{}}  Main \\ Broad \end{tabular} & \begin{tabular}[]{@{}c@{}} $-34\pm3$ \\ $+3\pm7$ \end{tabular} & \begin{tabular}[]{@{}c@{}} $94\pm6$ \\ $218\pm18$ \end{tabular} & 820  &  17.1$\pm1.7$  & $7.4\pm1.8$  \\ 
\hline
\end{tabular}
\parbox[t]{1\textwidth}{ \textit{Notes.} $-$ Columns: (1) Molecular transition. (2) Line component. (3) Best-fitting velocity shift of the mean of the corresponding line component with respect to the systemic velocity, as inferred from the performed Gaussian fits (see Section~\ref{sec:line_profiles}). The associated statistical errors are also reported. (4) Best-fitting FWHM of the corresponding line component with associated statistical error. (5) Full velocity range covered by the line channels (identified interactively in the channel maps) with intensities $\geq$3$\sigma$ (shaded regions in Figure~\ref{fig:CO_spectra}). (6) Flux density integrated over all the channels in the range defined by the FWZI. (7) Average ratio of the CO(3-2) line over the 1-0 transition (i.e.\,ratio of the integrated flux density at each transition).\\ 
}
\end{table*}

\subsection{Molecular gas mass}\label{sec:mol_masses}
An estimate of the molecular gas mass of I00183 was already provided in \citet{Mao14} based on an Australia Telescope Compact Array (ATCA) CO(1-0) detection, and in \cite{Ruffa18} from the ten times lower-resolution Cycle 0 ALMA CO(1-0) data. However, it is worth repeating the exercise to estimate the potential amount of flux that has been filtered out at the higher resolution of the data presented in this work.

Following \citet{Carilli13}, we calculated the CO(1-0) luminosity using
\begin{eqnarray}
L'_{\rm CO}=3.25\times10^{7}~(\dfrac{S_{\rm CO}}{\rm Jy~km s^{-1}})~(\dfrac{\nu_{\rm obs}}{\rm GHz})^{-2}~(\dfrac{D_{\rm L}}{\rm Mpc})^{2}~(1+z)^{-3}
,\end{eqnarray}\label{eq:luminosity}
where $\nu_{\rm obs}$ is the observing frequency (i.e.\,$\nu_{\rm sky}$ reported in Table~\ref{tab:ALMA observations summary}), $z$ is the redshift, D$_{\rm L}$ is the luminosity distance, and $S_{\rm CO}$ is the integrated flux density (as reported in Table~\ref{tab:line_parameters}). We obtained $L'_{\rm CO}=1.8\pm0.2\times10^{10}$~K km s$^{-1}$ pc$^{2}$. From this luminosity we derived the H$_{\rm 2}$ mass using the conversion equation, expressed as in \citet{Bolatto13},
\begin{displaymath}
M(H_{\rm 2})=\alpha~L'_{\rm CO}~M_{\odot},
\end{displaymath}
where $\alpha$ is the  H$_{2}$ mass-to-CO luminosity conversion factor. This depends on the molecular gas conditions (e.g.\ excitation, dynamics, geometry and stability of single molecular clouds) and other ISM properties (e.g.\,metallicity), and is therefore likely to vary systematically between different galaxy types. In this case, we used the typical conversion factor associated to ULIRGs, $\alpha=0.8$ M$_{\odot}$~(K~km~s$^{-1}$~pc$^{2}$)$^{-1}$ \citep[see][]{Downes98,Bolatto13}. The obtained molecular gas mass is $M_{\rm mol}=(1.0\pm0.1)\times10^{10}$~M\textsubscript{$\odot$}. This is fully consistent (within the uncertainties) with that estimated from previous CO(1-0) observations by \citet[][reporting $M_{\rm mol}=1.0\times10^{10}$~M\textsubscript{$\odot$}; no errors provided]{Mao14} and \citet[][reporting $M_{\rm mol}=(1.1\pm0.1)\times10^{10}$~M\textsubscript{$\odot$}]{Ruffa18}, indicating that the higher-resolution data presented here do not lead us to resolve out a relevant fraction of the total line flux.

\begin{figure*}
\centering
\includegraphics[clip=true, trim={4 120 5 120}, scale=0.60]{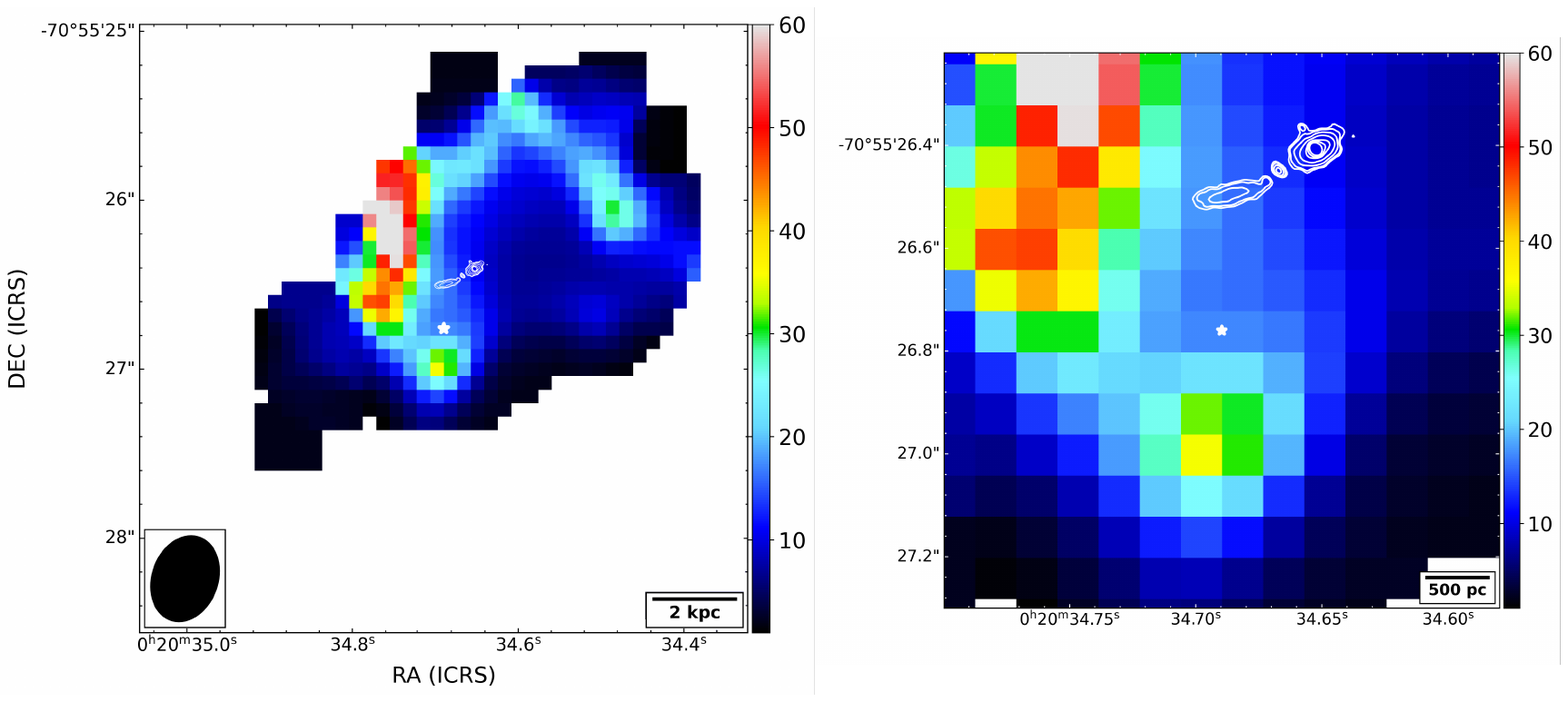}
\caption[]{{\bf Left panel:} $R_{\rm 31} \equiv S_{\rm CO(3-2)}/S_{\rm CO(1-0)}$ map of I00183 with overlaid in white the 2.3~GHz continuum contours from the VLBI observations of I00183 (from \citealp{Norris12}). {\bf Right panel:} Zoom-in of the same $R_{\rm 31}$ map in the central $0.54\arcsec \times 0.54\arcsec$ ($\approx2.6\times2.6$~kpc$^{2}$). In each panel, the bar to the right shows the colour scale. The beam size is illustrated in the bottom-left corner of the left panel, while a scale bar is shown in the bottom-right corner of each panel. The phase centre of each map is set to the sky coordinates reported in Table~\ref{tab:I00183_properties}, corresponding to the observed optical peak of the system (see Section~\ref{sec:VST_obs} for details). In each panel, this position is also marked with a white star. East is to the left and North to the top.\label{fig:CO_ratios}}
\end{figure*}

\subsection{Line ratio}\label{sec:ratios_results}
We carried out the analysis of the strength of the CO(3-2) transition relative to the CO(1-0), with the aim of obtaining a map of their flux density ratio. Ratios of molecular lines are indeed powerful tools to investigate the physical conditions of the gas, as different molecules, isotopologues, and transitions of the same species trace different gas components within the same galaxy. In particular, ratios of multiple CO lines are typically used to probe the excitation temperature and density of the molecular gas (see e.g. \citealp{Ruffa22} for details).

To study the relative strength of two gas transitions in the most reliable way possible, the data cubes used to make the ratio maps need to be identical. We thus re-imaged the two CO datasets making new data cubes with the same number of channels, channel width (i.e.\,20~km~s$^{-1}$), pixel and image size. We also corrected for the different $uv$ coverages by smoothing to a common resolution (i.e.\,selecting a common $uv$ range and convolving the images to a common synthesized beam). This resulted in data cubes with synthesized beams of $0.42\arcsec \times 0.32\arcsec$ (corresponding to $\approx2.0 \times 1.5$~kpc$^{2}$). Masked integrated intensity maps of these smoothed data cubes were produced using the technique described in Section~\ref{sec:mom_maps}, and then used to make ratio map using the {\sc CASA} task {\tt immath}. The resulting $R_{\rm 31} \equiv S_{\rm CO(3-2)}/S_{\rm CO(1-0)}$ map\footnote{We note that in this work we use integrated flux density ratios (expressed in Jy~km~s$^{-1}$), which are different from brightness temperature ratios (expressed in K~km~s$^{-1}$). See \citet{Ruffa22} for details.} is shown in Figure~\ref{fig:CO_ratios}, with overlaid in white the 2.3~GHz continuum contours from the VLBI observations of I00183 presented by \citet {Norris12}. As clearly illustrated by the Figure, the obtained beam size is quite large with respect to the size of the complex gas sub-structures, thus impeding a fully spatially-resolved analysis and inevitably leading us to miss the most reliable details on the gas physics. Nevertheless, the map in Figure~\ref{fig:CO_ratios} can provide us with useful hints on its conditions. Indeed, as already clear from the qualitative analysis presented in Section~\ref{sec:mom_maps}, the gas can be overall considered as highly excited, with average $R_{\rm 31}\approx10$, reaching values $\geq60$ around the location of the SE radio jet (as particularly visible from the right panel of Figure~\ref{fig:CO_ratios}). Such large values imply that the gas is thermalised (i.e.\,at densities above 3000~cm$^{-3}$) and overall optically thin, since they are slightly to significantly above the maximum values of $R_{\rm 31}=9$ for optically thick emission \citep[$\approx1$ for ratios in brightness temperature; see e.g.][]{Dasyra16,Oosterloo17}. Assuming local thermodynamic equilibrium (LTE) and optically thin conditions, CO ratios $>20$ correspond to gas excitation temperatures $T_{\rm ex} \gg50$~K \citep[e.g.][]{Oosterloo17}. Such extreme molecular gas conditions and - in particular - the location of the most excited gas fraction relative to the radio jet component are very similar to those observed in the other few objects with an ongoing jet-ISM interaction on (sub-)kpc scales which have been studied using a multi-line approach, e.g.\,NGC\,1068 \citep[][]{Garcia14}, IC\,5063 \citep{Dasyra16,Oosterloo17}, NGC\,3100 \citep{Ruffa22}, and the Teacup galaxy \citep{Audibert23}. It is thus tempting to claim that also in I00183 the expanding radio plasma contributes to changing the gas conditions around the centre of the remnant system. This is further discussed in Section~\ref{sec:discussion}.

\section{Photometric analysis}\label{sec:phot}
\begin{figure}
    \centering
    \includegraphics[width=\columnwidth]{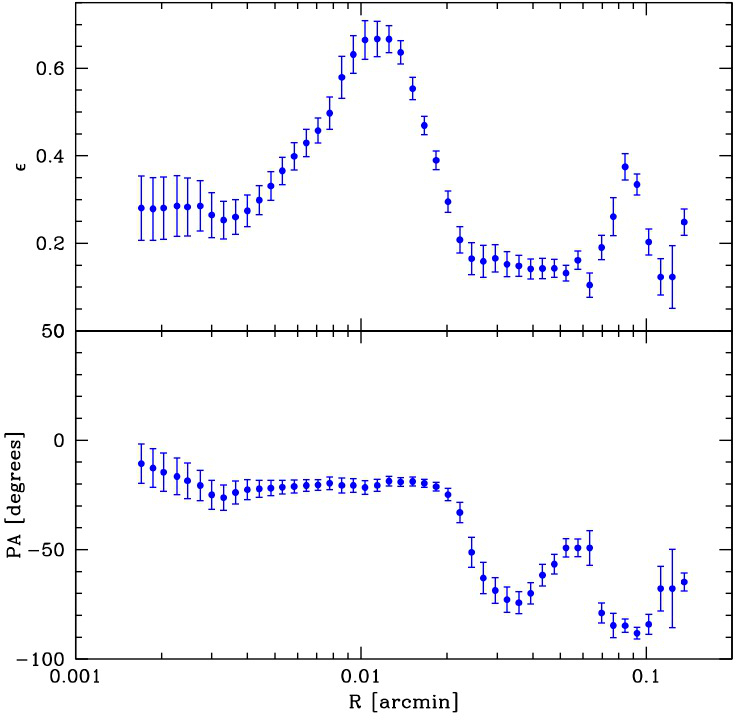}
    \includegraphics[width=\columnwidth]{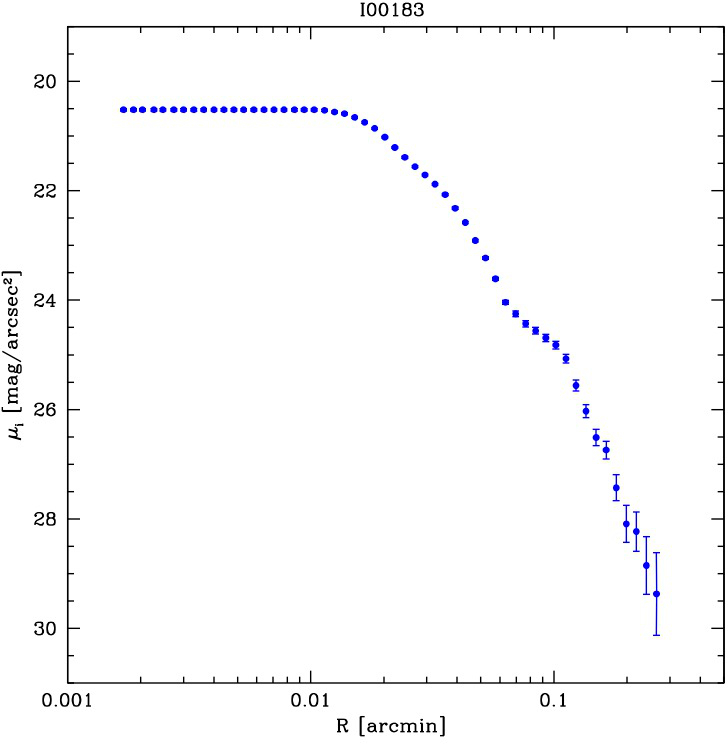}
    \caption{Ellipticity ($\epsilon$) and Position Angle (PA) profiles (top panels) and azimuthally-averaged surface brightness profile (bottom panel) of I00183 obtained from the {\it i}-band VST map using the method described in Section~\ref{sec:phot}.}
    \label{fig:VST_profiles}
\end{figure}

The photometric analysis has been performed on the sky-subtracted {\it i}-band VST image of I00183. 
The first step of this analysis consisted in carefully estimating any residual background fluctuation, by using the same procedure described in papers of the VEGAS series (see e.g. \citealt{Spavone2017}). On the final map, we masked all the bright sources in the field (stars and background galaxies), and performed the isophotal analysis by using the task {\tt ELLIPSE} of the {\sc IRAF} (Image Reduction and Analysis Facility) software \citep[][]{Tody86}. This method provides geometrical parameters (ellipticity $\epsilon$ and position angle PA) and the azimuthally-averaged light distribution within isophotal annuli of specified thickness. The resulting profiles are shown in Figure~\ref{fig:VST_profiles}.

Both the $\epsilon$ and the PA profiles show a change of trend at different radii, in line with the disturbed state of the system. A first change is observed at a radius of $\approx0.007$~arcmin ($\approx2.0$~kpc). This is especially visible as an abrupt increase in the ellipticity, which reaches its maximum at $\approx0.01$~arcmin ($\approx2.8$~kpc; where instead the PA remains almost constant around $-20$~deg) and then progressively decreases. Another abrupt change of both the ellipticity and the PA is observed at a radius of $\approx0.06$~arcmin ($\approx17$~kpc). At the same radius, also the surface brightness profile presents a bump. 

\begin{figure*}
    \centering
    \includegraphics[clip=true, trim={20 15 10 5}, scale=0.66]{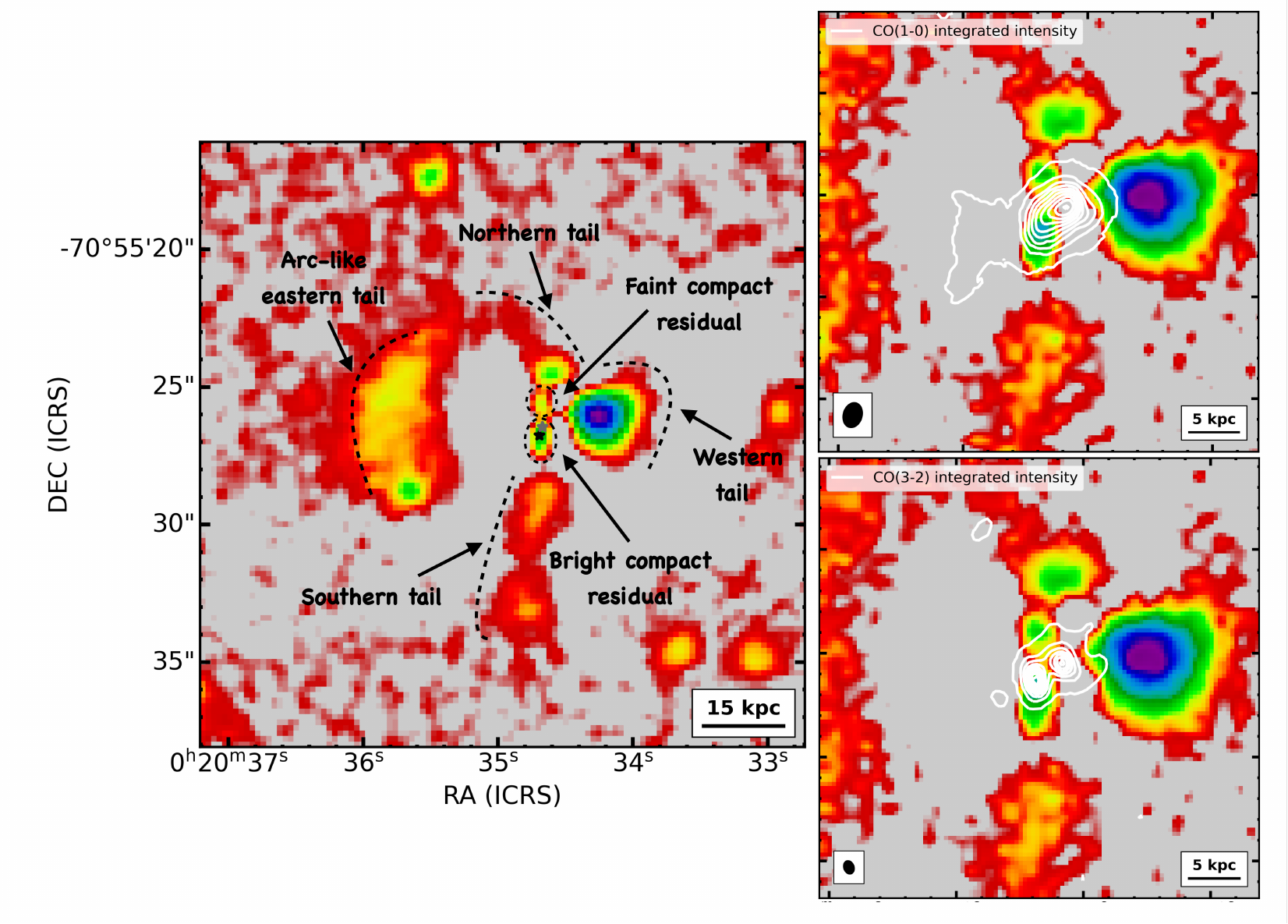}
    \caption{{\bf Left panel:} Residual map obtained by subtracting from the original VST {\it i}-band sky-subtracted image presented in Figure~\ref{fig:VST_map} the galaxy model described in Section~\ref{sec:phot}. The image has been smoothed for visualization purposes using a Gaussian filter with a standard deviation $\sigma=1$, and its size is $23\arcsec \times 23\arcsec$ ($\approx110\times110$~kpc$^{2}$). Labels and dashed contours are overlaid in black to highlight all the relevant residual structures. The black star indicates the preferred position assumed so far for the optical galaxy centre (corresponding to the coordinates reported in Table~\ref{tab:I00183_properties}), while the gray star indicates the position of the radio core as inferred from the VLBI observations presented in \citet{Norris12} (see also Section~\ref{sec:ratios_results}). {\bf Right panel, top:} Zoom-in of the residual VST map in the central $8\arcsec \times 8\arcsec$ ($\approx38\times38$~kpc$^{2}$) with overlaid in white CO(1-0) integrated intensity contours from the Cycle 7 ALMA observations presented in this paper. {\bf Right panel, bottom:} As above, but with overlaid CO(3-2) integrated intensity contours from the Cycle 2 ALMA observations presented in this paper. A scale bar is shown in the bottom-right corner of each panel, the CO synthesised beam sizes are shown in the bottom-left corners of the right panels. For each CO transition, we show 10 contours, which are equally spaced between the minimum and the maximum significant values of the integrated intensity map (as illustrated by the colour scale in the left-hand panels of Figure~\ref{fig:CO_moments}).}
    \label{fig:VST_residual}
\end{figure*}

We also produced a two-dimensional (2D) model of the galaxy's light distribution by using the {\sc IRAF} task {\tt BMODEL}. With this procedure, we created a noiseless 2D model of the galaxy, which is built from the results of the photometric analysis described above (thus taking into account the variations of $\epsilon$ and PA). We note that on both the {\tt ELLIPSE} and {\tt BMODEL} procedures, we fixed the centre of the system at the position that - when looking at the purple area in Figure~\ref{fig:VST_map} - one would identify by-eye as the geometrical galaxy centre (thus slightly to the north of both the observed optical peak and the radio core). This is done in order to obtain a simple, axisymmetric model of the mean light distribution of the system, which then allows us to clearly highlight any asymmetry once such model is subtracted from the original image. In Figure~\ref{fig:VST_residual} we show the obtained residual map, smoothed for visualization purposes using a Gaussian filter with a standard deviation $\sigma=1$ and with labels describing all the emerged structures.
Prominent tails to both the North and the West sides of the system clearly stand out from the residuals, in addition to the eastern arc-like and the southern elongated tidal structures already distinguishable from the observed map (see Figure~\ref{fig:VST_map} and Section~\ref{sec:VST_obs}). Moreover, Figure~\ref{fig:VST_residual} clearly shows the presence of two distinct residual compact (i.e.\,point-like) structures in the central regions of the system, which are separated from each other by a projected linear distance of $\approx1.1\arcsec$ ($\approx5.2$~kpc). This, in first instance, would provide support to the idea that I00183 hosts two nuclei from the progenitor galaxies which have not yet completely merged (see also Section~\ref{sec:VST_obs}). However, in a case like this and with the quality of the data currently available for this source, we would expect to find additional evidence in this regard. For instance, in confirmed cases of double (or triple) nuclei in ULIRGs, a clear spatial correlation between the nuclei and the optical peak(s), radio core and/or molecular gas distribution is observed \citep[e.g.][]{Pereira18,Pereira-Santaella21,Lamperti22,Ceci25}. Instead, in the case of I00183, both the optical peak and the radio core are located well within the brighter and southern compact residual (as respectively illustrated by the black and gray stars in the left panel of Figure~\ref{fig:VST_residual}). In addition, both molecular gas transitions do not show any obvious evidence for a morphological correlation with the northern and fainter compact residual (as illustrated by the right panels of Figure~\ref{fig:VST_residual}), whereas the brightest peak of the CO(3-2) transition is clearly spatially correlated with the southern compact structure (and thus with the optical peak and/or the radio core). Furthermore, the $\approx1$~kpc-resolution ALMA continuum maps presented in Figure~\ref{fig:continuum} show the presence of single compact unresolved structures around the position of the radio core (black star in both panels of Figure~\ref{fig:continuum}), without any clear sign of peak doubling or elongation in the direction of the two putative nuclei. We therefore conclude that it is unlikely that the two residual point-like structures in Figure~\ref{fig:VST_residual} trace the presence of close binary nuclei. Another hypothesis would be that both compact residuals arise from a prominent nuclear dust lane. The presence of dust in those areas is indeed confirmed by the (unresolved) spatial distribution of the $\sim250$~GHz continuum and its spectral analysis (see Section~\ref{sec:cont_discuss} for details), and could also provide an explanation to the observed displacement between the radio core and the optical peak (as already mentioned in Section~\ref{sec:VST_obs}). However, colour (dust extinction) maps obtained from multi-band optical photometry would be needed to properly test this scenario and draw solid conclusions. All of the above findings and their implications are further discussed in Section~\ref{sec:discuss_evolution}. 

\section{Discussion}\label{sec:discussion}

\subsection{ALMA continuum emission}\label{sec:cont_discuss}
A detailed analysis of the radio to sub-mm continuum of I00183, including a fitting of the far infrared part of its spectral energy distribution (SED), has been presented in \citet{Ruffa18} and thus not repeated here. In summary, we found that thermal emission from dust contributes for less than 4\% to the $\approx90$~GHz continuum, while it is the dominant emission mechanism at frequencies $\geq250$~GHz. The continuum spectrum up to at least 100~GHz turned out to be dominated by non-thermal synchrotron emission from the core-double lobe radio structure extending for $\approx1.7$~kpc in the nuclear regions of I00183. Such structure is nicely resolved at VLBI scales in the 2.3~GHz observations presented by \citet[][]{Norris12} and overlaid as white contours in Figure~\ref{fig:CO_ratios}. Both the radio source and the dust distribution are instead totally unresolved in all the ALMA continuum observations obtained so far (including those presented in this paper and illustrated in Figure~\ref{fig:continuum}), thus making difficult to carry out a morphological study that can corroborate the results obtained from the spectral analysis. As recently shown in hydrodynamic simulations of jets interacting with the multi-phase ISM \citep[][]{Young25}, it is also possible that existing high-resolution VLBI observations do not capture the full extent of the radio emission. Higher-resolution mm/sub-mm observations would be needed to carry out a spatially-resolved  study of both components and complement the results obtained so far on the continuum of I00183.  

\subsection{Galaxy evolutionary stage}\label{sec:discuss_evolution}
Thanks to a combination of high spatial resolution, large field-of-view and very deep surface brightness limit (this latter really exceptional for optical imaging with a ground-based telescope in the $i$-band; see Section~\ref{sec:VST_obs}), the VST map presented in Figure~\ref{fig:VST_map} provides us with tight constraints on the evolutionary stage of I00183. We confirm the disturbed dynamical state of the system by mapping - for the first time with such level of detail - faint asymmetric features extending up to projected distances from the core of $\approx8.5\arcsec$ ($\approx40$~kpc) to the East and $\approx7.4\arcsec$ ($\approx35$~kpc) to the South. In addition to those, prominent tails to both the North and the West sides of the system clearly stand out from the map illustrated in Figure~\ref{fig:VST_residual}, which is the residual between the observed and the modelled stellar light distribution (see Section~\ref{sec:phot}). The most likely explanation for such structures is that they are collisional debris, that are the outcome of gravity tides in galaxy encounters and thus the classic signatures of a recent (and possibly still ongoing) merger event \citep{Toomre72}. Typically, these type of debris is mainly made up of stellar streams and/or shells, and tidal tails \citep[e.g.][]{Veilleux02, Janowiecki10,Pereira18,Serra19,Mancillas19,Beaulieu22}. Stellar streams are thin ($\ll300$~pc in radius), very faint, elongated stellar structures which look like narrow, long filaments that emanate from the main system. Shells are density wave-like stellar accumulations with typical circular concentric shapes. When aligned to the major axis of the remnant system, shells are interleaved, thus appearing to accumulate alternatively on each side of the object and progressively propagating towards larger distances \citep[while increasing in number and decreasing their surface brightness; see e.g.][]{Mancillas19}. Tidal tails are thick ($>300$~pc), radially-elongated structures connected to the main body of the remnant and appearing to be emanating from it. Tails are produced as a result of tidal forces acting within the interacting objects, and can be associated with both intermediate-mass and major merger events \citep[with mass ratios between 3:1 and 7:1; e.g.][]{Mancillas19}. Given their appearance and the definitions above, we conclude that all the asymmetric and residual structures visible in Figures~\ref{fig:VST_map} and \ref{fig:VST_residual} are likely tidal tails. 

The detection of such collisional debris is essential to reconstruct the assembly history of a system, as they trace the last merger event(s) and store information about their progenitors. For instance, the strength and length of tidal tails are strong functions of the encounter geometry and the merger phase. A strong resonance between the orbital and rotational motions of stars and gas produces strong tidal tails in prograde encounters, whereas the lack of resonance in retrograde ones inhibits the formation of prominent tidal tails \citep[e.g.][]{Toomre72,Mihos96,Mihos05}. Systems in the pre-merger phase, where the involved objects have gone through the first encounter(s) but still have to completely merge, are expected to have very prominent and extended tidal features. As the merger proceed, the tidal tails gradually disappear, as the material in the tails is accreted back onto the remnant or escapes the system. The typical maximum survival time of tidal tails is about 2~Gyr \citep[e.g.][]{Veilleux02,Mancillas19}. 

All of the above, when compared with the properties of I00183, suggests that this system formed through a major merger event, likely characterised by a prograde encounter. Furthermore, as witnessed in particular by its evident tidal tails, I00183 is likely no older than 2~Gyr. 
The consistent amount of gas and dust observed in the central regions of the remnant \citep[e.g.][this paper]{Spoon09,Mao14,Iwasawa17,Ruffa18} also indicates that I00183 formed through the merger between two gas-rich spirals. Indeed, during mergers of this type, non-axisymmetries generated by tidal forces can drive (at least part of) the gas from the two gas-rich progenitors towards the centre of the gravitational potential of the newly-formed system (e.g. \citealp{Holden24}). It then takes time for this gas to settle down in the new environment, dynamically-relaxing around the main (or merged) nucleus. This is consistent with the complex CO kinematics observed in the middle panels of Figure~\ref{fig:CO_moments}, characterised by no clear patterns of ordered (circular) rotation. While in the process of settling, at least part of the cold gas can lose sufficient angular momentum to accrete onto a pre-existing SMBH, thus feeding the nuclear activity \citep[e.g.][]{Alexander12}. This is clearly the case of I00183, where the SMBH located within the brightest compact residual in Figure~\ref{fig:VST_residual} not only switched on, but has been also already able to give rise to the powerful radio source visible in Figure~\ref{fig:CO_ratios}. The brightest peak of the CO(3-2) transition is also found to be spatially-correlated with the brightest compact residual (wherein the radio core is located), making likely a connection between the AGN energetic output and a molecular gas over-density/high-excitation around it (see also Sections~\ref{sec:ratios_results} and \ref{sec:discuss_outflow}).

\begin{figure}
    \centering
    \includegraphics[clip=true, trim={3 3 3 3}, scale=0.47]{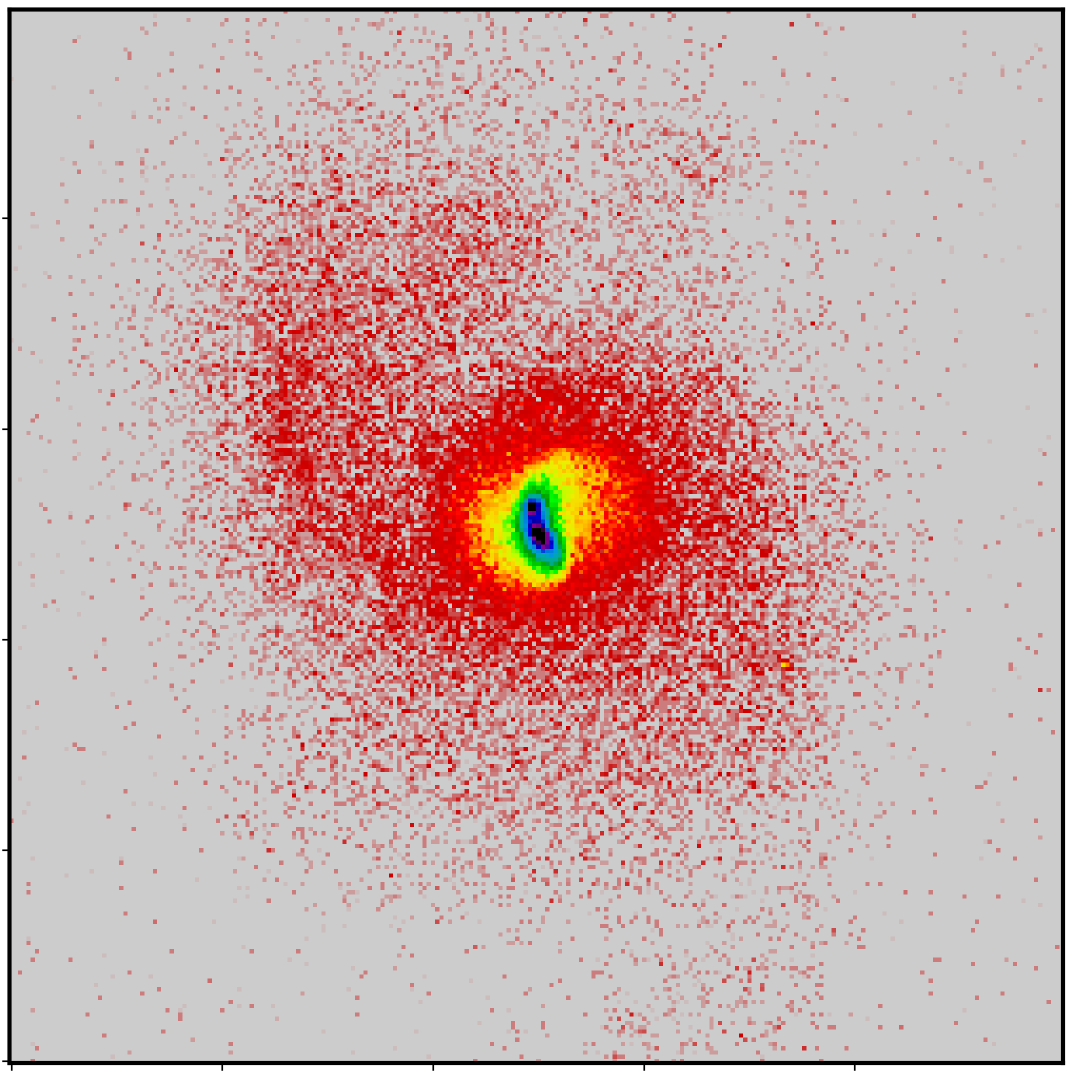}
    \caption{Snapshot of \textsc{GalMer} merger \#431 taken 1.05~Gyr after the start of the encounter and 650~Myr after the first pericentral passage. The image shows the distribution of stellar light on a logarithmic scale.}
    \label{fig:GalMer_map}
\end{figure}

We further investigated the consistency between our observational results and the expectation for a gas-rich major merger using the database of simulated galaxy mergers \textsc{GalMer} \citep[][]{Chilingarian10}. This includes two-galaxy mergers over a range of morphologies, mass ratios and orbital parameters of the progenitors. We visually-inspected all the possible \textsc{GalMer} combinations of gas-rich Sa$+$Sb (the most likely progenitors of ULIRGs) mergers, checking for consistency with the features observed in I00183. We found some qualitative similarities with the \textsc{GalMer} run \#431, where the two spiral progenitors have a prograde spin, an initial distance of 100~kpc and a pericentral distance of 16~kpc. 
In Figure~\ref{fig:GalMer_map} we present a snapshot of such \textsc{GalMer} run, which is taken 1.05~Gyr after the start of the encounter and 650~Myr after the first pericentral passage. Similarly to what we observe in I00183 (see Figures~\ref{fig:VST_map} and \ref{fig:VST_residual}), at this stage the simulated merger exhibits a clear asymmetric elongation in the South East-North West direction, as well as prominent stellar tidal tails on the East and South-West sides of the central structure. The exact shape and size of the tails vary with viewing angle, but overall they are always clearly visible.
 We note, however, that the \textsc{GalMer} database is quite limited in terms of its size and explored properties. For example, the range of orbital parameters explored in the database is narrow, the resolution of the \textsc{GalMer} simulations does not allow a thorough treatment of the complex gas physics, and no AGN feedback is included. It would thus be very hard to find an exact match to the observed properties of I00183. Similarly, it is possible that other potential merger conditions/ages (not included in the inspected simulations) could qualitatively reproduce the features observed in I00183. Despite these caveats, our comparison is as useful excercise, which provides some additional support to the scenario inferred from the optical analysis illustrated above.

\subsection{CO kinematics}\label{sec:discuss_outflow}
The analysis of the multiple CO transitions presented in this paper allows us to complement the results obtained from the study at ten times lower spatial resolution (i.e. beam FWHM $= 3.10\arcsec \times 2.20\arcsec$, equivalent to $\approx 14.7 \times 10.4$~kpc) of the sole CO(1-0) transition presented in \citet[][]{Ruffa18}. In particular, it allows us to put more constraints on the gas dynamical state, revealing a complex gas kinematics, without clear signs of ordered (circular) rotation, but with at least one non-circular kinematic component that can be clearly identified. Given the evolutionary stage of the system, this latter may simply trace gas that is tidally perturbed by the recent merger and/or still in the process of settling (i.e. relaxing) within the potential well of the newly-formed system. On the other hand, it may trace the presence of non-circular motions such as a gas outflow. We favour this latter interpretation, as disclosed in Section~\ref{sec:mom_maps} and discussed in detail in the following.

In the past two decades, various studies have shown the presence of prominent outflows in ULIRGs \citep[see e.g.][for recent reviews on this subject]{Veilleux20,U22}. Up until $\sim10$ years ago, however, such detections were mostly restricted to the ionized gas component (which accounts only for a minor fraction of the total gas mass in galaxies), and were usually found to be confined on pc-scales or within the AGN photoionization cones. In I00183, the presence of such type of features has been indeed known since long ago. For instance, a first signature of ionised gas outflow was observed by \citet{Heckman90}, identifying a blueshifted [O{\sc iii}]$\lambda5007$ emission extending about $10\arcsec$ to the east of the nucleus. Outflows traced by the 12.81~$\mu$m [Ne{\sc ii}] and 15.51~$\mu$m [Ne{\sc iii}] lines were then observed by \citet{Spoon09}, in a region $<3\arcsec$ east of the nucleus. 

It is instead relatively recent the finding that kpc-scale outflows in ULIRGs contain a large (in terms of the dominant mass) component of atomic and molecular gas \citep[see e.g.][for a collection of cases in typical ULIRGs]{Feruglio10,Cicone14,Pereira16,Calderon16,Fiore17,Saito18,Pereira18,Lamperti22,Holden24}, which is often massive enough to play a relevant role in the regulation of the SF processes within the host system. Crucially, massive cold gas outflows in ULIRGs are believed to trace the disruption of the thick obscuring nuclear layers, before the newly-formed system evolves as a bright (unobscured) quasar hosted in a passive spheroid \citep[e.g.][]{Hopkins11,Alexander12,U22}. As described in Section~\ref{sec:intro}, based on a number of multi-wavelength constraints, I00183 appears to have been caught exactly in this brief transitional period \citep[e.g.][]{Spoon07,Spoon09}. The presence of multi-phase gas outflows is thus highly expected in this source. This is one of the arguments supporting our hypothesis that the non-circular kinematic component identified in the CO kinematics and described in Sections~\ref{sec:mom_maps} and \ref{sec:line_profiles} traces the presence of an outflow. This was already tentatively inferred from the analysis of the ten-times lower-resolution CO(1-0) data, showing the presence of a low-surface brightness blueshifted tail extending up to about 35~kpc to the East of the galaxy centre and with spectral velocities up to $\approx-800$~km~s$^{-1}$ \citep{Ruffa18}. At the higher-resolution of the CO data presented here, such blueshifted tail is much less extended in the 1-0 transition (while it is only barely detected in the 3-2), indicating that the most extreme part of this diffuse component is resolved out (although we demonstrate that the contribution of this latter is negligible in terms of total molecular gas mass; see Section~\ref{sec:mol_masses}). The hypothesis that such CO plume (and its associated velocities) traces an outflow is also supported by the fact that similar blueshifted features have been already observed in a number of ionized gas components in I00183, and securely attributed to outflowing gas (see also Section~\ref{sec:intro}). In particular, it is worth highlighting that the blueshifted CO plume is clearly spatially-correlated with the [O{\sc iii}]$\lambda5007$ outflow, as illustrated in the middle panel of Figure~\ref{fig:CO10_moms} \citep[see also][Spoon et al.\,in prep]{Iwasawa17}. This suggests that the two gas phases are related and form part of a large-scale multi-phase outflow. If this hypothesis is correct, in the high-resolution data presented here, we are likely seeing only the bright and relatively compact base of the molecular gas outflow, of which only the blueshifted component is distinguishable by-eye from the moment 1 maps in Figure~\ref{fig:CO_moments}. Indeed, if we assume a bi-conical geometry for the outflow, with the same axis for both gas phases, its redshifted component is likely embedded within the complex (unresolved) kinematics of the main CO structure (to the NW of the blueshifted plume). Nevertheless, this is clearly identifiable in both the CO spectral profiles (see Section~\ref{sec:line_profiles}).

Based on the above, investigating molecular outflows in ULIRGs and characterise their properties (i.e. the outflow mass, energy, and momentum rate) is crucial to determine their impact onto their host galaxies. Nevertheless, studies of this type have been so far limited to a few objects, mostly (if not exclusively) located at redshift $z\ll0.1$ \citep[e.g.][]{Garcia15,Pereira18,Fotopoulou19,Lamperti22,Su23,Holden24}. Analysing molecular outflows in objects like I00183 can thus complement and/or add new insights into the evolutionary scenario associated to ULIRGs, putting also fundamental constraints on theoretical models and on galaxy evolution in general.

\begin{figure}
    \centering
    \includegraphics[clip=true, trim={3 3 3 3}, scale=0.38]{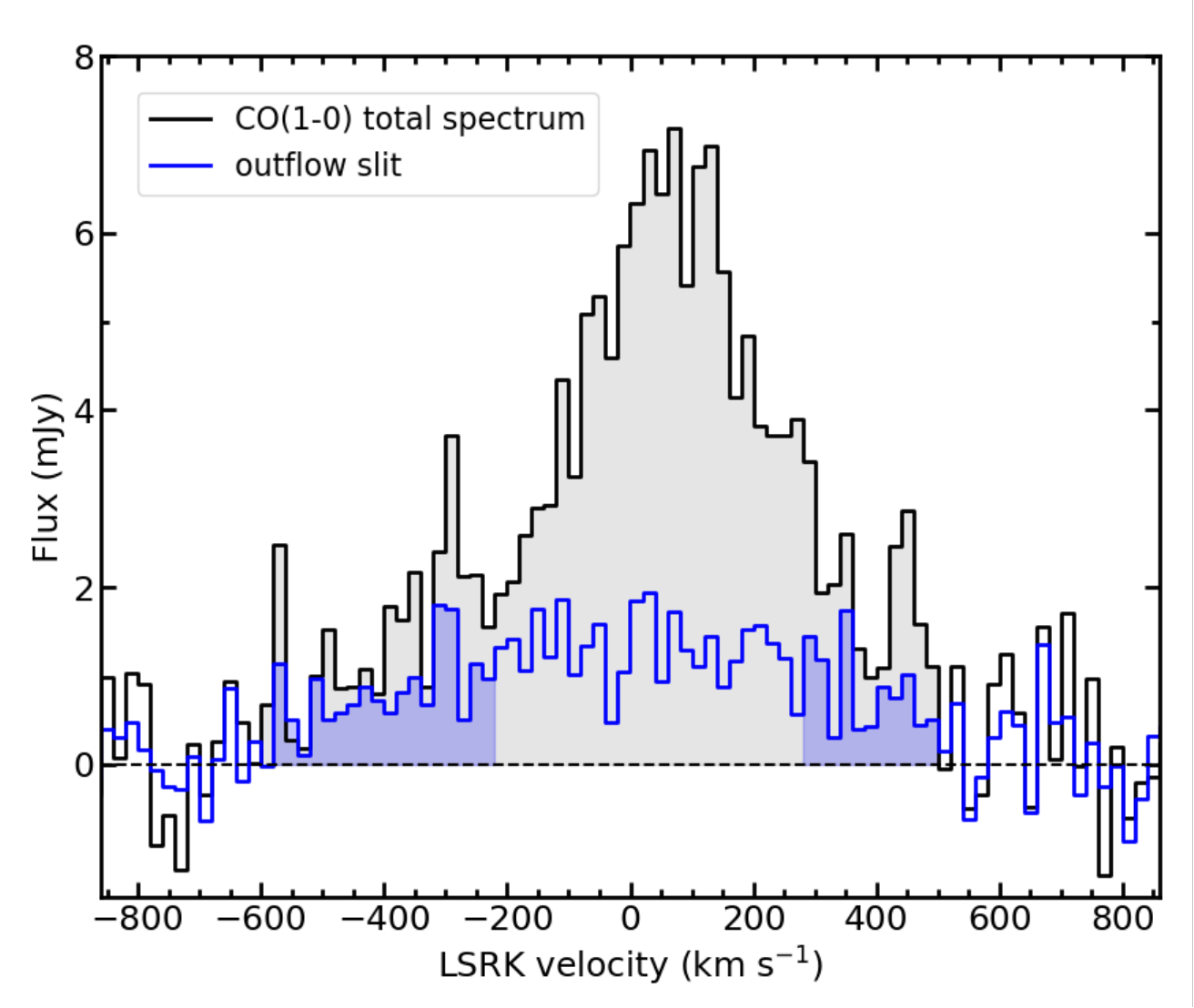}
    \caption{Spectral profile of the broad CO(1-0) component (blue histogram), extracted from the cleaned (unsmoothed) CO(1-0) data cube within a box of $1.0\arcsec\times0.3\arcsec$ in size, placed along the C slit illustrated in the bottom-left panel of Figure~\ref{fig:CO10_PVDs} (i.e. along the direction of the putative outflow). The outflow spectrum is overlaid on the total CO(1-0) integrated spectrum (black histogram; see also Figure~\ref{fig:CO10_spectrum}). The channels highlighted by the blue shaded areas are those used to estimate the mass of the outflow. The black dashed horizontal line indicates the zero flux level. See Section~\ref{sec:discuss_outflow} for details.}
    \label{fig:outflow_spectrum}
\end{figure}

As a common practice, a rough \textbf{(i.e. approximate)} estimate of the outflowing gas mass can be obtained by calculating the integrated flux density of the high-velocity spectral component. To avoid as much as possible contamination from non-outflowing gas and rather obtain a conservative estimate, we extracted from the cleaned CO(1-0) data cube a spectrum within a box of $1.0\arcsec\times0.3\arcsec$ in size, placed along the C slit illustrated in the bottom-left panel of Figure~\ref{fig:CO10_PVDs} (i.e. along the direction of the putative outflow). The obtained spectral profile is shown as a blue histogram in Figure~\ref{fig:outflow_spectrum}.
From this spectrum, we calculated the integrated flux density of the high-velocity component, including only the channels highlighted by the blue shaded areas in Figure~\ref{fig:outflow_spectrum}. We then estimated the mass of the outflow ($M_\mathrm{out}$) using the same methods described in Sections~\ref{sec:line_profiles} and \ref{sec:mol_masses}, and obtaining $M_\mathrm{out}\approx2.2\times10^{9}$~M$_{\odot}$ (about 20\% of the total estimated molecular gas mass; see Section~\ref{sec:mol_masses}).
The mass outflow rate can then be calculated from the relation:
\begin{equation}
    \dot{M}_\mathrm{out} \approx C  \frac{M_\mathrm{out}V_\mathrm{out}}{r_\mathrm{out}}.
    \label{eq:outflow_mass_rate}
\end{equation}
Here $V_{\rm out}$ is the outflow velocity and is commonly defined as $V_{\rm out}=\Delta V_{\rm broad}+2\sigma_{\rm broad}$, where $\Delta V_{\rm broad}$ is the shift between the velocity peak of the broad spectral component and the gas systemic velocity, $\sigma_{\rm broad}=FWHM_{\rm broad}/2.35$ and its numerator is the FWHM of the broad Gaussian component of the spectrum \citep[see e.g.][]{Rupke13,Feruglio15,Murthy22,Su23}. We note that this definition of the outflow velocity assumes that the blueshifted component of the outflow represents gas moving directly toward the observer's line-of-sight, thus implicitly accounting for projection effects. The parameters needed to estimate $V_{\rm out}$ are adopted as obtained from our CO(1-0) Gaussian spectral fitting and reported in Table~\ref{tab:line_parameters}. The estimated $V_{\rm out}$ is thus $\approx439$~km~s$^{-1}$. The parameter $r_{\rm out}$ is the outflow extension. Based on the analysis carried out in Section~\ref{sec:cube_analysis}, we assume $r_{\rm max}\approx1.0\arcsec\simeq5$~kpc. $C$ is a factor that depends on the outflow history. We use a typical value of $C=3$, which takes into account that the outflowing gas has a constant average volume density in a cone or a sphere \citep[see e.g.][]{Maiolino12,Cicone14}. 
Using these values, we \textbf{estimated} a mass outflow rate of $\dot{M}_{\rm out}\approx609$~M$_{\odot}$~yr$^{-1}$ and a gas depletion timescale of $t_{\rm dep}=M_{\rm tot}/\dot{M}_{\rm out}\approx16$~Myr. Both these quantities are quite extreme, but well within the range of those typically estimated in similar types of objects \citep[e.g.][]{Cicone14,Pereira18}. From these estimations, we also derived the outflow momentum rate ($\dot{P}_{\rm out}$) and kinetic power ($\dot{E}$). The former has been determined as $\dot{P}_{\rm out}=\dot{M}_{\rm out} \times V_{\rm out}\approx1.2\times10^{36}$~g~cm~s$^{-2}$ \citep[see e.g.][]{Garcia15}. Following \citet[][see also \citealp{Murthy22}]{Willott99}, we estimated the outflow kinetic power (in erg~s$^{-1}$) as:
\begin{equation}
    \dot{E} = 6.34 \times 10^{35} \dfrac{\dot{M}_{\rm out}}{2} \left(\Delta V_{\rm broad}^{2} + \dfrac{FWHM^{2}_{\rm broad}}{1.85}\right)
    \label{eq:outflow_kin_power}
\end{equation}
where $\Delta V_{\rm broad}$ and $FWHM_{\rm broad}$ are as in Table~\ref{tab:line_parameters}, and $\dot{M}_{\rm out}$ and $V_{\rm out}$ as estimated above. We obtained $\dot{E}\approx1.82\times10^{43}$~erg~s$^{-1}$.

We note that, as mentioned in Section~\ref{sec:cube_analysis}, by means of the high-resolution CO data presented in this paper we are probably detecting just the compact base of the outflow component. If we take into account also the diffuse component observed in the much lower resolution data presented in \citet{Ruffa18}, we have to assume $V_{\rm out}\approx556$~km~s$^{-1}$, $r_{\rm out}\approx7\arcsec$ ($\simeq34$~kpc), and  $M_\mathrm{out}\approx3.0\times10^{9}$~M$_{\odot}$. This latter has been obtained by adding to the $M_\mathrm{out}$ estimated above the mass of the diffuse outflow component, which was in turn calculated from the spectrum extracted within such region (i.e. blue component in Figure~4 of \citealp{Ruffa18}). Using these values, we obtain $\dot{M}_{\rm out}\approx150$~M$_{\odot}$~yr$^{-1}$, $t_{\rm dep}\approx67$~Myr, $\dot{P}_{\rm out}\approx5.28\times10^{35}$~g~cm~s$^{-2}$, and $\dot{E}\approx6.54\times10^{42}$~erg~s$^{-1}$. 

\subsubsection{What powers the outflow?}
In objects like ULIRGs, both star formation and AGN can drive powerful gas outflows through the huge amounts of energy these processes are able to release onto the surroundings. In particular, AGN energetic feedback is currently believed to operate in two (non-exclusive) main modes: radiative (or quasar) mode and kinetic mode \citep[e.g.][]{Morganti17,Morganti21}. In radiative-mode AGN (such as those typically hosted in ULIRGs), the dominant energetic output is the electromagnetic radiation produced by the efficient conversion of the potential energy of the matter accreted onto the SMBH. The radiation pressure generated by this process is thus typically found to be the main outflow-acceleration mechanisms in ULIRGs, driving either direct `in-situ' acceleration or relativistic winds \citep[e.g.][]{Hopkins08,Cicone14,Cicone18}. As observed in I00183, however, radiative and kinetic-mode feedback can co-exist, the latter being in the form of collimated outflows of non-thermal plasma (i.e.\,the radio jets). While previously believed to operate mostly on tens of kpc scales \citep[see, e.g.][for a review]{McNamara12}, in the past decade, radio jets that expand from the vicinity of the SMBH have been increasingly observed to alter the distribution, kinematics and physics of the surrounding gaseous medium also on (sub-)kpc scales (\citealp[e.g.][]{Combes13,Garcia14,Morganti15,Mahony16,Zovaro19,Santoro20,Venturi20,Ruffa20,Murthy22,Ruffa22,Audibert23,Pap23}). These observational findings are supported by 3D hydrodynamical simulations showing that radio jets expanding through the surrounding layers of matter can produce turbulent cocoons of shocked gas that can be accelerated up to $1000$~km~s$^{-1}$ and over a wide range of directions \citep[e.g.][]{Wagner12,Wagner16,Mukherjee16,Mukherjee18a,Mukherjee18b}.
The analysis carried out in Section~\ref{sec:ratios_results} allows us to show that high gas excitation conditions are occurring around the location of the radio lobes (see Figure~\ref{fig:CO_ratios}). Such conditions are very similar to those observed in the few existing spatially resolved, multi-line studies of jet–ISM interactions \citep[see e.g.][]{Oosterloo17,Ruffa22,Audibert23}. Also the properties of the jet flow (as observed at VLBI scales; see \citealp{Norris12} and Figure~\ref{fig:CO_ratios}) point in this direction. Indeed, the two jets are rather asymmetric, both in surface brightness and morphology: the SE jet is fainter and also undergoes an evident bend, the NW jet is instead more compact and shows a bright knot as if it is being compressed.
In general, the observed radio features suggest that the jet flow decollimates abruptly and decreases in surface brightness at short distances from the core, qualitatively consistent with rapid deceleration \citep[e.g.][]{Laing14}. All of these characteristics (both on the jet and molecular gas properties) are remarkably similar to those observed in a low-redshift radio galaxy with an ongoing jet-ISM interaction (i.e.\,NGC\,3100; see \citealp{Ruffa20,Ruffa22}). This supports a scenario in which I00183 hosts a radio source that is either very young (i.e. $t_{\rm age}\lesssim2$~Myr) or frustrated, and in either case is strongly interacting with the surrounding gaseous medium. It is thus interesting to verify whether the energetics of the putative molecular outflow is compatible with being jet-induced.

In star-forming systems such as I00183, the kinetic power released by supernovae (SNe) is capable of driving outflows. Following \citet{Veilleux05}, we estimated the mass outflow rate driven by SNe as \.{M}$_{\rm SNe} = 0.26$ (SFR/M$_{\odot}$ yr$^{-1}$)~M$_{\odot}$~yr$^{-1}$. Adopting a SFR of 260~M$_{\rm \odot}$~yr$^{-1}$ (as estimated in \citealp{Ruffa18} from the FIR luminosity), we obtain \.{M}$_{\rm SNe}=68$~M$_{\rm \odot}$~yr$^{-1}$, which is much lower than the mass outflow rates estimated for I00183. 
This excludes the power from supernovae as the main driver of the molecular outflow.


To investigate the role of the AGN radiation pressure as the main driver for the CO outflow we first need to estimate the bolometric luminosity of the active nucleus. Following \citet{Ho08}, the AGN bolometric luminosity of a quasar can be calculated as $L_{\rm BOL} \equiv L_{\rm AGN} = 83\times L_{\rm X,2-10}$, where $L_{\rm X,2-10}$ is the intrinsic (i.e. absorption-corrected) X-ray luminosity in the $2-10$~keV energy range\footnote{We are aware that the hard X-ray bolometric correction, $K_{\rm X}$, reported by \citet{Ho08} is much larger than more recent estimates. Among these, there is the one presented in \citet{Duras20}, where $K_{\rm X}$ for type 2 (i.e. obscured) AGN like the one in I00183 is $\sim10$. We nevertheless prefer to adopt the former, as our aim here is to obtain an upper limit on the radiation pressure from the AGN and verify whether or not - even in the most extreme possible case - this is compatible with the estimated energetics of the molecular outflow.}. We take this latter from the {\it Chandra} spectral fitting presented in \citet{Ruffa18}, whereby $L_{\rm X,2-10}=7.46^{+1.29}_{-1.15}\times10^{43}$~erg~s$^{-1}$, obtaining $L_{\rm AGN}=6.19^{+1.07}_{-0.95}\times10^{45}$~erg~s$^{-1}$. The force exerted on the gas due to the AGN radiation pressure is thus $L_{\rm AGN}/c=2.06^{+0.36}_{-0.32}\times10^{35}$~g~cm~s$^{-2}$, which is at least a factor of about two lower than the estimated outflow momentum rates. We thus conclude that the AGN radiation pressure is unlikely to be the only driver of the molecular outflow, at least at present times.

As discussed above, the coupling between (sub-)kpc scale radio jets and the surrounding cold gas has recently been shown to be very effective in driving molecular gas outflows, even in objects where radiative-mode AGN feedback typically dominates \citep[e.g.][]{Oosterloo17,Mukherjee21}. We use the dynamical radio source models of \citet[][]{Shabala08,ShabalaGodfrey13} to estimate the kinetic power and age of the jets in our target. Assuming a gas density of $10^4$\,cm$^{-3}$ (realistically close to the actual gas density around the radio jets in I00183; \citealp[see e.g.][]{Spoon07,Ruffa18}), we obtain a jet power of $\sim10^{43}$~erg~s$^{-1}$ and age of $14$~Myr, for the observed radio source size and luminosity (see Section~\ref{sec:intro} and Table~\ref{tab:I00183_properties}). These numbers are consistent with the outflow energetics inferred above. We note that using the well-known scaling relations of \citet{Cavagnolo10} results in an unrealistic jet power which is two orders of magnitude higher, and age correspondingly shorter -- because this relation is not well-suited to compact radio sources which evolve rapidly through the size-luminosity space \citep[][]{Alexander00,TurnerShabala15,Young25}. Our results are only moderately sensitive to model assumptions: increasing the gas density by a factor of five will decrease the jet power by a factor of two, and correspondingly double the jet age. Hence both the energetics and timescale analysis suggest that the radio jets can plausibly drive the putative CO outflow.

All together, our results suggest a scenario in which the molecular (and possibly also the ionised) outflow was triggered by either radiation from the AGN at an earlier stage of evolution, when the accretion rate and thus the radiation pressure were likely much higher; or by the jets, assuming that these are not intrinsically young by rather frustrated within the dense nuclear regions and/or that the full extent of the radio emission is not visible in the existing VLBI images (as discussed in Section~\ref{sec:cont_discuss}). Our analysis illustrates how the high gas excitation conditions in I00183 are plausibly due to an ongoing jet-ISM interaction, further supporting the idea that radio jets play a significant role in the overall AGN's impact on its surroundings and thus on the subsequent evolution of the host system.

\section{Summary and conclusions}\label{sec:conclusions}
We presented newly-acquired ALMA observations of the $^{12}$CO(1-0) line and 94~GHz continuum, along with very deep $i$-band VLT Survey Telescope (VST) imaging of the ULIRG IRAS 00183-7111 (I00183), located at $z=0.328$. These data have resolutions of $0.22\arcsec$ ($\approx1$~kpc) and $0.85\arcsec$ ($\approx4$~kpc), respectively, and are analysed in combination with archival ALMA data of the $^{12}$CO(3-2) line and 250~GHz continuum and with sub-arcsec resolution VLBI continuum data at 2.3~GHz. Such detailed multi-wavelength analysis allows us to explore the link between galaxy merger, AGN ignition, radio jet expansion and galactic-scale molecular outflows, for the first time with a high level of detail in an ULIRG at $z>0.3$. The main obtained results can be summarized as follows:
\begin{itemize}
\item Thanks to a combination of high spatial resolution, large field-of-view, long integration time (i.e. 6 hours) and thus a very deep surface brightness limit ($\mu_{i} \sim$ 29.5 mag~arcsec$^{-2}$ in the final stacked mosaic), the VST imaging provides us with tight constraints on the assembly history of I00183. We confirm the disturbed dynamical state of the system by mapping - for the first time with a high level of detail - faint asymmetric features extending up to projected distances from the core of $\approx8.5\arcsec$ ($\approx40$~kpc) to the East and $\approx7.4\arcsec$ ($\approx35$~kpc) to the South. In addition to those, prominent tails to both the North and the West sides of the system clearly stand out from our photometric modelling (see Section~\ref{sec:phot}). We claim that the most likely explanation for such structures is that they are tidal tails, whose characteristics lead us to conclude that I00183 probably formed through a major merger between two gas-rich spirals, characterised by a prograde encounter and no older than 2~Gyr.
\item The recent merger channelled consistent amounts of cold gas in the central regions of the remnant, as nicely traced by the $^{12}$CO(1-0) and $^{12}$CO(3-2) detections. From these detections, we estimated $M_{\rm mol}=(1.0\pm0.1)\times10^{10}$~M\textsubscript{$\odot$}, which was obtained assuming a H$_{2}$ mass-to-CO luminosity conversion factor typical for ULIRGs (i.e. $\alpha=0.8$). The spatial correlation between the CO distrubution and the radio core (as inferred from the available VLBI observation at 2.3~GHz) suggests that this gas likely contributed to the ignition of the AGN and thus to the launch of the radio jets (see Section~\ref{sec:discuss_outflow}).
\item The analysis of the $R_{\rm 31} \equiv S_{\rm CO(3-2)}/S_{\rm CO(1-0)}$ flux density ratio shows the presence of a high-excitation gas component with  $R_{\rm 31}$ reaching values up to $60$ (see Section~\ref{sec:ratios_results}). The latter is consistent with gas in an optically thin regime and excitation temperatures $T_{\rm ex}\gg50$~K. Such conditions are very similar to those observed in the few existing spatially-resolved, multi-line studies of jet-ISM interactions. The spatial correlation between the regions where the largest line ratios are observed and the location of the radio lobes suggests that also in I00183 the expanding radio plasma is likely responsible for the extreme gas conditions in the circumnuclear regions.
\item A qualitative analysis of the CO kinematics suggests a disturbed gas dynamical state, with no clear signs of regular rotation. Nevertheless, the study of the gas PVDs and integrated spectra (see Sections~\ref{sec:mom_maps} and \ref{sec:line_profiles}) allows us to clearly identify at least one non-rotational kinematic component that we interpret as an outflow with $v_{\rm out}\approx 439$~km~s$^{-1}$ and $\dot{M_{\rm out}}\approx609$~M$_{\odot}$~yr$^{-1}$. By comparing the energetics and geometrical properties of the outflow with that of possible driving mechanisms (i.e. SNe, AGN radiation pressure, jet-induced shocks), we conclude that this may be triggered by either radiation from the AGN in the past or the kinetic energy input from a frustrated radio jet.
\end{itemize}
Our work shows how studies like the one presented here are crucial not only to better understand the nature of ULIRGs, but also to improve our knowledge about the interplay between galaxy mergers, star formation and black hole accretion, AGN feedback and galactic-scale outflows, which are all basic ingredients of galaxy formation and evolution and thus crucial to test theories in this field.

\section*{Acknowledgements}
IR warmly acknowledges Ray Norris, for kindly providing the VLBI continuum map of I00183 presented in this work. IR also acknowledges support from grant ST/S00033X/1 through the UK Science and Technology Facilities Council (STFC). MS and EI acknowledge the support by the Italian Ministry for Education University and Research (MIUR) grant PRIN 2022 2022383WFT “SUNRISE”, CUP C53D23000850006 and by VST funds. SGB acknowledges support from the Spanish grant PID2022-138560NB-I00, funded by MCIN/AEI/10.13039/501100011033/FEDER, EU. This paper makes use of the following ALMA data: ADS/JAO.ALMA\#[2013.1.00826.S], and \#[2019.1.00851.S]. ALMA is a partnership of ESO (representing its member states), NSF (USA) and NINS (Japan), together with NRC (Canada), NSC and ASIAA (Taiwan), and KASI (Republic of Korea), in cooperation with the Republic of Chile. The Joint ALMA Observatory is operated by ESO, AUI/NRAO and NAOJ. The National Radio Astronomy Observatory is a facility of the National Science Foundation operated under cooperative agreement by Associated Universities, Inc. This paper has also made use of Image Reduction and Analysis Facility (IRAF) software, distributed by the National Optical Astronomy Observatories, which is operated by the Associated Universities for Research in Astronomy, Inc. under cooperative agreement with the National Science Foundation. This work has also made use of the NASA/IPAC Extragalactic Database (NED), which is operated by the Jet Propulsion Laboratory, California Institute of Technology under contract with NASA. We acknowledge the usage of the HyperLeda database (\url{http://leda.univ-lyon1.fr}). This research made also use of {\tt Astropy} \citep[][]{Astropy13,Astropy18}, {\tt Matplotlib} \citep[][]{Hunter07}, and {\tt NumPy} \citep[][]{Walt11,Harris20}.

\section*{Data Availability}
The ALMA data used in this article are available to download in the ALMA archive (\url{https://almascience.nrao.edu/asax/}; project codes: 2013.1.00826.S and 2019.1.00851.S). Part of the reduced VEGAS data have been made available through the ESO Phase3, unpublished VEGAS data are available upon request to the survey's PIs (E.\,Iodice \& M.\,Spavone). The VLBI data are publicily-available on European VLBI Network Data Archive (\url{https://www.re3data.org/repository/r3d100012641}). The GalMer simulations are available at the Horizon GalMer Database: \url{http://galmer.obspm.fr}. The calibrated data, final data products, and original plots generated for the research study underlying this article will be shared upon reasonable request to the first author.



\bibliographystyle{mnras}
\bibliography{mybibliography} 

@article{Sanders96,
   author = {{Sanders}, D.~B. and {Mirabel}, I.~F.},
    title = "{Luminous Infrared Galaxies}",
  journal = {\araa},
     year = 1996,
   volume = 34,
    pages = {749},
      doi = {10.1146/annurev.astro.34.1.749},
   adsurl = {http://adsabs.harvard.edu/abs/1996ARA%26A..34..749S}
}

@article{Nandra07,
   author = {{Nandra}, K. and {Iwasawa}, K.},
    title = "{A Compton-thick active galactic nucleus powering the hyperluminous infrared galaxy IRAS 00182-7112}",
  journal = {\mnras},
archivePrefix = "arXiv",
   eprint = {0708.0158},
     year = 2007,
    month = nov,
   volume = 382,
    pages = {L1-L5},
      doi = {10.1111/j.1745-3933.2007.00372.x},
   adsurl = {http://adsabs.harvard.edu/abs/2007MNRAS.382L...1N}
}

@ARTICLE{Veilleux02,
   author = {{Veilleux}, S. and {Kim}, D.-C. and {Sanders}, D.~B.},
    title = "{Optical and Near-Infrared Imaging of the IRAS 1 Jy Sample of Ultraluminous Infrared Galaxies. II. The Analysis}",
  journal = {\apjs},
   eprint = {astro-ph/0207401},
     year = 2002,
    month = dec,
   volume = 143,
    pages = {315-376},
      doi = {10.1086/343844},
   adsurl = {http://adsabs.harvard.edu/abs/2002ApJS..143..315V}
}

@ARTICLE{Lipari03,
   author = {{L{\'{\i}}pari}, S. and {Terlevich}, R. and {D{\'{\i}}az}, R.~J. and 
	{Taniguchi}, Y. and {Zheng}, W. and {Tsvetanov}, Z. and {Carranza}, G. and 
	{Dottori}, H.},
    title = "{Extreme galactic wind and Wolf-Rayet features in infrared mergers and infrared quasi-stellar objects}",
  journal = {\mnras},
   eprint = {astro-ph/0007316},
     year = 2003,
    month = mar,
   volume = 340,
    pages = {289-303},
      doi = {10.1046/j.1365-8711.2003.06309.x},
   adsurl = {http://adsabs.harvard.edu/abs/2003MNRAS.340..289L}
}

@ARTICLE{Hopkins08,
   author = {{Hopkins}, P.~F. and {Hernquist}, L. and {Cox}, T.~J. and {Kere{\v s}}, D.
	},
    title = "{A Cosmological Framework for the Co-Evolution of Quasars, Supermassive Black Holes, and Elliptical Galaxies. I. Galaxy Mergers and Quasar Activity}",
  journal = {\apjs},
archivePrefix = "arXiv",
   eprint = {0706.1243},
     year = 2008,
    month = apr,
   volume = 175,
      eid = {356-389},
    pages = {356-389},
      doi = {10.1086/524362},
   adsurl = {http://adsabs.harvard.edu/abs/2008ApJS..175..356H}
}

@ARTICLE{Spoon09,
   author = {{Spoon}, H.~W.~W. and {Armus}, L. and {Marshall}, J.~A. and 
	{Bernard-Salas}, J. and {Farrah}, D. and {Charmandaris}, V. and 
	{Kent}, B.~R.},
    title = "{High-Velocity Neon Line Emission From the ULIRG IRAS F00183-7111: Revealing the Optically Obscured Base of a Nuclear Outflow}",
  journal = {\apj},
archivePrefix = "arXiv",
   eprint = {0811.1562},
     year = 2009,
    month = mar,
   volume = 693,
    pages = {1223-1235},
      doi = {10.1088/0004-637X/693/2/1223},
   adsurl = {http://adsabs.harvard.edu/abs/2009ApJ...693.1223S}
}

@ARTICLE{Armus89,
   author = {{Armus}, L. and {Heckman}, T.~M. and {Miley}, G.~K.},
    title = "{Long-slit optical spectroscopy of powerful far-infrared galaxies - The nature of the nuclear energy source}",
  journal = {\apj},
     year = 1989,
    month = dec,
   volume = 347,
    pages = {727-742},
      doi = {10.1086/168164},
   adsurl = {http://adsabs.harvard.edu/abs/1989ApJ...347..727A}
}

@ARTICLE{Drake04,
   author = {{Drake}, C.~L. and {McGregor}, P.~J. and {Dopita}, M.~A.},
    title = "{Radio-Excess IRAS Galaxies. II. Host Galaxies}",
  journal = {\aj},
     year = 2004,
    month = sep,
   volume = 128,
    pages = {955-968},
      doi = {10.1086/422921},
   adsurl = {http://adsabs.harvard.edu/abs/2004AJ....128..955D}
}

@ARTICLE{Spoon04,
   author = {{Spoon}, H.~W.~W. and {Armus}, L. and {Cami}, J. and {Tielens}, A.~G.~G.~M. and 
	{Chiar}, J.~E. and {Peeters}, E. and {Keane}, J.~V. and {Charmandaris}, V. and 
	{Appleton}, P.~N. and {Teplitz}, H.~I. and {Burgdorf}, M.~J.
	},
    title = "{Fire and Ice: Spitzer Infrared Spectrograph (IRS) Mid-Infrared Spectroscopy of IRAS F00183-7111}",
  journal = {\apjs},
     year = 2004,
    month = sep,
   volume = 154,
    pages = {184-187},
      doi = {10.1086/422813},
   adsurl = {http://adsabs.harvard.edu/abs/2004ApJS..154..184S}
}

@ARTICLE{Norris12,
   author = {{Norris}, R.~P. and {Lenc}, E. and {Roy}, A.~L. and {Spoon}, H.
	},
    title = "{The radio core of the ultraluminous infrared galaxy F00183-7111: watching the birth of a quasar}",
  journal = {\mnras},
archivePrefix = "arXiv",
   eprint = {1107.3895},
     year = 2012,
    month = may,
   volume = 422,
    pages = {1453-1459},
      doi = {10.1111/j.1365-2966.2012.20717.x},
   adsurl = {http://adsabs.harvard.edu/abs/2012MNRAS.422.1453N}
}

@ARTICLE{Mao14,
   author = {{Mao}, M.~Y. and {Norris}, R.~P. and {Emonts}, B. and {Sharp}, R. and 
	{Feain}, I. and {Chow}, K. and {Lenc}, E. and {Stevens}, J.},
    title = "{Star formation in the ultraluminous infrared galaxy F00183-7111}",
  journal = {\mnras},
archivePrefix = "arXiv",
   eprint = {1401.5121},
     year = 2014,
    month = may,
   volume = 440,
    pages = {L31-L35},
      doi = {10.1093/mnrasl/slu015},
   adsurl = {http://adsabs.harvard.edu/abs/2014MNRAS.440L..31M}
}

@ARTICLE{Spoon07,
   author = {{Spoon}, H.~W.~W. and {Marshall}, J.~A. and {Houck}, J.~R. and 
	{Elitzur}, M. and {Hao}, L. and {Armus}, L. and {Brandl}, B.~R. and 
	{Charmandaris}, V.},
    title = "{Mid-Infrared Galaxy Classification Based on Silicate Obscuration and PAH Equivalent Width}",
  journal = {\apjl},
   eprint = {astro-ph/0611918},
     year = 2007,
    month = jan,
   volume = 654,
    pages = {L49-L52},
      doi = {10.1086/511268},
   adsurl = {http://adsabs.harvard.edu/abs/2007ApJ...654L..49S}
}

@ARTICLE{Bolatto13,
   author = {{Bolatto}, A.~D. and {Wolfire}, M. and {Leroy}, A.~K.},
    title = "{The CO-to-H$_{2}$ Conversion Factor}",
  journal = {\araa},
archivePrefix = "arXiv",
   eprint = {1301.3498},
     year = 2013,
    month = aug,
   volume = 51,
    pages = {207-268},
      doi = {10.1146/annurev-astro-082812-140944},
   adsurl = {http://adsabs.harvard.edu/abs/2013ARA%26A..51..207B}
}

@ARTICLE{Downes98,
   author = {{Downes}, D. and {Solomon}, P.~M.},
    title = "{Rotating Nuclear Rings and Extreme Starbursts in Ultraluminous Galaxies}",
  journal = {\apj},
   eprint = {astro-ph/9806377},
     year = 1998,
    month = nov,
   volume = 507,
    pages = {615-654},
      doi = {10.1086/306339},
   adsurl = {http://adsabs.harvard.edu/abs/1998ApJ...507..615D}
}

@ARTICLE{Carilli13,
   author = {{Carilli}, C.~L. and {Walter}, F.},
    title = "{Cool Gas in High-Redshift Galaxies}",
  journal = {\araa},
archivePrefix = "arXiv",
   eprint = {1301.0371},
     year = 2013,
    month = aug,
   volume = 51,
    pages = {105-161},
      doi = {10.1146/annurev-astro-082812-140953},
   adsurl = {http://adsabs.harvard.edu/abs/2013ARA%26A..51..105C}
}

@ARTICLE{Heckman90,
   author = {{Heckman}, T.~M. and {Armus}, L. and {Miley}, G.~K.},
    title = "{On the nature and implications of starburst-driven galactic superwinds}",
  journal = {\apjs},
     year = 1990,
    month = dec,
   volume = 74,
    pages = {833-868},
      doi = {10.1086/191522},
   adsurl = {http://adsabs.harvard.edu/abs/1990ApJS...74..833H}
}

@ARTICLE{Rau11,
   author = {{Rau}, U. and {Cornwell}, T.~J.},
    title = "{A multi-scale multi-frequency deconvolution algorithm for synthesis imaging in radio interferometry}",
  journal = {\aap},
archivePrefix = "arXiv",
   eprint = {1106.2745},
 primaryClass = "astro-ph.IM",
     year = 2011,
    month = aug,
   volume = 532,
      eid = {A71},
    pages = {A71},
    doi = {10.1051/0004-6361/201117104},
   adsurl = {http://adsabs.harvard.edu/abs/2011A%26A...532A..71R}
}

@ARTICLE{Calderon16,
   author = {{Calder{\'o}n}, D. and {Bauer}, F.~E. and {Veilleux}, S. and 
	{Graci{\'a}-Carpio}, J. and {Sturm}, E. and {Lira}, P. and {Schulze}, S. and 
	{Kim}, S.},
    title = "{Searching for molecular outflows in hyperluminous infrared galaxies}",
  journal = {\mnras},
archivePrefix = "arXiv",
   eprint = {1605.05364},
     year = 2016,
    month = aug,
   volume = 460,
    pages = {3052-3062},
      doi = {10.1093/mnras/stw1210},
   adsurl = {http://adsabs.harvard.edu/abs/2016MNRAS.460.3052C}
}

@ARTICLE{Iwasawa17,
   author = {{Iwasawa}, K. and {Spoon}, H.~W.~W. and {Comastri}, A. and {Gilli}, R. and 
	{Lanzuisi}, G. and {Piconcelli}, E. and {Vignali}, C. and {Brusa}, M. and 
	{Puccetti}, S.},
    title = "{The active nucleus of the ULIRG IRAS F00183-7111 viewed by NuSTAR}",
  journal = {\aap},
archivePrefix = "arXiv",
   eprint = {1709.01708},
     year = 2017,
    month = oct,
   volume = 606,
      eid = {A117},
    pages = {A117},
      doi = {10.1051/0004-6361/201730950},
   adsurl = {http://adsabs.harvard.edu/abs/2017A%26A...606A.117I}
}

@ARTICLE{Dame11,
   author = {{Dame}, T.~M.},
    title = "{Optimization of Moment Masking for CO Spectral Line Surveys}",
  journal = {ArXiv e-prints},
archivePrefix = "arXiv",
   eprint = {1101.1499},
 primaryClass = "astro-ph.IM",
     year = 2011,
    month = jan,
   adsurl = {http://adsabs.harvard.edu/abs/2011arXiv1101.1499D}
}

@INPROCEEDINGS{McMullin07,
   author = {{McMullin}, J.~P. and {Waters}, B. and {Schiebel}, D. and {Young}, W. and 
	{Golap}, K.},
    title = "{CASA Architecture and Applications}",
booktitle = {Astronomical Data Analysis Software and Systems XVI},
     year = 2007,
   series = {Astronomical Society of the Pacific Conference Series},
   volume = 376,
   editor = {{Shaw}, R.~A. and {Hill}, F. and {Bell}, D.~J.},
    month = oct,
    pages = {127},
   adsurl = {http://adsabs.harvard.edu/abs/2007ASPC..376..127M}
}

@ARTICLE{Dasyra16,
   author = {{Dasyra}, K.~M. and {Combes}, F. and {Oosterloo}, T. and {Oonk}, J.~B.~R. and 
	{Morganti}, R. and {Salom{\'e}}, P. and {Vlahakis}, N.},
    title = "{ALMA reveals optically thin, highly excited CO gas in the jet-driven winds of the galaxy IC 5063}",
  journal = {\aap},
archivePrefix = "arXiv",
   eprint = {1609.03421},
     year = 2016,
    month = nov,
   volume = 595,
      eid = {L7},
    pages = {L7},
      doi = {10.1051/0004-6361/201629689},
   adsurl = {http://adsabs.harvard.edu/abs/2016A%26A...595L...7D}
}

@ARTICLE{Laing14,
   author = {{Laing}, R.~A. and {Bridle}, A.~H.},
    title = "{Systematic properties of decelerating relativistic jets in low-luminosity radio galaxies}",
  journal = {\mnras},
archivePrefix = "arXiv",
   eprint = {1311.1015},
     year = 2014,
    month = feb,
   volume = 437,
    pages = {3405-3441},
      doi = {10.1093/mnras/stt2138},
   adsurl = {http://adsabs.harvard.edu/abs/2014MNRAS.437.3405L}
}

@ARTICLE{Alexander12,
   author = {{Alexander}, D.~M. and {Hickox}, R.~C.},
    title = "{What drives the growth of black holes?}",
  journal = {\nar},
archivePrefix = "arXiv",
   eprint = {1112.1949},
     year = 2012,
    month = jun,
   volume = 56,
    pages = {93-121},
      doi = {10.1016/j.newar.2011.11.003},
   adsurl = {http://adsabs.harvard.edu/abs/2012NewAR..56...93A}
}

@ARTICLE{Kimball15,
   author = {{Kimball}, A.~E. and {Lacy}, M. and {Lonsdale}, C.~J. and {Macquart}, J.-P.
	},
    title = "{ALMA detection of a disc-dominated [C II] emission line at z=4.6 in the luminous QSO J1554+1937}",
  journal = {\mnras},
archivePrefix = "arXiv",
   eprint = {1505.05262},
     year = 2015,
    month = sep,
   volume = 452,
    pages = {88-98},
      doi = {10.1093/mnras/stv1160},
   adsurl = {http://adsabs.harvard.edu/abs/2015MNRAS.452...88K}
}

@ARTICLE{Oosterloo17,
   author = {{Oosterloo}, T. and {Raymond Oonk}, J.~B. and {Morganti}, R. and 
	{Combes}, F. and {Dasyra}, K. and {Salom{\'e}}, P. and {Vlahakis}, N. and 
	{Tadhunter}, C.},
    title = "{Properties of the molecular gas in the fast outflow in the Seyfert galaxy IC 5063}",
  journal = {\aap},
archivePrefix = "arXiv",
   eprint = {1710.01570},
     year = 2017,
    month = dec,
   volume = 608,
      eid = {A38},
    pages = {A38},
      doi = {10.1051/0004-6361/201731781},
   adsurl = {http://adsabs.harvard.edu/abs/2017A%26A...608A..38O}
}

@ARTICLE{Combes13,
   author = {{Combes}, F. and {Garc{\'{\i}}a-Burillo}, S. and {Casasola}, V. and 
	{Hunt}, L. and {Krips}, M. and {Baker}, A.~J. and {Boone}, F. and 
	{Eckart}, A. and {Marquez}, I. and {Neri}, R. and {Schinnerer}, E. and 
	{Tacconi}, L.~J.},
    title = "{ALMA observations of feeding and feedback in nearby Seyfert galaxies: an AGN-driven outflow in NGC 1433}",
  journal = {\aap},
archivePrefix = "arXiv",
   eprint = {1309.7486},
     year = 2013,
    month = oct,
   volume = 558,
      eid = {A124},
    pages = {A124},
      doi = {10.1051/0004-6361/201322288},
   adsurl = {http://adsabs.harvard.edu/abs/2013A%26A...558A.124C}
}

@INPROCEEDINGS{Rupen99,
   author = {{Rupen}, M.~P.},
    title = "{Spectral Line Observing II: Calibration and Analysis}",
booktitle = {Synthesis Imaging in Radio Astronomy II},
     year = 1999,
   series = {Astronomical Society of the Pacific Conference Series},
   volume = 180,
   editor = {{Taylor}, G.~B. and {Carilli}, C.~L. and {Perley}, R.~A.},
    pages = {229},
   adsurl = {http://adsabs.harvard.edu/abs/1999ASPC..180..229R}
}

@ARTICLE{Bosma81a,
   author = {{Bosma}, A.},
    title = "{21-cm line studies of spiral galaxies. I - Observations of the galaxies NGC 5033, 3198, 5055, 2841, and 7331}",
  journal = {\aj},
     year = 1981,
    month = dec,
   volume = 86,
    pages = {1791-1824},
      doi = {10.1086/113062},
   adsurl = {http://adsabs.harvard.edu/abs/1981AJ.....86.1791B}
}

@ARTICLE{Bosma81b,
   author = {{Bosma}, A.},
    title = "{21-cm line studies of spiral galaxies. II. The distribution and kinematics of neutral hydrogen in spiral galaxies of various morphological types.}",
  journal = {\aj},
     year = 1981,
    month = dec,
   volume = 86,
    pages = {1825-1846},
      doi = {10.1086/113063},
   adsurl = {http://adsabs.harvard.edu/abs/1981AJ.....86.1825B}
}

@ARTICLE{Kruit82,
   author = {{van der Kruit}, P.~C. and {Shostak}, G.~S.},
    title = "{Studies of nearly face-on spiral galaxies. I - The velocity dispersion of the H I gas in NGC 3938}",
  journal = {\aap},
     year = 1982,
    month = jan,
   volume = 105,
    pages = {351-358},
   adsurl = {http://adsabs.harvard.edu/abs/1982A%26A...105..351V}
}

@ARTICLE{Veilleux05,
   author = {{Veilleux}, S. and {Cecil}, G. and {Bland-Hawthorn}, J.},
    title = "{Galactic Winds}",
  journal = {\araa},
   eprint = {astro-ph/0504435},
     year = 2005,
    month = sep,
   volume = 43,
    pages = {769-826},
      doi = {10.1146/annurev.astro.43.072103.150610},
   adsurl = {http://adsabs.harvard.edu/abs/2005ARA%26A..43..769V}
}

@ARTICLE{Hopkins11,
   author = {{Hopkins}, P.~F. and {Quataert}, E.},
    title = "{An analytic model of angular momentum transport by gravitational torques: from galaxies to massive black holes}",
  journal = {\mnras},
archivePrefix = "arXiv",
   eprint = {1007.2647},
 primaryClass = "astro-ph.CO",
     year = 2011,
    month = aug,
   volume = 415,
    pages = {1027-1050},
      doi = {10.1111/j.1365-2966.2011.18542.x},
   adsurl = {http://adsabs.harvard.edu/abs/2011MNRAS.415.1027H}
}

@ARTICLE{Garcia14,
   author = {{Garc{\'{\i}}a-Burillo}, S. and {Combes}, F. and {Usero}, A. and 
	{Aalto}, S. and {Krips}, M. and {Viti}, S. and {Alonso-Herrero}, A. and 
	{Hunt}, L.~K. and {Schinnerer}, E. and {Baker}, A.~J. and {Boone}, F. and 
	{Casasola}, V. and {Colina}, L. and {Costagliola}, F. and {Eckart}, A. and 
	{Fuente}, A. and {Henkel}, C. and {Labiano}, A. and {Mart{\'{\i}}n}, S. and 
	{M{\'a}rquez}, I. and {Muller}, S. and {Planesas}, P. and {Ramos Almeida}, C. and 
	{Spaans}, M. and {Tacconi}, L.~J. and {van der Werf}, P.~P.},
    title = "{Molecular line emission in NGC 1068 imaged with ALMA. I. An AGN-driven outflow in the dense molecular gas}",
  journal = {\aap},
archivePrefix = "arXiv",
   eprint = {1405.7706},
     year = 2014,
    month = jul,
   volume = 567,
      eid = {A125},
    pages = {A125},
      doi = {10.1051/0004-6361/201423843},
   adsurl = {http://adsabs.harvard.edu/abs/2014A%26A...567A.125G}
}

@ARTICLE{Wagner12,
   author = {{Wagner}, A.~Y. and {Bicknell}, G.~V. and {Umemura}, M.},
    title = "{Driving Outflows with Relativistic Jets and the Dependence of Active Galactic Nucleus Feedback Efficiency on Interstellar Medium Inhomogeneity}",
  journal = {\apj},
archivePrefix = "arXiv",
   eprint = {1205.0542},
     year = 2012,
    month = oct,
   volume = 757,
      eid = {136},
    pages = {136},
      doi = {10.1088/0004-637X/757/2/136},
   adsurl = {http://adsabs.harvard.edu/abs/2012ApJ...757..136W}
}

@ARTICLE{Ho08,
       author = {{Ho}, L.~C.},
        title = "{Nuclear activity in nearby galaxies.}",
      journal = {\araa},
         year = "2008",
        month = "Sep",
       volume = {46},
        pages = {475-539},
          doi = {10.1146/annurev.astro.45.051806.110546},
archivePrefix = {arXiv},
       eprint = {0803.2268},
 primaryClass = {astro-ph},
       adsurl = {https://ui.adsabs.harvard.edu/abs/2008ARA&A..46..475H}
}

@ARTICLE{Feruglio10,
       author = {{Feruglio}, C. and {Maiolino}, R. and {Piconcelli}, E. and {Menci}, N. and
         {Aussel}, H. and {Lamastra}, A. and {Fiore}, F.},
        title = "{Quasar feedback revealed by giant molecular outflows}",
      journal = {\aap},
         year = "2010",
        month = "Jul",
       volume = {518},
          eid = {L155},
        pages = {L155},
          doi = {10.1051/0004-6361/201015164},
archivePrefix = {arXiv},
       eprint = {1006.1655},
 primaryClass = {astro-ph.CO},
       adsurl = {https://ui.adsabs.harvard.edu/abs/2010A&A...518L.155F}
}

@ARTICLE{Morganti15,
       author = {{Morganti}, Raffaella and {Oosterloo}, Tom and {Oonk}, J.~B. Raymond and
         {Frieswijk}, Wilfred and {Tadhunter}, Clive},
        title = "{The fast molecular outflow in the Seyfert galaxy IC 5063 as seen by ALMA}",
      journal = {\aap},
         year = "2015",
        month = "Aug",
       volume = {580},
          eid = {A1},
        pages = {A1},
          doi = {10.1051/0004-6361/201525860},
archivePrefix = {arXiv},
       eprint = {1505.07190},
 primaryClass = {astro-ph.GA},
       adsurl = {https://ui.adsabs.harvard.edu/abs/2015A&A...580A...1M}
}

@INPROCEEDINGS{Tody86,
       author = {{Tody}, Doug},
        title = "{The IRAF Data Reduction and Analysis System}",
    booktitle = {\procspie},
         year = "1986",
       editor = {{Crawford}, David L.},
       series = {Society of Photo-Optical Instrumentation Engineers (SPIE) Conference Series},
       volume = {627},
        month = "Jan",
        pages = {733},
          doi = {10.1117/12.968154},
       adsurl = {https://ui.adsabs.harvard.edu/abs/1986SPIE..627..733T}
}

@ARTICLE{Ruffa19a,
       author = {{Ruffa}, Ilaria and {Prandoni}, Isabella and {Laing}, Robert A. and
         {Paladino}, Rosita and {Parma}, Paola and {de Ruiter}, Hans and
         {Mignano}, Arturo and {Davis}, Timothy A. and {Bureau}, Martin and
         {Warren}, Joshua},
        title = "{The AGN fuelling/feedback cycle in nearby radio galaxies I. ALMA observations and early results}",
      journal = {\mnras},
         year = "2019",
        month = "Apr",
       volume = {484},
       number = {3},
        pages = {4239-4259},
          doi = {10.1093/mnras/stz255},
archivePrefix = {arXiv},
       eprint = {1901.07513},
 primaryClass = {astro-ph.GA},
       adsurl = {https://ui.adsabs.harvard.edu/abs/2019MNRAS.484.4239R}
}

@ARTICLE{Wagner16,
       author = {{Wagner}, A.~Y. and {Bicknell}, G.~V. and {Umemura}, M. and {Sutherland
        }, R.~S. and {Silk}, J.},
        title = "{Galaxy-scale AGN feedback - theory}",
      journal = {Astronomische Nachrichten},
         year = 2016,
        month = feb,
       volume = {337},
       number = {1-2},
        pages = {167},
          doi = {10.1002/asna.201512287},
archivePrefix = {arXiv},
       eprint = {1510.03594},
 primaryClass = {astro-ph.GA},
       adsurl = {https://ui.adsabs.harvard.edu/abs/2016AN....337..167W}
}

@ARTICLE{Morganti17,
       author = {{Morganti}, Raffaella},
        title = "{The many routes to AGN feedback}",
      journal = {Frontiers in Astronomy and Space Sciences},
         year = 2017,
        month = nov,
       volume = {4},
          eid = {42},
        pages = {42},
          doi = {10.3389/fspas.2017.00042},
archivePrefix = {arXiv},
       eprint = {1712.05301},
 primaryClass = {astro-ph.GA},
       adsurl = {https://ui.adsabs.harvard.edu/abs/2017FrASS...4...42M}
}

@ARTICLE{McNamara12,
       author = {{McNamara}, B.~R. and {Nulsen}, P.~E.~J.},
        title = "{Mechanical feedback from active galactic nuclei in galaxies, groups and clusters}",
      journal = {New Journal of Physics},
         year = 2012,
        month = may,
       volume = {14},
       number = {5},
          eid = {055023},
        pages = {055023},
          doi = {10.1088/1367-2630/14/5/055023},
archivePrefix = {arXiv},
       eprint = {1204.0006},
 primaryClass = {astro-ph.CO},
       adsurl = {https://ui.adsabs.harvard.edu/abs/2012NJPh...14e5023M}
}

@ARTICLE{Zovaro19,
       author = {{Zovaro}, Henry R.~M. and {Sharp}, Robert and {Nesvadba}, Nicole P.~H. and
         {Bicknell}, Geoffrey V. and {Mukherjee}, Dipanjan and {Wagner}, Alexander Y. and {Groves}, Brent and {Krishna}, Shreyam},
        title = "{Jets blowing bubbles in the young radio galaxy 4C 31.04}",
      journal = {\mnras},
         year = 2019,
        month = apr,
       volume = {484},
       number = {3},
        pages = {3393-3409},
          doi = {10.1093/mnras/stz233},
archivePrefix = {arXiv},
       eprint = {1811.08971},
 primaryClass = {astro-ph.GA},
       adsurl = {https://ui.adsabs.harvard.edu/abs/2019MNRAS.484.3393Z}
}

@ARTICLE{Mukherjee16,
       author = {{Mukherjee}, Dipanjan and {Bicknell}, Geoffrey V. and {Sutherland
        }, Ralph and {Wagner}, Alex},
        title = "{Relativistic jet feedback in high-redshift galaxies - I. Dynamics}",
      journal = {\mnras},
         year = 2016,
        month = sep,
       volume = {461},
       number = {1},
        pages = {967-983},
          doi = {10.1093/mnras/stw1368},
archivePrefix = {arXiv},
       eprint = {1606.01143},
 primaryClass = {astro-ph.HE},
       adsurl = {https://ui.adsabs.harvard.edu/abs/2016MNRAS.461..967M}
}

@ARTICLE{Mukherjee18a,
       author = {{Mukherjee}, Dipanjan and {Wagner}, Alexander Y. and
         {Bicknell}, Geoffrey V. and {Morganti}, Raffaella and {Oosterloo}, Tom and
         {Nesvadba}, Nicole and {Sutherland}, Ralph S.},
        title = "{The jet-ISM interactions in IC 5063}",
      journal = {\mnras},
         year = 2018,
        month = may,
       volume = {476},
       number = {1},
        pages = {80-95},
          doi = {10.1093/mnras/sty067},
archivePrefix = {arXiv},
       eprint = {1801.06875},
 primaryClass = {astro-ph.HE},
       adsurl = {https://ui.adsabs.harvard.edu/abs/2018MNRAS.476...80M}
}

@ARTICLE{Mukherjee18b,
       author = {{Mukherjee}, Dipanjan and {Bicknell}, Geoffrey V. and {Wagner}, Alexander Y. and {Sutherland}, Ralph S. and {Silk}, Joseph},
        title = "{Relativistic jet feedback - III. Feedback on gas discs}",
      journal = {\mnras},
         year = 2018,
        month = oct,
       volume = {479},
       number = {4},
        pages = {5544-5566},
          doi = {10.1093/mnras/sty1776},
archivePrefix = {arXiv},
       eprint = {1803.08305},
 primaryClass = {astro-ph.HE},
       adsurl = {https://ui.adsabs.harvard.edu/abs/2018MNRAS.479.5544M}
}

@ARTICLE{Mahony16,
       author = {{Mahony}, E.~K. and {Oonk}, J.~B.~R. and {Morganti}, R. and
         {Tadhunter}, C. and {Bessiere}, P. and {Short}, P. and
         {Emonts}, B.~H.~C. and {Oosterloo}, T.~A.},
        title = "{Jet-driven outflows of ionized gas in the nearby radio galaxy 3C 293}",
      journal = {\mnras},
         year = 2016,
        month = jan,
       volume = {455},
       number = {3},
        pages = {2453-2460},
          doi = {10.1093/mnras/stv2456},
archivePrefix = {arXiv},
       eprint = {1510.06498},
 primaryClass = {astro-ph.GA},
       adsurl = {https://ui.adsabs.harvard.edu/abs/2016MNRAS.455.2453M}
}

@ARTICLE{Cavagnolo10,
       author = {{Cavagnolo}, K.~W. and {McNamara}, B.~R. and {Nulsen}, P.~E.~J. and
         {Carilli}, C.~L. and {Jones}, C. and {B{\^\i}rzan}, L.},
        title = "{A Relationship Between AGN Jet Power and Radio Power}",
      journal = {\apj},
         year = 2010,
        month = sep,
       volume = {720},
       number = {2},
        pages = {1066-1072},
          doi = {10.1088/0004-637X/720/2/1066},
archivePrefix = {arXiv},
       eprint = {1006.5699},
 primaryClass = {astro-ph.CO},
       adsurl = {https://ui.adsabs.harvard.edu/abs/2010ApJ...720.1066C}
}

@ARTICLE{Santoro20,
       author = {{Santoro}, F. and {Tadhunter}, C. and {Baron}, D. and {Morganti}, R. and
         {Holt}, J.},
        title = "{AGN-driven outflows and the AGN feedback efficiency in young radio galaxies}",
      journal = {arXiv e-prints},
         year = 2020,
        month = sep,
          eid = {arXiv:2009.11175},
        pages = {arXiv:2009.11175},
archivePrefix = {arXiv},
       eprint = {2009.11175},
 primaryClass = {astro-ph.GA},
       adsurl = {https://ui.adsabs.harvard.edu/abs/2020arXiv200911175S}
}

@ARTICLE{Venturi20,
       author = {{Venturi}, G. and {Cresci}, G. and {Marconi}, A. and {Mingozzi}, M. and
         {Nardini}, E. and {Carniani}, S. and {Mannucci}, F. and {Marasco}, A. and
         {Maiolino}, R. and {Perna}, M. and {Treister}, E. and {Bland
        -Hawthorn}, J. and {Gallimore}, J.},
        title = "{MAGNUM survey: compact jets causing large turmoil in galaxies -- Enhanced line widths perpendicular to radio jets as tracers of jet-ISM interaction}",
      journal = {arXiv e-prints},
         year = 2020,
        month = nov,
          eid = {arXiv:2011.04677},
        pages = {arXiv:2011.04677},
archivePrefix = {arXiv},
       eprint = {2011.04677},
 primaryClass = {astro-ph.GA},
       adsurl = {https://ui.adsabs.harvard.edu/abs/2020arXiv201104677V}
}

@ARTICLE{Feruglio15,
       author = {{Feruglio}, C. and {Fiore}, F. and {Carniani}, S. and {Piconcelli}, E. and
         {Zappacosta}, L. and {Bongiorno}, A. and {Cicone}, C. and
         {Maiolino}, R. and {Marconi}, A. and {Menci}, N. and {Puccetti}, S. and
         {Veilleux}, S.},
        title = "{The multi-phase winds of Markarian 231: from the hot, nuclear, ultra-fast wind to the galaxy-scale, molecular outflow}",
      journal = {\aap},
         year = 2015,
        month = nov,
       volume = {583},
          eid = {A99},
        pages = {A99},
          doi = {10.1051/0004-6361/201526020},
archivePrefix = {arXiv},
       eprint = {1503.01481},
 primaryClass = {astro-ph.GA},
       adsurl = {https://ui.adsabs.harvard.edu/abs/2015A&A...583A..99F}
}

@ARTICLE{Ruffa20,
       author = {{Ruffa}, Ilaria and {Laing}, Robert A. and {Prandoni}, Isabella and
         {Paladino}, Rosita and {Parma}, Paola and {Davis}, Timothy A. and
         {Bureau}, Martin},
        title = "{The AGN fuelling/feedback cycle in nearby radio galaxies - III. 3D relative orientations of radio jets and CO discs and their interaction}",
      journal = {\mnras},
         year = 2020,
        month = oct,
       volume = {499},
       number = {4},
        pages = {5719-5731},
          doi = {10.1093/mnras/staa3166},
archivePrefix = {arXiv},
       eprint = {2010.04685},
 primaryClass = {astro-ph.GA},
       adsurl = {https://ui.adsabs.harvard.edu/abs/2020MNRAS.499.5719R}
}

@MISC{Newville14,
       author = {{Newville}, Matthew and {Stensitzki}, Till and {Allen}, Daniel B. and {Ingargiola}, Antonino},
        title = "{LMFIT: Non-Linear Least-Square Minimization and Curve-Fitting for Python}",
         year = 2014,
        month = sep,
          eid = {10.5281/zenodo.11813},
          doi = {10.5281/zenodo.11813},
      version = {0.8.0},
    publisher = {Zenodo},
       adsurl = {https://ui.adsabs.harvard.edu/abs/2014zndo.....11813N}
}

@ARTICLE{ODea20,
       author = {{O'Dea}, Christopher P. and {Saikia}, D.~J.},
        title = "{Compact steep-spectrum and peaked-spectrum radio sources}",
      journal = {arXiv e-prints},
         year = 2020,
        month = sep,
          eid = {arXiv:2009.02750},
        pages = {arXiv:2009.02750},
archivePrefix = {arXiv},
       eprint = {2009.02750},
 primaryClass = {astro-ph.GA},
       adsurl = {https://ui.adsabs.harvard.edu/abs/2020arXiv200902750O}
}

@ARTICLE{Dominguez20,
       author = {{Dom{\'\i}nguez-Fern{\'a}ndez}, A.~J. and {Alonso-Herrero}, A. and {Garc{\'\i}a-Burillo}, S. and {Davies}, R.~I. and {Usero}, A. and {Labiano}, A. and {Levenson}, N.~A. and {Pereira-Santaella}, M. and {Imanishi}, M. and {Ramos Almeida}, C. and {Rigopoulou}, D.},
        title = "{Searching for molecular gas inflows and outflows in the nuclear regions of five Seyfert galaxies}",
      journal = {\aap},
         year = 2020,
        month = nov,
       volume = {643},
          eid = {A127},
        pages = {A127},
          doi = {10.1051/0004-6361/201936961},
archivePrefix = {arXiv},
       eprint = {2003.05663},
 primaryClass = {astro-ph.GA},
       adsurl = {https://ui.adsabs.harvard.edu/abs/2020A&A...643A.127D}
}

@ARTICLE{Maiolino12,
       author = {{Maiolino}, R. and {Gallerani}, S. and {Neri}, R. and {Cicone}, C. and {Ferrara}, A. and {Genzel}, R. and {Lutz}, D. and {Sturm}, E. and {Tacconi}, L.~J. and {Walter}, F. and {Feruglio}, C. and {Fiore}, F. and {Piconcelli}, E.},
        title = "{Evidence of strong quasar feedback in the early Universe}",
      journal = {\mnras},
         year = 2012,
        month = sep,
       volume = {425},
       number = {1},
        pages = {L66-L70},
          doi = {10.1111/j.1745-3933.2012.01303.x},
archivePrefix = {arXiv},
       eprint = {1204.2904},
 primaryClass = {astro-ph.CO},
       adsurl = {https://ui.adsabs.harvard.edu/abs/2012MNRAS.425L..66M}
}

@INPROCEEDINGS{Morganti21,
       author = {{Morganti}, Raffaella and {Oosterloo}, Tom and {Tadhunter}, Clive N.},
        title = "{Taking snapshots of the jet-ISM interplay with ALMA}",
    booktitle = {Galaxy Evolution and Feedback across Different Environments},
         year = 2021,
       editor = {{Storchi Bergmann}, Thaisa and {Forman}, William and {Overzier}, Roderik and {Riffel}, Rog{\'e}rio},
       series    = {IAU Symposium},
       volume = {359},
        month = {jan},
        pages = {243-248},
          doi = {10.1017/S1743921320001775},
archivePrefix = {arXiv},
       eprint = {2005.04765},
 primaryClass = {astro-ph.HE},
       adsurl = {https://ui.adsabs.harvard.edu/abs/2021IAUS..359..243M}
}

@ARTICLE{Veilleux20,
       author = {{Veilleux}, Sylvain and {Maiolino}, Roberto and {Bolatto}, Alberto D. and {Aalto}, Susanne},
        title = "{Cool outflows in galaxies and their implications}",
      journal = {\aapr},
         year = 2020,
        month = apr,
       volume = {28},
       number = {1},
          eid = {2},
        pages = {2},
          doi = {10.1007/s00159-019-0121-9},
archivePrefix = {arXiv},
       eprint = {2002.07765},
 primaryClass = {astro-ph.GA},
       adsurl = {https://ui.adsabs.harvard.edu/abs/2020A&ARv..28....2V}
}

@ARTICLE{Mukherjee21,
       author = {{Mukherjee}, Dipanjan and {Bicknell}, Geoffrey V. and {Wagner}, Alexander Y.},
        title = "{Resolved simulations of jet-ISM interaction: Implications for gas dynamics and star formation}",
      journal = {arXiv e-prints},
         year = 2021,
        month = oct,
          eid = {arXiv:2110.11900},
        pages = {arXiv:2110.11900},
archivePrefix = {arXiv},
       eprint = {2110.11900},
 primaryClass = {astro-ph.GA},
       adsurl = {https://ui.adsabs.harvard.edu/abs/2021arXiv211011900M}
}

@ARTICLE{Faber76,
       author = {{Faber}, S.~M. and {Jackson}, R.~E.},
        title = "{Velocity dispersions and mass-to-light ratios for elliptical galaxies.}",
      journal = {\apj},
         year = 1976,
        month = mar,
       volume = {204},
        pages = {668-683},
          doi = {10.1086/154215},
       adsurl = {https://ui.adsabs.harvard.edu/abs/1976ApJ...204..668F}
}

@ARTICLE{Murray05,
       author = {{Murray}, Norman and {Quataert}, Eliot and {Thompson}, Todd A.},
        title = "{On the Maximum Luminosity of Galaxies and Their Central Black Holes: Feedback from Momentum-driven Winds}",
      journal = {\apj},
         year = 2005,
        month = jan,
       volume = {618},
       number = {2},
        pages = {569-585},
          doi = {10.1086/426067},
archivePrefix = {arXiv},
       eprint = {astro-ph/0406070},
 primaryClass = {astro-ph},
       adsurl = {https://ui.adsabs.harvard.edu/abs/2005ApJ...618..569M}
}

@ARTICLE{Fiore17,
       author = {{Fiore}, F. and {Feruglio}, C. and {Shankar}, F. and {Bischetti}, M. and {Bongiorno}, A. and {Brusa}, M. and {Carniani}, S. and {Cicone}, C. and {Duras}, F. and {Lamastra}, A. and {Mainieri}, V. and {Marconi}, A. and {Menci}, N. and {Maiolino}, R. and {Piconcelli}, E. and {Vietri}, G. and {Zappacosta}, L.},
        title = "{AGN wind scaling relations and the co-evolution of black holes and galaxies}",
      journal = {\aap},
         year = 2017,
        month = may,
       volume = {601},
          eid = {A143},
        pages = {A143},
          doi = {10.1051/0004-6361/201629478},
archivePrefix = {arXiv},
       eprint = {1702.04507},
 primaryClass = {astro-ph.GA},
       adsurl = {https://ui.adsabs.harvard.edu/abs/2017A&A...601A.143F}
}

@ARTICLE{Cicone18,
       author = {{Cicone}, Claudia and {Brusa}, Marcella and {Ramos Almeida}, Cristina and {Cresci}, Giovanni and {Husemann}, Bernd and {Mainieri}, Vincenzo},
        title = "{The largely unconstrained multiphase nature of outflows in AGN host galaxies}",
      journal = {Nature Astronomy},
         year = 2018,
        month = feb,
       volume = {2},
        pages = {176-178},
          doi = {10.1038/s41550-018-0406-3},
archivePrefix = {arXiv},
       eprint = {1802.10308},
 primaryClass = {astro-ph.GA},
       adsurl = {https://ui.adsabs.harvard.edu/abs/2018NatAs...2..176C}
}

@ARTICLE{Ruffa18,
       author = {{Ruffa}, I. and {Vignali}, C. and {Mignano}, A. and {Paladino}, R. and {Iwasawa}, K.},
        title = "{The role of molecular gas in the nuclear regions of IRAS 00183-7111. ALMA and X-ray investigations of an ultraluminous infrared galaxy}",
      journal = {\aap},
         year = 2018,
        month = sep,
       volume = {616},
          eid = {A127},
        pages = {A127},
          doi = {10.1051/0004-6361/201732268},
archivePrefix = {arXiv},
       eprint = {1805.06477},
 primaryClass = {astro-ph.GA},
       adsurl = {https://ui.adsabs.harvard.edu/abs/2018A&A...616A.127R}
}

@ARTICLE{Lister03,
       author = {{Lister}, M.~L. and {Kellermann}, K.~I. and {Vermeulen}, R.~C. and {Cohen}, M.~H. and {Zensus}, J.~A. and {Ros}, E.},
        title = "{4C +12.50: A Superluminal Precessing Jet in the Recent Merger System IRAS 13451+1232}",
      journal = {\apj},
         year = 2003,
        month = feb,
       volume = {584},
       number = {1},
        pages = {135-146},
          doi = {10.1086/345666},
archivePrefix = {arXiv},
       eprint = {astro-ph/0210372},
 primaryClass = {astro-ph},
       adsurl = {https://ui.adsabs.harvard.edu/abs/2003ApJ...584..135L}
}

@ARTICLE{Holt06,
       author = {{Holt}, J. and {Tadhunter}, C. and {Morganti}, R. and {Bellamy}, M. and {Gonz{\'a}lez Delgado}, R.~M. and {Tzioumis}, A. and {Inskip}, K.~J.},
        title = "{The co-evolution of the obscured quasar PKS 1549-79 and its host galaxy: evidence for a high accretion rate and warm outflow}",
      journal = {\mnras},
         year = 2006,
        month = aug,
       volume = {370},
       number = {4},
        pages = {1633-1650},
          doi = {10.1111/j.1365-2966.2006.10604.x},
archivePrefix = {arXiv},
       eprint = {astro-ph/0606304},
 primaryClass = {astro-ph},
       adsurl = {https://ui.adsabs.harvard.edu/abs/2006MNRAS.370.1633H}
}

@ARTICLE{Capaccioli2015,
   author = {{Capaccioli}, M. and {Spavone}, M. and {Grado}, A. and {Iodice}, E. and 
	{Limatola}, L. and {Napolitano}, N.~R. and {Cantiello}, M. and 
	{Paolillo}, M. and {Romanowsky}, A.~J. and {Forbes}, D.~A. and 
	{Puzia}, T.~H. and {Raimondo}, G. and {Schipani}, P.},
    title = "{VEGAS: A VST Early-type GAlaxy Survey. I. Presentation, wide-field surface photometry, and substructures in NGC 4472}",
  journal = {\aap},
archivePrefix = "arXiv",
   eprint = {1507.01336},
     year = 2015,
    month = sep,
   volume = 581,
      eid = {A10},
    pages = {A10},
      doi = {10.1051/0004-6361/201526252},
   adsurl = {http://adsabs.harvard.edu/abs/2015A%26A...581A..10C}
}

@ARTICLE{Iodice2021,
       author = {{Iodice}, E. and {Spavone}, M. and {Raj}, M.~A. and {Capaccioli}, M. and {Cantiello}, M. and {VEGAS science team}},
        title = "{The VST Early-type GAlaxy Survey (VEGAS) data release 1}",
      journal = {arXiv e-prints},
         year = 2021,
        month = feb,
          eid = {arXiv:2102.04950},
        pages = {arXiv:2102.04950},
          doi = {10.48550/arXiv.2102.04950},
archivePrefix = {arXiv},
       eprint = {2102.04950},
 primaryClass = {astro-ph.GA},
       adsurl = {https://ui.adsabs.harvard.edu/abs/2021arXiv210204950I}
}

@ARTICLE{Schipani2012,
   author = {{Schipani}, P. and {Noethe}, L. and {Arcidiacono}, C. and {Argomedo}, J. and 
	{Dall'Ora}, M. and {D'Orsi}, S. and {Farinato}, J. and {Magrin}, D. and 
	{Marty}, L. and {Ragazzoni}, R. and {Umbriaco}, G.},
    title = "{Removing static aberrations from the active optics system of a wide-field telescope}",
  journal = {Journal of the Optical Society of America A},
     year = 2012,
    month = jul,
   volume = 29,
    pages = {1359},
      doi = {10.1364/JOSAA.29.001359},
   adsurl = {http://adsabs.harvard.edu/abs/2012JOSAA..29.1359S}
}

@ARTICLE{Spavone2017,
       author = {{Spavone}, Marilena and {Capaccioli}, Massimo and {Napolitano}, Nicola R. and {Iodice}, Enrichetta and {Grado}, Aniello and {Limatola}, Luca and {Cooper}, Andrew P. and {Cantiello}, Michele and {Forbes}, Duncan A. and {Paolillo}, Maurizio and {Schipani}, Pietro},
        title = "{VEGAS: A VST Early-type GAlaxy Survey. II. Photometric study of giant ellipticals and their stellar halos}",
      journal = {\aap},
         year = 2017,
        month = jul,
       volume = {603},
          eid = {A38},
        pages = {A38},
          doi = {10.1051/0004-6361/201629111},
archivePrefix = {arXiv},
       eprint = {1703.10835},
 primaryClass = {astro-ph.GA},
       adsurl = {https://ui.adsabs.harvard.edu/abs/2017A&A...603A..38S}
}

@ARTICLE{McFarland2013,
       author = {{McFarland}, John P. and {Verdoes-Kleijn}, Gijs and {Sikkema}, Gert and
         {Helmich}, Ewout M. and {Boxhoorn}, Danny R. and {Valentijn}, Edwin A.},
        title = "{The Astro-WISE optical image pipeline. Development and implementation}",
      journal = {Experimental Astronomy},
         year = 2013,
        month = jan,
       volume = {35},
       number = {1-2},
        pages = {45-78},
          doi = {10.1007/s10686-011-9266-x},
       adsurl = {https://ui.adsabs.harvard.edu/abs/2013ExA....35...45M}
}

@ARTICLE{Venhola2017,
   author = {{Venhola}, A. and {Peletier}, R. and {Laurikainen}, E. and {Salo}, H. and 
	{Lisker}, T. and {Iodice}, E. and {Capaccioli}, M. and {Kleijn}, G.~V. and 
	{Valentijn}, E. and {Mieske}, S. and {Hilker}, M. and {Wittmann}, C. and 
	{van de Ven}, G. and {Grado}, A. and {Spavone}, M. and {Cantiello}, M. and 
	{Napolitano}, N. and {Paolillo}, M. and {Falc{\'o}n-Barroso}, J.
	},
    title = "{The Fornax Deep Survey with VST. III. Low surface brightness dwarfs and ultra diffuse galaxies in the center of the Fornax cluster}",
  journal = {\aap},
archivePrefix = "arXiv",
   eprint = {1710.04616},
     year = 2017,
    month = dec,
   volume = 608,
      eid = {A142},
    pages = {A142},
      doi = {10.1051/0004-6361/201730696},
   adsurl = {http://adsabs.harvard.edu/abs/2017A%26A...608A.142V}
}

@ARTICLE{Astropy13,
       author = {{Astropy Collaboration} and {Robitaille}, Thomas P. and {Tollerud}, Erik J. and {Greenfield}, Perry and {Droettboom}, Michael and {Bray}, Erik and {Aldcroft}, Tom and {Davis}, Matt and {Ginsburg}, Adam and {Price-Whelan}, Adrian M. and {Kerzendorf}, Wolfgang E. and {Conley}, Alexander and {Crighton}, Neil and {Barbary}, Kyle and {Muna}, Demitri and {Ferguson}, Henry and {Grollier}, Fr{\'e}d{\'e}ric and {Parikh}, Madhura M. and {Nair}, Prasanth H. and {Unther}, Hans M. and {Deil}, Christoph and {Woillez}, Julien and {Conseil}, Simon and {Kramer}, Roban and {Turner}, James E.~H. and {Singer}, Leo and {Fox}, Ryan and {Weaver}, Benjamin A. and {Zabalza}, Victor and {Edwards}, Zachary I. and {Azalee Bostroem}, K. and {Burke}, D.~J. and {Casey}, Andrew R. and {Crawford}, Steven M. and {Dencheva}, Nadia and {Ely}, Justin and {Jenness}, Tim and {Labrie}, Kathleen and {Lim}, Pey Lian and {Pierfederici}, Francesco and {Pontzen}, Andrew and {Ptak}, Andy and {Refsdal}, Brian and {Servillat}, Mathieu and {Streicher}, Ole},
        title = "{Astropy: A community Python package for astronomy}",
      journal = {\aap},
         year = 2013,
        month = oct,
       volume = {558},
          eid = {A33},
        pages = {A33},
          doi = {10.1051/0004-6361/201322068},
archivePrefix = {arXiv},
       eprint = {1307.6212},
 primaryClass = {astro-ph.IM},
       adsurl = {https://ui.adsabs.harvard.edu/abs/2013A&A...558A..33A}
}

@ARTICLE{Astropy18,
       author = {{Astropy Collaboration} and {Price-Whelan}, A.~M. and {Sip{\H{o}}cz}, B.~M. and {G{\"u}nther}, H.~M. and {Lim}, P.~L. and {Crawford}, S.~M. and {Conseil}, S. and {Shupe}, D.~L. and {Craig}, M.~W. and {Dencheva}, N. and {Ginsburg}, A. and {VanderPlas}, J.~T. and {Bradley}, L.~D. and {P{\'e}rez-Su{\'a}rez}, D. and {de Val-Borro}, M. and {Aldcroft}, T.~L. and {Cruz}, K.~L. and {Robitaille}, T.~P. and {Tollerud}, E.~J. and {Ardelean}, C. and {Babej}, T. and {Bach}, Y.~P. and {Bachetti}, M. and {Bakanov}, A.~V. and {Bamford}, S.~P. and {Barentsen}, G. and {Barmby}, P. and {Baumbach}, A. and {Berry}, K.~L. and {Biscani}, F. and {Boquien}, M. and {Bostroem}, K.~A. and {Bouma}, L.~G. and {Brammer}, G.~B. and {Bray}, E.~M. and {Breytenbach}, H. and {Buddelmeijer}, H. and {Burke}, D.~J. and {Calderone}, G. and {Cano Rodr{\'\i}guez}, J.~L. and {Cara}, M. and {Cardoso}, J.~V.~M. and {Cheedella}, S. and {Copin}, Y. and {Corrales}, L. and {Crichton}, D. and {D'Avella}, D. and {Deil}, C. and {Depagne}, {\'E}. and {Dietrich}, J.~P. and {Donath}, A. and {Droettboom}, M. and {Earl}, N. and {Erben}, T. and {Fabbro}, S. and {Ferreira}, L.~A. and {Finethy}, T. and {Fox}, R.~T. and {Garrison}, L.~H. and {Gibbons}, S.~L.~J. and {Goldstein}, D.~A. and {Gommers}, R. and {Greco}, J.~P. and {Greenfield}, P. and {Groener}, A.~M. and {Grollier}, F. and {Hagen}, A. and {Hirst}, P. and {Homeier}, D. and {Horton}, A.~J. and {Hosseinzadeh}, G. and {Hu}, L. and {Hunkeler}, J.~S. and {Ivezi{\'c}}, {\v{Z}}. and {Jain}, A. and {Jenness}, T. and {Kanarek}, G. and {Kendrew}, S. and {Kern}, N.~S. and {Kerzendorf}, W.~E. and {Khvalko}, A. and {King}, J. and {Kirkby}, D. and {Kulkarni}, A.~M. and {Kumar}, A. and {Lee}, A. and {Lenz}, D. and {Littlefair}, S.~P. and {Ma}, Z. and {Macleod}, D.~M. and {Mastropietro}, M. and {McCully}, C. and {Montagnac}, S. and {Morris}, B.~M. and {Mueller}, M. and {Mumford}, S.~J. and {Muna}, D. and {Murphy}, N.~A. and {Nelson}, S. and {Nguyen}, G.~H. and {Ninan}, J.~P. and {N{\"o}the}, M. and {Ogaz}, S. and {Oh}, S. and {Parejko}, J.~K. and {Parley}, N. and {Pascual}, S. and {Patil}, R. and {Patil}, A.~A. and {Plunkett}, A.~L. and {Prochaska}, J.~X. and {Rastogi}, T. and {Reddy Janga}, V. and {Sabater}, J. and {Sakurikar}, P. and {Seifert}, M. and {Sherbert}, L.~E. and {Sherwood-Taylor}, H. and {Shih}, A.~Y. and {Sick}, J. and {Silbiger}, M.~T. and {Singanamalla}, S. and {Singer}, L.~P. and {Sladen}, P.~H. and {Sooley}, K.~A. and {Sornarajah}, S. and {Streicher}, O. and {Teuben}, P. and {Thomas}, S.~W. and {Tremblay}, G.~R. and {Turner}, J.~E.~H. and {Terr{\'o}n}, V. and {van Kerkwijk}, M.~H. and {de la Vega}, A. and {Watkins}, L.~L. and {Weaver}, B.~A. and {Whitmore}, J.~B. and {Woillez}, J. and {Zabalza}, V. and {Astropy Contributors}},
        title = "{The Astropy Project: Building an Open-science Project and Status of the v2.0 Core Package}",
      journal = {\aj},
         year = 2018,
        month = sep,
       volume = {156},
       number = {3},
          eid = {123},
        pages = {123},
          doi = {10.3847/1538-3881/aabc4f},
archivePrefix = {arXiv},
       eprint = {1801.02634},
 primaryClass = {astro-ph.IM},
       adsurl = {https://ui.adsabs.harvard.edu/abs/2018AJ....156..123A}
}

@ARTICLE{Hunter07,
       author = {{Hunter}, John D.},
        title = "{Matplotlib: a 2D graphics environment}",
      journal = {Computing in Science \& Engineering},
         year = 2007,
        month = jun,
       volume = {9},
       number = {3},
        pages = {90-95},
          doi = {10.1109/MCSE.2007.55},
}

@ARTICLE{Walt11,
       author = {{van der Walt}, Stefan and {Colbert} Chris S. and {Varoquaux} Gael},
        title = "{The NumPy Array: A Structure for Efficient Numerical Computation}",
      journal = {Computing in Science \& Engineering},
         year = 2011,
        month = mar,
       volume = {13},
       number = {2},
        pages = {22-30},
          doi = {10.1109/MCSE.2011.37},
}

@ARTICLE{Harris20,
       author = {{Harris}, C.~R. and {Millman}, K.~J. and {van der Walt}, S.~J. and {Gommers}, R. and {Virtanen}, P. and {Cournapeau}, D. and {Wieser}, E. and {Taylor}, J. and {Berg}, S. and {Smith}, N.~J. and {Kern}, R. and {Picus}, M. and {Hoyer}, S. and {van Kerkwijk}, M.~H. and {Brett}, M. and {Haldane}, A. and {del Río}, J.~F. and {Wiebe}, M. and {Peterson}, P. and {Gérard-Marchant}, P. and {Sheppard}, K. and {Reddy}, T. and {Weckesser}, W. and {Abbasi}, H. and {Gohlke}, C. and {Oliphant}, T.~E.},
        title = "{Array programming with NumPy}",
      journal = {\nat},
         year = 2020,
        month = sept,
       volume = {585},
        pages = {357-362},
          doi = {https://doi.org/10.1038/s41586-020-2649-2},
}

@ARTICLE{Ruffa22,
       author = {{Ruffa}, Ilaria and {Prandoni}, Isabella and {Davis}, Timothy A. and {Laing}, Robert A. and {Paladino}, Rosita and {Casasola}, Viviana and {Parma}, Paola and {Bureau}, Martin},
        title = "{The AGN fuelling/feedback cycle in nearby radio galaxies - IV. Molecular gas conditions and jet-ISM interaction in NGC 3100}",
      journal = {\mnras},
         year = 2022,
        month = mar,
       volume = {510},
       number = {3},
        pages = {4485-4503},
          doi = {10.1093/mnras/stab3541},
archivePrefix = {arXiv},
       eprint = {2112.00755},
 primaryClass = {astro-ph.GA},
       adsurl = {https://ui.adsabs.harvard.edu/abs/2022MNRAS.510.4485R}
}

@ARTICLE{Fotopoulou19,
       author = {{Fotopoulou}, C.~M. and {Dasyra}, K.~M. and {Combes}, F. and {Salom{\'e}}, P. and {Papachristou}, M.},
        title = "{Complex molecular gas kinematics in the inner 5 kpc of 4C12.50 as seen by ALMA}",
      journal = {\aap},
         year = 2019,
        month = sep,
       volume = {629},
          eid = {A30},
        pages = {A30},
          doi = {10.1051/0004-6361/201834416},
archivePrefix = {arXiv},
       eprint = {1908.01011},
 primaryClass = {astro-ph.GA},
       adsurl = {https://ui.adsabs.harvard.edu/abs/2019A&A...629A..30F}
}

@ARTICLE{Mancillas19,
       author = {{Mancillas}, Brisa and {Duc}, Pierre-Alain and {Combes}, Fran{\c{c}}oise and {Bournaud}, Fr{\'e}d{\'e}ric and {Emsellem}, Eric and {Martig}, Marie and {Michel-Dansac}, Leo},
        title = "{Probing the merger history of red early-type galaxies with their faint stellar substructures}",
      journal = {\aap},
         year = 2019,
        month = dec,
       volume = {632},
          eid = {A122},
        pages = {A122},
          doi = {10.1051/0004-6361/201936320},
archivePrefix = {arXiv},
       eprint = {1909.07500},
 primaryClass = {astro-ph.GA},
       adsurl = {https://ui.adsabs.harvard.edu/abs/2019A&A...632A.122M}
}

@ARTICLE{Chilingarian10,
       author = {{Chilingarian}, I.~V. and {Di Matteo}, P. and {Combes}, F. and {Melchior}, A. -L. and {Semelin}, B.},
        title = "{The GalMer database: galaxy mergers in the virtual observatory}",
      journal = {\aap},
         year = 2010,
        month = jul,
       volume = {518},
          eid = {A61},
        pages = {A61},
          doi = {10.1051/0004-6361/200912938},
archivePrefix = {arXiv},
       eprint = {1003.3243},
 primaryClass = {astro-ph.IM},
       adsurl = {https://ui.adsabs.harvard.edu/abs/2010A&A...518A..61C}
}

@ARTICLE{Toomre72,
       author = {{Toomre}, Alar and {Toomre}, Juri},
        title = "{Galactic Bridges and Tails}",
      journal = {\apj},
         year = 1972,
        month = dec,
       volume = {178},
        pages = {623-666},
          doi = {10.1086/151823},
       adsurl = {https://ui.adsabs.harvard.edu/abs/1972ApJ...178..623T}
}

@ARTICLE{Mihos05,
       author = {{Mihos}, J. Christopher and {Harding}, Paul and {Feldmeier}, John and {Morrison}, Heather},
        title = "{Diffuse Light in the Virgo Cluster}",
      journal = {\apjl},
         year = 2005,
        month = sep,
       volume = {631},
       number = {1},
        pages = {L41-L44},
          doi = {10.1086/497030},
archivePrefix = {arXiv},
       eprint = {astro-ph/0508217},
 primaryClass = {astro-ph},
       adsurl = {https://ui.adsabs.harvard.edu/abs/2005ApJ...631L..41M}
}

@ARTICLE{Janowiecki10,
       author = {{Janowiecki}, Steven and {Mihos}, J. Christopher and {Harding}, Paul and {Feldmeier}, John J. and {Rudick}, Craig and {Morrison}, Heather},
        title = "{Diffuse Tidal Structures in the Halos of Virgo Ellipticals}",
      journal = {\apj},
         year = 2010,
        month = jun,
       volume = {715},
       number = {2},
        pages = {972-985},
          doi = {10.1088/0004-637X/715/2/972},
archivePrefix = {arXiv},
       eprint = {1004.1473},
 primaryClass = {astro-ph.CO},
       adsurl = {https://ui.adsabs.harvard.edu/abs/2010ApJ...715..972J}
}

@ARTICLE{Serra19,
       author = {{Serra}, P. and {Maccagni}, F.~M. and {Kleiner}, D. and {de Blok}, W.~J.~G. and {van Gorkom}, J.~H. and {Hugo}, B. and {Iodice}, E. and {J{\'o}zsa}, G.~I.~G. and {Kamphuis}, P. and {Kraan-Korteweg}, R. and {Loni}, A. and {Makhathini}, S. and {Moln{\'a}r}, D. and {Oosterloo}, T. and {Peletier}, R. and {Ramaila}, A. and {Ramatsoku}, M. and {Smirnov}, O. and {Smith}, M. and {Spavone}, M. and {Thorat}, K. and {Trager}, S.~C. and {Venhola}, A.},
        title = "{Neutral hydrogen gas within and around NGC 1316}",
      journal = {\aap},
         year = 2019,
        month = aug,
       volume = {628},
          eid = {A122},
        pages = {A122},
          doi = {10.1051/0004-6361/201936114},
archivePrefix = {arXiv},
       eprint = {1907.08265},
 primaryClass = {astro-ph.GA},
       adsurl = {https://ui.adsabs.harvard.edu/abs/2019A&A...628A.122S}
}

@INPROCEEDINGS{Beaulieu22,
       author = {{Beaulieu}, Damien and {Petric}, Andreea and {Robert}, Carmelle and {Alatalo}, Katherine and {Heckman}, Timothy and {Merhi}, Maya and {Rousseau-Nepton}, Laurie and {Rowlands}, Kate},
        title = "{Forming Stars in a Dual AGN Host: Molecular and Ionized Gas in the Nearby, Luminous Infrared Merger, Mrk 266}",
    booktitle = {American Astronomical Society Meeting Abstracts},
         year = 2022,
       series = {American Astronomical Society Meeting Abstracts},
       volume = {54},
        month = jun,
          eid = {213.05},
        pages = {213.05},
       adsurl = {https://ui.adsabs.harvard.edu/abs/2022AAS...24021305B}
}

@ARTICLE{Su23,
       author = {{Su}, Renzhi and {Mahony}, Elizabeth K. and {Gu}, Minfeng and {Sadler}, Elaine M. and {Curran}, S.~J. and {Allison}, James R. and {Yoon}, Hyein and {Aditya}, J.~N.~H.~S. and {Chandola}, Yogesh and {Chen}, Yongjun and {Moss}, Vanessa A. and {Wu}, Zhongzu and {Shao}, Xi and {Liu}, Xiang and {Glowacki}, Marcin and {Whiting}, Matthew T. and {Weng}, Simon},
        title = "{Does a radio jet drive the massive multiphase outflow in the ultra-luminous infrared galaxy IRAS 10565 + 2448?}",
      journal = {\mnras},
         year = 2023,
        month = apr,
       volume = {520},
       number = {4},
        pages = {5712-5723},
          doi = {10.1093/mnras/stad370},
archivePrefix = {arXiv},
       eprint = {2302.00943},
 primaryClass = {astro-ph.GA},
       adsurl = {https://ui.adsabs.harvard.edu/abs/2023MNRAS.520.5712S}
}

@ARTICLE{Mihos96,
       author = {{Mihos}, J. Christopher and {Hernquist}, Lars},
        title = "{Gasdynamics and Starbursts in Major Mergers}",
      journal = {\apj},
         year = 1996,
        month = jun,
       volume = {464},
        pages = {641},
          doi = {10.1086/177353},
archivePrefix = {arXiv},
       eprint = {astro-ph/9512099},
 primaryClass = {astro-ph},
       adsurl = {https://ui.adsabs.harvard.edu/abs/1996ApJ...464..641M}
}

@ARTICLE{Ruffa23,
       author = {{Ruffa}, Ilaria and {Davis}, Timothy A. and {Cappellari}, Michele and {Bureau}, Martin and {Elford}, Jacob and {Iguchi}, Satoru and {Lelli}, Federico and {Liang}, Fu-Heng and {Liu}, Lijie and {Lu}, Anan and {Sarzi}, Marc and {Williams}, Thomas G.},
        title = "{WISDOM project - XIV. SMBH mass in the early-type galaxies NGC 0612, NGC 1574, and NGC 4261 from CO dynamical modelling}",
      journal = {\mnras},
         year = 2023,
        month = jul,
       volume = {522},
       number = {4},
        pages = {6170-6195},
          doi = {10.1093/mnras/stad1119},
archivePrefix = {arXiv},
       eprint = {2304.06117},
 primaryClass = {astro-ph.GA},
       adsurl = {https://ui.adsabs.harvard.edu/abs/2023MNRAS.522.6170R}
}

@ARTICLE{Murthy22,
       author = {{Murthy}, Suma and {Morganti}, Raffaella and {Wagner}, Alexander Y. and {Oosterloo}, Tom and {Guillard}, Pierre and {Mukherjee}, Dipanjan and {Bicknell}, Geoffrey},
        title = "{Cold gas removal from the centre of a galaxy by a low-luminosity jet}",
      journal = {Nature Astronomy},
         year = 2022,
        month = feb,
       volume = {6},
        pages = {488-495},
          doi = {10.1038/s41550-021-01596-6},
archivePrefix = {arXiv},
       eprint = {2202.05222},
 primaryClass = {astro-ph.GA},
       adsurl = {https://ui.adsabs.harvard.edu/abs/2022NatAs...6..488M}
}

@ARTICLE{Audibert23,
       author = {{Audibert}, A. and {Ramos Almeida}, C. and {Garc{\'\i}a-Burillo}, S. and {Combes}, F. and {Bischetti}, M. and {Meenakshi}, M. and {Mukherjee}, D. and {Bicknell}, G. and {Wagner}, A.~Y.},
        title = "{Jet-induced molecular gas excitation and turbulence in the Teacup}",
      journal = {\aap},
         year = 2023,
        month = mar,
       volume = {671},
          eid = {L12},
        pages = {L12},
          doi = {10.1051/0004-6361/202345964},
archivePrefix = {arXiv},
       eprint = {2302.13884},
 primaryClass = {astro-ph.GA},
       adsurl = {https://ui.adsabs.harvard.edu/abs/2023A&A...671L..12A}
}

@ARTICLE{Cicone14,
       author = {{Cicone}, C. and {Maiolino}, R. and {Sturm}, E. and {Graci{\'a}-Carpio}, J. and {Feruglio}, C. and {Neri}, R. and {Aalto}, S. and {Davies}, R. and {Fiore}, F. and {Fischer}, J. and {Garc{\'\i}a-Burillo}, S. and {Gonz{\'a}lez-Alfonso}, E. and {Hailey-Dunsheath}, S. and {Piconcelli}, E. and {Veilleux}, S.},
        title = "{Massive molecular outflows and evidence for AGN feedback from CO observations}",
      journal = {\aap},
         year = 2014,
        month = feb,
       volume = {562},
          eid = {A21},
        pages = {A21},
          doi = {10.1051/0004-6361/201322464},
archivePrefix = {arXiv},
       eprint = {1311.2595},
 primaryClass = {astro-ph.CO},
       adsurl = {https://ui.adsabs.harvard.edu/abs/2014A&A...562A..21C}
}

@ARTICLE{Saito18,
       author = {{Saito}, T. and {Iono}, D. and {Ueda}, J. and {Espada}, D. and {Sliwa}, K. and {Nakanishi}, K. and {Lu}, N. and {Xu}, C.~K. and {Michiyama}, T. and {Kaneko}, H. and {Yamashita}, T. and {Ando}, M. and {Yun}, M.~S. and {Motohara}, K. and {Kawabe}, R.},
        title = "{Imaging the molecular outflows of the prototypical ULIRG NGC 6240 with ALMA}",
      journal = {\mnras},
         year = 2018,
        month = mar,
       volume = {475},
       number = {1},
        pages = {L52-L56},
          doi = {10.1093/mnrasl/slx207},
archivePrefix = {arXiv},
       eprint = {1712.07660},
 primaryClass = {astro-ph.GA},
       adsurl = {https://ui.adsabs.harvard.edu/abs/2018MNRAS.475L..52S}
}

@ARTICLE{Pereira16,
       author = {{Pereira-Santaella}, M. and {Colina}, L. and {Garc{\'\i}a-Burillo}, S. and {Alonso-Herrero}, A. and {Arribas}, S. and {Cazzoli}, S. and {Emonts}, B. and {Piqueras L{\'o}pez}, J. and {Planesas}, P. and {Storchi Bergmann}, T. and {Usero}, A. and {Villar-Mart{\'\i}n}, M.},
        title = "{High-velocity extended molecular outflow in the star-formation dominated luminous infrared galaxy ESO 320-G030}",
      journal = {\aap},
         year = 2016,
        month = oct,
       volume = {594},
          eid = {A81},
        pages = {A81},
          doi = {10.1051/0004-6361/201628875},
archivePrefix = {arXiv},
       eprint = {1607.03674},
 primaryClass = {astro-ph.GA},
       adsurl = {https://ui.adsabs.harvard.edu/abs/2016A&A...594A..81P}
}

@ARTICLE{Pereira18,
       author = {{Pereira-Santaella}, M. and {Colina}, L. and {Garc{\'\i}a-Burillo}, S. and {Combes}, F. and {Emonts}, B. and {Aalto}, S. and {Alonso-Herrero}, A. and {Arribas}, S. and {Henkel}, C. and {Labiano}, A. and {Muller}, S. and {Piqueras L{\'o}pez}, J. and {Rigopoulou}, D. and {van der Werf}, P.},
        title = "{Spatially resolved cold molecular outflows in ULIRGs}",
      journal = {\aap},
         year = 2018,
        month = aug,
       volume = {616},
          eid = {A171},
        pages = {A171},
          doi = {10.1051/0004-6361/201833089},
archivePrefix = {arXiv},
       eprint = {1805.03667},
 primaryClass = {astro-ph.GA},
       adsurl = {https://ui.adsabs.harvard.edu/abs/2018A&A...616A.171P}
}

@ARTICLE{Garcia15,
       author = {{Garc{\'\i}a-Burillo}, S. and {Combes}, F. and {Usero}, A. and {Aalto}, S. and {Colina}, L. and {Alonso-Herrero}, A. and {Hunt}, L.~K. and {Arribas}, S. and {Costagliola}, F. and {Labiano}, A. and {Neri}, R. and {Pereira-Santaella}, M. and {Tacconi}, L.~J. and {van der Werf}, P.~P.},
        title = "{High-resolution imaging of the molecular outflows in two mergers: <ASTROBJ>IRAS 17208-0014</ASTROBJ> and <ASTROBJ>NGC 1614</ASTROBJ>}",
      journal = {\aap},
         year = 2015,
        month = aug,
       volume = {580},
          eid = {A35},
        pages = {A35},
          doi = {10.1051/0004-6361/201526133},
archivePrefix = {arXiv},
       eprint = {1505.04705},
 primaryClass = {astro-ph.GA},
       adsurl = {https://ui.adsabs.harvard.edu/abs/2015A&A...580A..35G}
}

@ARTICLE{Willott99,
       author = {{Willott}, Chris J. and {Rawlings}, Steve and {Blundell}, Katherine M. and {Lacy}, Mark},
        title = "{The emission line-radio correlation for radio sources using the 7C Redshift Survey}",
      journal = {\mnras},
         year = 1999,
        month = nov,
       volume = {309},
       number = {4},
        pages = {1017-1033},
          doi = {10.1046/j.1365-8711.1999.02907.x},
archivePrefix = {arXiv},
       eprint = {astro-ph/9905388},
 primaryClass = {astro-ph},
       adsurl = {https://ui.adsabs.harvard.edu/abs/1999MNRAS.309.1017W}
}

@ARTICLE{Swinbank14,
       author = {{Swinbank}, A.~M. and {Simpson}, J.~M. and {Smail}, Ian and {Harrison}, C.~M. and {Hodge}, J.~A. and {Karim}, A. and {Walter}, F. and {Alexander}, D.~M. and {Brandt}, W.~N. and {de Breuck}, C. and {da Cunha}, E. and {Chapman}, S.~C. and {Coppin}, K.~E.~K. and {Danielson}, A.~L.~R. and {Dannerbauer}, H. and {Decarli}, R. and {Greve}, T.~R. and {Ivison}, R.~J. and {Knudsen}, K.~K. and {Lagos}, C.~D.~P. and {Schinnerer}, E. and {Thomson}, A.~P. and {Wardlow}, J.~L. and {Wei{\ss}}, A. and {van der Werf}, P.},
        title = "{An ALMA survey of sub-millimetre Galaxies in the Extended Chandra Deep Field South: the far-infrared properties of SMGs}",
      journal = {\mnras},
         year = 2014,
        month = feb,
       volume = {438},
       number = {2},
        pages = {1267-1287},
          doi = {10.1093/mnras/stt2273},
archivePrefix = {arXiv},
       eprint = {1310.6362},
 primaryClass = {astro-ph.CO},
       adsurl = {https://ui.adsabs.harvard.edu/abs/2014MNRAS.438.1267S}
}

@ARTICLE{Ruffa24,
       author = {{Ruffa}, Ilaria and {Davis}, Timothy A.},
        title = "{Molecular Gas Kinematics in Local Early-Type Galaxies with ALMA}",
      journal = {Galaxies},
         year = 2024,
        month = jul,
       volume = {12},
       number = {4},
          eid = {36},
        pages = {36},
          doi = {10.3390/galaxies12040036},
archivePrefix = {arXiv},
       eprint = {2405.10683},
 primaryClass = {astro-ph.GA},
       adsurl = {https://ui.adsabs.harvard.edu/abs/2024Galax..12...36R}
}

@ARTICLE{Holden24,
       author = {{Holden}, Luke R. and {Tadhunter}, Clive and {Audibert}, Anelise and {Oosterloo}, Tom and {Ramos Almeida}, Cristina and {Morganti}, Raffaella and {Pereira-Santaella}, Miguel and {Lamperti}, Isabella},
        title = "{ALMA reveals a compact and massive molecular outflow driven by the young AGN in a nearby ULIRG}",
      journal = {\mnras},
         year = 2024,
        month = may,
       volume = {530},
       number = {1},
        pages = {446-456},
          doi = {10.1093/mnras/stae810},
archivePrefix = {arXiv},
       eprint = {2403.08869},
 primaryClass = {astro-ph.GA},
       adsurl = {https://ui.adsabs.harvard.edu/abs/2024MNRAS.530..446H}
}

@ARTICLE{Pereira-Santaella21,
       author = {{Pereira-Santaella}, M. and {Colina}, L. and {Garc{\'\i}a-Burillo}, S. and {Lamperti}, I. and {Gonz{\'a}lez-Alfonso}, E. and {Perna}, M. and {Arribas}, S. and {Alonso-Herrero}, A. and {Aalto}, S. and {Combes}, F. and {Labiano}, A. and {Piqueras-L{\'o}pez}, J. and {Rigopoulou}, D. and {van der Werf}, P.},
        title = "{Physics of ULIRGs with MUSE and ALMA: The PUMA project. II. Are local ULIRGs powered by AGN? The subkiloparsec view of the 220 GHz continuum}",
      journal = {\aap},
         year = 2021,
        month = jul,
       volume = {651},
          eid = {A42},
        pages = {A42},
          doi = {10.1051/0004-6361/202140955},
archivePrefix = {arXiv},
       eprint = {2104.08238},
 primaryClass = {astro-ph.GA},
       adsurl = {https://ui.adsabs.harvard.edu/abs/2021A&A...651A..42P}
}

@ARTICLE{Lamperti22,
       author = {{Lamperti}, I. and {Pereira-Santaella}, M. and {Perna}, M. and {Colina}, L. and {Arribas}, S. and {Garc{\'\i}a-Burillo}, S. and {Gonz{\'a}lez-Alfonso}, E. and {Aalto}, S. and {Alonso-Herrero}, A. and {Combes}, F. and {Labiano}, A. and {Piqueras-L{\'o}pez}, J. and {Rigopoulou}, D. and {van der Werf}, P.},
        title = "{Physics of ULIRGs with MUSE and ALMA: The PUMA project. IV. No tight relation between cold molecular outflow rates and AGN luminosities}",
      journal = {\aap},
         year = 2022,
        month = dec,
       volume = {668},
          eid = {A45},
        pages = {A45},
          doi = {10.1051/0004-6361/202244054},
archivePrefix = {arXiv},
       eprint = {2209.03380},
 primaryClass = {astro-ph.GA},
       adsurl = {https://ui.adsabs.harvard.edu/abs/2022A&A...668A..45L}
}

@ARTICLE{Duras20,
       author = {{Duras}, F. and {Bongiorno}, A. and {Ricci}, F. and {Piconcelli}, E. and {Shankar}, F. and {Lusso}, E. and {Bianchi}, S. and {Fiore}, F. and {Maiolino}, R. and {Marconi}, A. and {Onori}, F. and {Sani}, E. and {Schneider}, R. and {Vignali}, C. and {La Franca}, F.},
        title = "{Universal bolometric corrections for active galactic nuclei over seven luminosity decades}",
      journal = {\aap},
         year = 2020,
        month = apr,
       volume = {636},
          eid = {A73},
        pages = {A73},
          doi = {10.1051/0004-6361/201936817},
archivePrefix = {arXiv},
       eprint = {2001.09984},
 primaryClass = {astro-ph.GA},
       adsurl = {https://ui.adsabs.harvard.edu/abs/2020A&A...636A..73D}
}

@ARTICLE{Skrutskie2006,
       author = {{Skrutskie}, M.~F. and {Cutri}, R.~M. and {Stiening}, R. and {Weinberg}, M.~D. and {Schneider}, S. and {Carpenter}, J.~M. and {Beichman}, C. and {Capps}, R. and {Chester}, T. and {Elias}, J. and {Huchra}, J. and {Liebert}, J. and {Lonsdale}, C. and {Monet}, D.~G. and {Price}, S. and {Seitzer}, P. and {Jarrett}, T. and {Kirkpatrick}, J.~D. and {Gizis}, J.~E. and {Howard}, E. and {Evans}, T. and {Fowler}, J. and {Fullmer}, L. and {Hurt}, R. and {Light}, R. and {Kopan}, E.~L. and {Marsh}, K.~A. and {McCallon}, H.~L. and {Tam}, R. and {Van Dyk}, S. and {Wheelock}, S.},
        title = "{The Two Micron All Sky Survey (2MASS)}",
      journal = {\aj},
         year = 2006,
        month = feb,
       volume = {131},
       number = {2},
        pages = {1163-1183},
          doi = {10.1086/498708},
       adsurl = {https://ui.adsabs.harvard.edu/abs/2006AJ....131.1163S}
}

@INPROCEEDINGS{Bertin2006,
       author = {{Bertin}, E.},
        title = "{Automatic Astrometric and Photometric Calibration with SCAMP}",
    booktitle = {Astronomical Data Analysis Software and Systems XV},
         year = 2006,
       editor = {{Gabriel}, C. and {Arviset}, C. and {Ponz}, D. and {Enrique}, S.},
       series = {Astronomical Society of the Pacific Conference Series},
       volume = {351},
        month = jul,
        pages = {112},
       adsurl = {https://ui.adsabs.harvard.edu/abs/2006ASPC..351..112B}
}

@ARTICLE{Ceci25,
       author = {{Ceci}, M. and {Cresci}, G. and {Arribas}, S. and {B{\"o}ker}, T. and {Bunker}, A.~J. and {Charlot}, S. and {Fahrion}, K. and {Lamperti}, I. and {Marconi}, A. and {Perna}, M. and {Tozzi}, G. and {Ulivi}, L.},
        title = "{The JWST/NIRSpec view of the nuclear region in the prototypical merging galaxy NGC 6240}",
      journal = {\aap},
         year = 2025,
        month = mar,
       volume = {695},
          eid = {A116},
        pages = {A116},
          doi = {10.1051/0004-6361/202452207},
archivePrefix = {arXiv},
       eprint = {2412.14685},
 primaryClass = {astro-ph.GA},
       adsurl = {https://ui.adsabs.harvard.edu/abs/2025A&A...695A.116C}
}

@ARTICLE{Rupke13,
       author = {{Rupke}, David S.~N. and {Veilleux}, Sylvain},
        title = "{Breaking the Obscuring Screen: A Resolved Molecular Outflow in a Buried QSO}",
      journal = {\apjl},
         year = 2013,
        month = sep,
       volume = {775},
       number = {1},
          eid = {L15},
        pages = {L15},
          doi = {10.1088/2041-8205/775/1/L15},
archivePrefix = {arXiv},
       eprint = {1308.4988},
 primaryClass = {astro-ph.CO},
       adsurl = {https://ui.adsabs.harvard.edu/abs/2013ApJ...775L..15R}
}

@ARTICLE{Pap23,
       author = {{Papachristou}, M. and {Dasyra}, K.~M. and {Fern{\'a}ndez-Ontiveros}, J.~A. and {Audibert}, A. and {Ruffa}, I. and {Combes}, F. and {Polkas}, M. and {Gkogkou}, A.},
        title = "{A plausible link between dynamically unsettled molecular gas and the radio jet in NGC 6328}",
      journal = {\aap},
         year = 2023,
        month = nov,
       volume = {679},
          eid = {A115},
        pages = {A115},
          doi = {10.1051/0004-6361/202346464},
archivePrefix = {arXiv},
       eprint = {2310.02033},
 primaryClass = {astro-ph.GA},
       adsurl = {https://ui.adsabs.harvard.edu/abs/2023A&A...679A.115P}
}

@ARTICLE{Bertin96,
       author = {{Bertin}, E. and {Arnouts}, S.},
        title = "{SExtractor: Software for source extraction.}",
      journal = {\aaps},
         year = 1996,
        month = jun,
       volume = {117},
        pages = {393-404},
          doi = {10.1051/aas:1996164},
       adsurl = {https://ui.adsabs.harvard.edu/abs/1996A&AS..117..393B}
}

@ARTICLE{Alexander00,
       author = {{Alexander}, P.},
        title = "{Evolutionary models for radio sources from compact sources to classical doubles}",
      journal = {\mnras},
         year = 2000,
        month = nov,
       volume = {319},
       number = {1},
        pages = {8-16},
          doi = {10.1046/j.1365-8711.2000.03711.x},
       adsurl = {https://ui.adsabs.harvard.edu/abs/2000MNRAS.319....8A}
}

@ARTICLE{Young25,
       author = {{Young}, Sophie A. and {Turner}, Ross J. and {Shabala}, Stanislav S. and {Stewart}, Georgia S.~C. and {Yates-Jones}, Patrick M.},
        title = "{Spectral signatures of young radio galaxies}",
      journal = {\pasa},
         year = 2025,
        month = jul,
       volume = {42},
          eid = {e100},
        pages = {e100},
          doi = {10.1017/pasa.2025.10068},
archivePrefix = {arXiv},
       eprint = {2412.14433},
 primaryClass = {astro-ph.GA},
       adsurl = {https://ui.adsabs.harvard.edu/abs/2025PASA...42..100Y}
}

@ARTICLE{ShabalaGodfrey13,
       author = {{Shabala}, S.~S. and {Godfrey}, L.~E.~H.},
        title = "{Size Dependence of the Radio-luminosity-Mechanical-power Correlation in Radio Galaxies}",
      journal = {\apj},
         year = 2013,
        month = jun,
       volume = {769},
       number = {2},
          eid = {129},
        pages = {129},
          doi = {10.1088/0004-637X/769/2/129},
archivePrefix = {arXiv},
       eprint = {1304.6089},
 primaryClass = {astro-ph.CO},
       adsurl = {https://ui.adsabs.harvard.edu/abs/2013ApJ...769..129S}
}

@ARTICLE{Shabala08,
       author = {{Shabala}, S.~S. and {Ash}, S. and {Alexander}, P. and {Riley}, J.~M.},
        title = "{The duty cycle of local radio galaxies}",
      journal = {\mnras},
         year = 2008,
        month = aug,
       volume = {388},
       number = {2},
        pages = {625-637},
          doi = {10.1111/j.1365-2966.2008.13459.x},
archivePrefix = {arXiv},
       eprint = {0805.4152},
 primaryClass = {astro-ph},
       adsurl = {https://ui.adsabs.harvard.edu/abs/2008MNRAS.388..625S}
}

@ARTICLE{TurnerShabala15,
       author = {{Turner}, Ross J. and {Shabala}, Stanislav S.},
        title = "{Energetics and Lifetimes of Local Radio Active Galactic Nuclei}",
      journal = {\apj},
         year = 2015,
        month = jun,
       volume = {806},
       number = {1},
          eid = {59},
        pages = {59},
          doi = {10.1088/0004-637X/806/1/59},
archivePrefix = {arXiv},
       eprint = {1504.05204},
 primaryClass = {astro-ph.GA},
       adsurl = {https://ui.adsabs.harvard.edu/abs/2015ApJ...806...59T}
}

@ARTICLE{Liddle07,
       author = {{Liddle}, Andrew R.},
        title = "{Information criteria for astrophysical model selection}",
      journal = {\mnras},
         year = 2007,
        month = may,
       volume = {377},
       number = {1},
        pages = {L74-L78},
          doi = {10.1111/j.1745-3933.2007.00306.x},
archivePrefix = {arXiv},
       eprint = {astro-ph/0701113},
 primaryClass = {astro-ph},
       adsurl = {https://ui.adsabs.harvard.edu/abs/2007MNRAS.377L..74L}
}

@ARTICLE{Leroy21,
       author = {{Leroy}, Adam K. and {Schinnerer}, Eva and {Hughes}, Annie and {Rosolowsky}, Erik and {Pety}, J{\'e}r{\^o}me and {Schruba}, Andreas and {Usero}, Antonio and {Blanc}, Guillermo A. and {Chevance}, M{\'e}lanie and {Emsellem}, Eric and {Faesi}, Christopher M. and {Herrera}, Cinthya N. and {Liu}, Daizhong and {Meidt}, Sharon E. and {Querejeta}, Miguel and {Saito}, Toshiki and {Sandstrom}, Karin M. and {Sun}, Jiayi and {Williams}, Thomas G. and {Anand}, Gagandeep S. and {Barnes}, Ashley T. and {Behrens}, Erica A. and {Belfiore}, Francesco and {Benincasa}, Samantha M. and {Be{\v{s}}li{\'c}}, Ivana and {Bigiel}, Frank and {Bolatto}, Alberto D. and {den Brok}, Jakob S. and {Cao}, Yixian and {Chandar}, Rupali and {Chastenet}, J{\'e}r{\'e}my and {Chiang}, I-Da and {Congiu}, Enrico and {Dale}, Daniel A. and {Deger}, Sinan and {Eibensteiner}, Cosima and {Egorov}, Oleg V. and {Garc{\'\i}a-Rodr{\'\i}guez}, Axel and {Glover}, Simon C.~O. and {Grasha}, Kathryn and {Henshaw}, Jonathan D. and {Ho}, I.-Ting and {Kepley}, Amanda A. and {Kim}, Jaeyeon and {Klessen}, Ralf S. and {Kreckel}, Kathryn and {Koch}, Eric W. and {Kruijssen}, J.~M. Diederik and {Larson}, Kirsten L. and {Lee}, Janice C. and {Lopez}, Laura A. and {Machado}, Josh and {Mayker}, Ness and {McElroy}, Rebecca and {Murphy}, Eric J. and {Ostriker}, Eve C. and {Pan}, Hsi-An and {Pessa}, Ismael and {Puschnig}, Johannes and {Razza}, Alessandro and {S{\'a}nchez-Bl{\'a}zquez}, Patricia and {Santoro}, Francesco and {Sardone}, Amy and {Scheuermann}, Fabian and {Sliwa}, Kazimierz and {Sormani}, Mattia C. and {Stuber}, Sophia K. and {Thilker}, David A. and {Turner}, Jordan A. and {Utomo}, Dyas and {Watkins}, Elizabeth J. and {Whitmore}, Bradley},
        title = "{PHANGS-ALMA: Arcsecond CO(2-1) Imaging of Nearby Star-forming Galaxies}",
      journal = {\apjs},
         year = 2021,
        month = dec,
       volume = {257},
       number = {2},
          eid = {43},
        pages = {43},
          doi = {10.3847/1538-4365/ac17f3},
archivePrefix = {arXiv},
       eprint = {2104.07739},
 primaryClass = {astro-ph.GA},
       adsurl = {https://ui.adsabs.harvard.edu/abs/2021ApJS..257...43L}
}

@ARTICLE{U22,
       author = {{U}, Vivian},
        title = "{The Role of AGN in Luminous Infrared Galaxies from the Multiwavelength Perspective}",
      journal = {Universe},
         year = 2022,
        month = jul,
       volume = {8},
       number = {8},
          eid = {392},
        pages = {392},
          doi = {10.3390/universe8080392},
archivePrefix = {arXiv},
       eprint = {2207.13690},
 primaryClass = {astro-ph.GA},
       adsurl = {https://ui.adsabs.harvard.edu/abs/2022Univ....8..392U}
}








\bsp	
\label{lastpage}
\end{document}